\documentclass[iop]{emulateapj}
\usepackage{lineno}
\usepackage{array}
\usepackage{booktabs}
\usepackage{tabularx}
\usepackage{ragged2e}
\usepackage{lineno}
\usepackage{appendix}
\usepackage{blindtext}
\usepackage{booktabs}
\usepackage{longtable}
\usepackage[referable]{threeparttablex}
\usepackage{wrapfig}
\usepackage{longtable,tabu}
\usepackage{booktabs}
\usepackage{amsmath}
\usepackage{longtable}
\usepackage{tabularx}
\usepackage{amsmath}
\usepackage{array}
\usepackage{booktabs}
\usepackage{ragged2e}
\usepackage{lineno}
\usepackage{graphicx}

\usepackage{natbib}
\usepackage{float}
\usepackage{tablefootnote}
\usepackage{apjfonts}
\usepackage{lipsum}
\usepackage{afterpage}

\newcommand\blfootnote[1]{%
  \begingroup
  \newcommand\thefootnote{}\footnote{#1}%
  \addtocounter{footnote}{-1}%
  \endgroup
}
\usepackage{hyperref}
\hypersetup{colorlinks=true, citecolor=blue, linkcolor=blue, urlcolor=blue}

\usepackage[T1]{fontenc}
\usepackage{beramono}
\usepackage{listings}

\DeclareUnicodeCharacter{2212}{-}
\DeclareUnicodeCharacter{0177}{\^y}
\shorttitle{TF Distances \& Dynamical Masses for 24 AGN Hosts}
\shortauthors{Robinson, et al.}

\begin{document}
\title{TULLY-FISHER DISTANCES AND Dynamical Mass Constraints for 24 Host Galaxies of Reverberation-Mapped AGN}

\author{Justin H. Robinson$^{1}$}
\author{Misty C. Bentz$^{1}$}
\author{H\'el\`ene M. Courtois$^{2}$}
\author{Megan C. Johnson$^{3}$}
\author{D. M. Crenshaw$^{1}$}
\author{Beena Meena$^{1}$}
\author{Garrett E. Polack$^{1}$}
\author{Michele L. Silverstein$^{4,5,6}$}
\author{Dading Chen$^{1}$}

\affiliation{\normalfont \centering $^{1}$Department of Physics and Astronomy, Georgia State University, Atlanta, GA 30303, USA; \url{jrob@astro.gsu.edu}}

\affiliation{\normalfont \centering $^{2}$University of Lyon; UCB Lyon 1/CNRS/IN2P3; IP2I Lyon, France}

\affiliation{\normalfont \centering $^{3}$United States Naval Observatory (USNO) 3450 Massachusetts Ave NW, Washington, DC 20392, USA}

\affiliation{\normalfont \centering $^{4}$NASA Goddard Space Flight Center, Greenbelt, MD 20771, USA}

\affiliation{\normalfont \centering $^{5}$RECONS Institute, Chambersburg, PA 17201, USA}

\altaffiliation{$^{6}$NASA Postdoctoral Program Fellow}

\begin{abstract}
We present Tully-Fisher distances for 24 AGN host galaxies with black hole mass ($M_{\textsc{BH}}$) measurements from reverberation mapping, as well as the first calibration of the $V-$band Tully-Fisher relation. Combining our measurements of HI 21\,cm emission with $HST$ and ground-based optical and near-infrared images allows multiple distance measurements for 19 galaxies and single measurements for the remaining 5. Separation of the nucleus from its host galaxy via surface brightness decomposition yields galaxy-only luminosities, thus allowing measurements of the distance moduli free of contamination from the AGN. For 14 AGN hosts, these are the first reported distances independent of redshift, and hence independent of peculiar velocities. For the remaining galaxies, we show good agreement between our distances and those previously reported from surface brightness fluctuations (SBF) and Cepheids. We also determine the total galaxy mass enclosed within the estimated HI radius, which when compared to the baryonic content allows for constraints on the dark matter masses. We find a typical mass fraction of $M_{\textsc{DM}}$/$M_{\textsc{DYN}}$ = 62\%, and find significant correlations between $M_{\textsc{BH}}$ $-$ $M_{\textsc{DYN}}$ and $M_{\textsc{BH}}$ $-$ $M_{\textsc{DM}}$. Finally, we scale our galaxy radii based on estimated relationships between visible and halo radii and assume a flat rotation curve out to the halo radius to approximate $M_{\textsc{HALO}}$. Over the range of $M_{\textsc{BH}}$ and $M_{\textsc{HALO}}$ in this sample, we find good agreement with observationally-constrained relationships between $M_{\textsc{BH}}$ and $M_{\textsc{HALO}}$ and with hydrodynamical simulations.
\end{abstract}


\received{}
\accepted{}
\section{Introduction} \label{sec:intro}
It has become apparent in the past two decades that supermassive black holes (SMBHs) and their host galaxies have a symbiotic relationship (see reviews by \citealt{kh2013} and \citealt{hb2014}), the discovery of which was unexpected given the vast difference in spatial and dynamical scales. The first indication of this was through empirical scaling relationships, for example the black hole mass $-$ bulge velocity dispersion relation ($M_{\textsc{BH}}$ $-$ $\sigma_\star$; \citealt{fm2000}, \citealt{gebhardt2000}, \citealt{kh2013}) and the black hole mass $-$ bulge luminosity relation ($M_{\textsc{BH}}$ $-$ $L_{\textsc{BULGE}}$; \citealt{kr1995,kh2013}). To further understand and explore black hole-galaxy evolution, these scaling relations also provide observational evidence to constrain the parameters of large cosmological simulations of galaxy and SMBH growth  \citep{steinborn2015,volonteri2016,mutlu2018}. The generally accepted interpretation of these scaling relations and simulations is that black hole-galaxy growth is regulated by active galactic nucleus (AGN) feedback \citep{sr1998,bower2006,ciotti2009,fanidakis2011}, and thus SMBHs play an important role in galaxy evolution. It is therefore vital to obtain accurate measurements of both galaxy and black hole characteristics for investigating these relationships.

AGN activity not only affects galaxy evolution, it also provides a mechanism for directly constraining the mass of the central black hole through reverberation mapping (RM; \citealt{bm1982,peterson1993}). Variability of the continuum emission is echoed through the variation in flux from the broad emission line region (BLR) gas. The time delay between the continuum signal and the BLR echo provides a measurement of the BLR radius ($R_{\textsc{BLR}}$), which when combined with the velocity of the BLR gas yields a constraint on the enclosed mass, or $M_{\textsc{BH}}$. In effect, RM relies on temporal resolution, as opposed to spatial resolution. Most other techniques that directly constrain $M_{\textsc{BH}}$, such as dynamical modeling, rely on spatial resolution and therefore cannot be applied beyond $\sim$100 Mpc \citep{Gltekin2009}. 

While RM measurements are effectively distance independent, they are both time consuming and resource intensive.  However, the discovery of the relationship between $R_{\textsc{BLR}}$ and the luminosity of the AGN ($R_{\textsc{BLR}}$ $-$ $L_{\textsc{AGN}}$; \citealt{kg1991,kaspi2000}), calibrated with the RM method, has led to an important shortcut for estimating SMBH masses at any redshift \citep{vp2006,Jiang2007,Kurk2007,Kurk2009,willott2010}. A single spectrum allows a measurement of $L_{\textsc{AGN}}$ to predict $R_{\textsc{BLR}}$, thus allowing $M_{\textsc{BH}}$ to be estimated without investing months or years of spectroscopic monitoring.

The largest known uncertainty in the $R_{\textsc{BLR}}$ $-$ $L_{\textsc{AGN}}$ relation is the lack of accurate distances for AGN hosts, which produces uncertainties in the derived luminosities upwards of a factor of $\sim$3 \citep{misty2013}. Only 9 galaxies in the complete sample of RM AGN systems have distances independent of redshift. Since the sample is dominated by AGNs with $z$ $\lesssim$ 0.1, the majority of distances estimated from redshift may be heavily affected by the velocity field from local gravitational interaction, rather than dominated by Hubble flow. These peculiar velocities ($V_{\textsc{PEC}}$) have been observed to be upwards of 500 km s$^{-1}$ in the local universe \citep{tully2008,tully2013}, which causes significant uncertainty in distances derived solely from spectroscopic redshift. Measurements of galaxy properties that directly depend on distance (e.g., luminosity) and are relied on for SMBH-galaxy scaling relationships are thus hindered by $z$-based distances. The reverberation sample serves as the basis for all secondary $M_{\textsc{BH}}$ estimates in distant AGNs, so accurate distance determinations are crucial for a large number of RM AGN hosts.

Accurate distance measurements to galaxies in the local universe have allowed tight construction of what is commonly referred to as the cosmic distance ladder. The methods that set the scale for nearly all extragalactic rungs of the ladder are Leavitt's Law (Cepheid period-luminosity relationship; \citealt{lp1912}) and the tip of the red giant branch (TRGB; \citealt{iben1983,salaris1997}). Both of these methods require individual stars to be resolved, demanding high resolving power only achievable by a few instruments, thus limiting their reach to $\sim$40 Mpc \citep{riess2016}. The surface brightness fluctuation (SBF; \citealt{tonry2001,Blakeslee2010}) and Fundamental Plane \citep{fj1976,dd1987,dressler1987} methods are only applicable to early-type galaxies, but most of the optically bright AGNs in the local universe are hosted by spirals.

One of the most widely utilized distance methods for disk galaxies is the Tully$-$Fisher (TF) relation \citep{tf1977}. The TF relation is an empirical correlation between the rotational velocity of a late-type galaxy and its brightness. A galaxy's mass is constrained by its rotation rate, and its luminosity traces the mass. Therefore, measurement of the maximum rotation rate yields the absolute magnitude of the galaxy, and the difference between the absolute and apparent magnitudes constrains the distance. 21\,cm emission from neutral hydrogen (HI), due to its overwhelming abundance in late-type galaxies and extension far past the stellar disk, is usually used to trace disk velocity. Resolved HI surveys (e.g., \citealt{walter2008,ott2012,koribalski2018,av2019,Koribalski2020,Maddox2020}) provide the most precise method for constraining maximum rotation rate ($V_{mx}$) through rotation curve analysis. For large samples of galaxies, the rotational broadening of unresolved, integrated HI emission may be used as a proxy for a rotation curve \citep{epstein1964,roberts1969}. The TF relation has traditionally been limited to $z \lesssim 0.1$ \citep{tully2008,tully2013}, even though HI has increasingly been detected at higher redshift \citep{jaffe2013, cc2015, fernandez2016}. Galaxies with $z \gtrsim 0.1$ are generally assumed to be within the Hubble flow.

We thus undertook a program to measure TF distances for as many RM AGN hosts as possible. In this paper, we use the HI emission lines from \citeauthor{me} (2019; hereafter Paper I) and present TF distance determinations for 24 Seyfert 1 galaxies. In Section \ref{sec:data}, we briefly discuss the HI spectral observations (see Paper I) and describe the imaging of our sample. In Section \ref{sec:modeling}, we describe the two-dimensional surface brightness modeling and separation of the central AGN from the host galaxy in the images. In Section \ref{measurements}, we describe the measurement methods for the calibrated galaxy magnitudes and HI 21\,cm line widths. In Section \ref{distances}, we detail our distance calculations and present the first calibration of the TF relation for the Johnson $V$ band. In Section \ref{sec:masses}, we report derived values of galaxy dynamical mass ($M_{\textsc{DYN}}$) and dark matter mass ($M_{\textsc{DM}}$) within the HI radius. Finally, in Section \ref{sec:discussion}, we explore the relationships between $M_{\textsc{DYN}}$ and $M_{\textsc{DM}}$ with $M_{\textsc{BH}}$.

Throughout this work we adopt a $\Lambda$CDM cosmology of H$_0 = 74$\,km\,s$^{-1}$\,Mpc$^{-1}$ \citep{riess2019}, $\Omega_{\textsc{M}}$=0.27, and $\Omega_{\normalfont{\Lambda}}$=0.73 \citep{bennett2014}.

\section{Spectroscopy and Imaging} \label{sec:data}
In Paper I, we presented HI spectroscopy of 31 AGN hosts with direct $M_{\textsc{BH}}$ measurements from the RM database of \cite{bhdatabase}, and constraints on gas mass ($M_{\textsc{GAS}}$) and total baryonic mass ($M_{\textsc{BARY}}$). Here we briefly summarize the 21\,cm observations and describe the optical and near-infrared imaging of the AGN hosts.

\subsection{HI 21\,cm Spectra}\label{radio}
Spectroscopy of the HI 21cm emission lines was acquired in 2013 (GBT13A-468; PI: Ou-Yang) and 2018/19 (GBT18B-258; PI: Robinson) with the Robert C. Byrd Green Bank Telescope\footnote{The Green Bank Observatory is a facility of the National Science Foundation operated under cooperative agreement by Associated Universities, Inc.} (GBT). The observational setups and instrument selections are detailed in Paper I. In brief, both data sets were observed in position-switched mode with equal on-off exposure pairs and typical exposures of 60\,s scans for GBT13A-468 and 120\,s scans for GBT18B-258. All scans were broken into 3\,s integrations to aid in radio frequency interference removal.  

Spectral reduction was carried out with the GBTIDL suite \citep{gbtidl} v2.8 for GBT13A-468 and v2.10.1 for GBT18B-258. Each on-off pair was combined with the standard (ON–OFF)/OFF procedure, and all exposures for one source were accumulated and averaged into a single spectrum. Low-order polynomials were fit to and subtracted from the baselines before spectral measurements were conducted.

While we detected HI emission lines from 31 of the 44 AGN host galaxies that were observed, we limit the analysis here to the 24 galaxies that exhibit a rotationally broadened dual-horned profile shape, as this is needed to recover the disk velocity information for use in TF distance determinations. Additionally, we note that although we limit this sample to dual-horned profiles, the galaxy inclinations tend to be oriented more face-on ($< 45^{\circ}$) than the typical galaxies targeted for TF-based distances. The 24 galaxies are listed in Table \ref{obs_setup}.

\subsection{Optical and Near-Infrared Imaging}\label{optical}
Optical and near-infrared images of the AGN host galaxies have been compiled from several observatories, with the goal of separating the AGN contribution from the galaxy via 2-dimensional surface brightness decomposition (described in Sec.\ \ref{sec:modeling}). For all ground-based data, images were reduced and combined in IRAF\footnote{IRAF is distributed by the National Optical Astronomy Observatory, which is operated by the Association of Universities for Research in Astronomy (AURA) under a cooperative agreement with the National Science Foundation.} following standard procedures.

\subsubsection{Previous Observations}
The majority of images used in this analysis were collected between 2003-2016 and have been previously described in publications. Table \ref{obs_setup} lists observation dates, instrument setups, and exposure times.

\textbf{\emph{HST}}: All 24 of our targets have medium-band $V$ images obtained with the $Hubble$ $Space$ $Telescope$ ($HST$). Galaxies were observed with either the Advanced Camera for Surveys (ACS) High Resolution Channel (HRC) through the F550M filter or the Wide Field Camera 3 (WFC3) through the F547M filter. The HRC has a field of view (FOV) of 29$''$ x 26$''$ and pixel scale of 0$\farcs$025 pixel$^{-1}$. The WFC3 observations utilized the UVIS channel, which has a 160$''$ x 160$''$ FOV and pixel scale of 0$\farcs$04 pixel$^{-1}$. In-depth descriptions of the $HST$ observations and reductions are available in \cite{misty2009a, misty2013, misty2018}.

\begin{longtable*}{@{\extracolsep{\fill}}llrlrrl@{}}
\caption{\\ Optical and Near-IR Imaging}\\
\toprule
\multicolumn{1}{c}{Target} & \multicolumn{1}{c}{RA}  & \multicolumn{1}{c}{Dec} & \multicolumn{1}{c}{$z$} &  \multicolumn{1}{c}{Date} & 
\multicolumn{1}{c}{Exp Time} &
\multicolumn{1}{c}{Obs Setup}\\
& \multicolumn{1}{c}{(hh mm ss.s)} & \multicolumn{1}{c}{(dd mm ss)} & &  \multicolumn{1}{c}{(yyyy-mm-dd)} & \multicolumn{1}{c}{(s)} & \\ 
\midrule
\endhead\hline
\endfoot 
\bottomrule 
\endlastfoot
Mrk 1044 & 02 30 05.5 & -08 59 53 & 0.01645 & 2011 Jan 10 & 725.0 & HST WFC3 F547M\\
& & & & 2019 Dec 20 & 810.0 & APO 3.5\,m ARCTIC B\\
& & & & 2019 Dec 20 & 240.0 & APO 3.5\,m ARCTIC V\\
& & & & 2019 Sep 28 & 90.0 & APO 3.5\,m ARCTIC R\\
& & & & 2019 Sep 28 & 60.0 & APO 3.5\,m ARCTIC I\\
Ark 120 & 05 16 11.4 & -00 08 66 & 0.03271 & 2006 Oct 30 & 2040.0 & HST ACS HRC F550M\\
MCG+08-11-011 & 05 54 53.6 & +46 26 22 & 0.02048 & 2016 May 21 & 2370.0 & HST WFC3 F547M\\
Mrk 6 &  06 52 12.2 & +74 25 37 & 0.01881 & 2014 Nov 06 & 2620.0 & HST WFC3 F547M\\
& & & & 2020 Feb 12 & 60.0 & APO 3.5\,m ARCTIC V\\
& & & & 2012 Jan 13 & 720.0 & WIYN WHIRC H\\
Mrk 374 & 06 59 38.1 & +54 11 48 & 0.04263 & 2016 May 21 & 2420.0 & HST WFC3 F547M\\
Mrk 79 & 07 42 32.8 & +49 48 35 & 0.02219 & 2006 Nov 08 & 2040.0 & HST ACS HRC F550M\\
& & & & 2003 Aug 12 & 1500.0 & MDM 1.3\,m Templeton B\\
& & & & 2003 Aug 12 & 1965.0 & MDM 1.3\,m Templeton V\\
& & & & 2003 Aug 12 & 1110.0 & MDM 1.3\,m Templeton R\\
& & & & 2012 Jan 13 & 4140.0 & WIYN WHIRC H\\
NGC 2617 & 08 35 38.8 & -04 05 18 & 0.01421 & 2016 May 21 & 2230.0 & HST WFC3 F547M\\
& & & & 2020 Feb 17 & 1020.0 & APO 3.5\,m ARCTIC B\\
& & & & 2020 Feb 17 & 840.0 & APO 3.5\,m ARCTIC V\\
& & & & 2020 Feb 17 & 450.0 & APO 3.5\,m ARCTIC R\\
NGC 3227 & 10 23 30.6 & +19 51 54 & 0.00386 & 2010 Mar 29 & 2250.0 & HST WFC3 F547M\\
& & & & 2003 Aug 12 & 1800.0 & MDM 1.3\,m Templeton B\\
& & & & 2003 Aug 12 & 2280.0 & MDM 1.3\,m Templeton V\\
& & & & 2003 Aug 12 & 1575.0 & MDM 1.3\,m Templeton R\\
& & & & 2013 Apr 26 & 1470.0 & WIYN WHIRC H\\
SBS 1116+583A & 11 18 57.7 & +58 03 24 & 0.02787 & 2010 Jun 06 & 2510.0 & HST WFC3 F547M\\
NGC 3783 & 11 39 01.7 & -37 44 19 & 0.00973 & 2011 Feb 09 & 2300.0 & HST WFC3 F547M\\
& & & & 2015 Apr 23 & 960.0 & SMARTS 0.9\,m Tek2K B\\
& & & & 2015 Apr 23 & 960.0 & SMARTS 0.9\,m Tek2K V\\
& & & & 2015 Apr 23 & 960.0 & SMARTS 0.9\,m Tek2K R\\
Mrk 1310 & 12 01 14.3  & -03 40 41 & 0.01956 & 2009 Dec 02 & 2240.0 & HST WFC3 F547M\\
& & & & 2015 Apr 23 & 960.0 & SMARTS 0.9\,m Tek2K B\\
& & & & 2015 Apr 23 & 960.0 & SMARTS 0.9\,m Tek2K V\\
& & & & 2015 Apr 23 & 960.0 & SMARTS 0.9\,m Tek2K R\\
& & & & 2013 Apr 27 & 4500.0 & WIYN WHIRC H\\
NGC 4051 & 12 03 09.6 & +44 31 53 & 0.00234 & 2010 Jul 17 & 2340.0 & HST WFC3 F547M\\
& & & & 2003 Aug 12 & 1250.0 & MDM 1.3\,m Templeton B\\
& & & & 2003 Aug 12 & 795.0 & MDM 1.3\,m Templeton V\\
& & & & 2003 Aug 12 & 690.0 & MDM 1.3\,m Templeton R\\
& & & & 2013 Apr 26 & 3060.0 & WIYN WHIRC H\\
NGC 4151 & 12 10 32.6 & +39 24 19&0.00332 & 2010 Jul 03 & 2310.0 & HST WFC3 F547M\\
& & & & 2003 Aug 12 & 1470.0 & MDM 1.3\,m Templeton B\\
& & & & 2003 Aug 12 & 1200.0 & MDM 1.3\,m Templeton V\\
& & & & 2003 Aug 12 & 1370.0 & MDM 1.3\,m Templeton R\\
& & & & 2019 May 26 & 1800.0 & ARCSAT SurveyCam I\\
& & & & 2013 Apr 27 & 1005.0 & WIYN WHIRC H\\
NGC 4593 & 12 39 39.4 & -05 20 39&0.00900 & 2010 Jul 10 & 2240.0 & HST WFC3 F547M\\
& & & & 2003 Aug 12 & 1650.0 & MDM 1.3\,m Templeton B\\
& & & & 2003 Aug 12 & 1860.0 & MDM 1.3\,m Templeton V\\
& & & & 2003 Aug 12 & 1380.0 & MDM 1.3\,m Templeton R\\
& & & & 2013 Apr 27 & 960.0 & WIYN WHIRC H\\
NGC 4748 & 12 52 12.4 & -13 24 53 &0.01463 & 2010 Jun 28 & 2250.0 & HST WFC3 F547M\\
& & & & 2015 Apr 23 & 960.0 & SMARTS 0.9\,m Tek2K B\\
& & & & 2015 Apr 23 & 960.0 & SMARTS 0.9\,m Tek2K V\\
& & & & 2015 Apr 23 & 960.0 & SMARTS 0.9\,m Tek2K R\\
& & & & 2013 Apr 27 & 3600.0 & WIYN WHIRC H\\
NGC 5548 & 14 17 59.5 & +25 08 12 &0.01718 & 2010 Jul 15 & 2260.0 & HST WFC3 F547M\\
& & & & 2003 Nov 17 & 4000.0 & MDM 1.3\,m Templeton B\\
& & & & 2003 Nov 17 & 2500.0 & MDM 1.3\,m Templeton V\\
& & & & 2003 Nov 17 & 1380.0 & MDM 1.3\,m Templeton R\\
Mrk 817 & 14 36 22.1 & +58 47 39 &0.03146 & 2003 Dec 08 & 1020.0 & HST ACS HRC F550M\\
& & & & 2003 Nov 04 & 2530.0 & MDM 1.3\,m Templeton B\\
& & & & 2003 Nov 04 & 1530.0 & MDM 1.3\,m Templeton V\\
& & & & 2003 Nov 04 & 930.0 & MDM 1.3\,m Templeton R\\
& & & & 2013 Apr 26 & 3520.0 & WIYN WHIRC H\\
NGC 5940 & 15 31 18.1 & +07 27 28 & 0.03393 & 2016 May 28 & 2230.0 & HST WFC3 F547M\\
Mrk 290 & 15 35 52.3 & +57 54 09 &0.02958 & 2010 Jul 25 & 2520.0 & HST WFC3 F547M\\
Zw 229-015 & 19 05 25.9 & +42 27 40 & 0.02788 & 2014 Nov 13 & 2320.0 & HST WFC3 F547M\\
1H1934-063 & 19 37 33.0 & -06 13 05 &0.01031 & 2016 May 27 & 2230.0 & HST WFC3 F547M\\
& & & & 2015 Apr 23 & 960.0 & SMARTS 0.9\,m Tek2K B\\
& & & & 2015 Apr 23 & 960.0 & SMARTS 0.9\,m Tek2K V\\
& & & & 2015 Apr 23 & 960.0 & SMARTS 0.9\,m Tek2K R\\
NGC 6814 & 19 42 40.6 & -10 19 25 & 0.00521 & 2010 May 06 & 2240.0 & HST WFC3 F547M\\
& & & & 2019 Jun 23 & 1800.0 & ARCSAT SurveyCam B\\
& & & & 2019 Jun 23 & 1800.0 & ARCSAT SurveyCam V\\
& & & & 2019 Jun 23 & 1800.0 & ARCSAT SurveyCam R\\
& & & & 2019 Jun 23 & 1800.0 & ARCSAT SurveyCam I\\
& & & & 2011 Sep 20 & 1200.0 & WIYN WHIRC H\\
NGC 7469 & 23 03 15.6 & +08 52 26 &0.01632 & 2009 Nov 11 & 2240.0 & HST WFC3 F547M\\
& & & & 2003 Nov 13 & 1440.0 & MDM 1.3\,m Templeton B\\
& & & & 2003 Nov 13 & 1260.0 & MDM 1.3\,m Templeton V\\
& & & & 2003 Nov 13 & 1260.0 & MDM 1.3\,m Templeton R\\
& & & & 2019 Sep 28 & 70.0 & APO 3.5\,m ARCTIC I\\
& & & & 2011 Sep 19 & 300.0 & WIYN WHIRC H\\
\label{obs_setup}
\end{longtable*}

\textbf{WIYN}: Eleven galaxies in our sample were imaged with the 3.5\,m WIYN telescope\footnote{The WIYN Observatory is a joint facility of the University of WisconsinMadison, Indiana University, the National Optical Astronomy Observatory and the University of Missouri.} (NOAO 2011B-0120; PI: Bentz; NOAO 2013A-0438; PI: Manne-Nicholas). The observations employed the High-Resolution Infrared Camera (WHIRC) and the $H$ filter. WHIRC is a 2048 x 2048 Raytheon Virgo HgCdTe, with a FOV of 202$''$ x 202$''$ and a pixel scale of 0$\farcs$0986 pixel$^{-1}$. Each target was observed with multiple short exposures and large dither patterns to aid in the removal of strong fringing, cosmic rays, and bad pixels. A full description of the observations and reduction is available in \cite{misty2018}.

\textbf{MDM}: Eight galaxies were observed with the MDM Observatory 1.3\,m McGraw-Hill Telescope. As described by \cite{misty2009b}, objects were observed through Johnson  $B$, $V$, and $R$ filters with the Templeton CCD, which has 1024 x 1024 pixels, a pixel scale of 0$\farcs$5 pixel$^{-1}$, and FOV of 8$\farcm$49 x 8$\farcm$49. 

\subsubsection{New Observations}
New imaging of galaxies in our sample was conducted between 2015-2020. Dates, instrument selections, and exposure times are listed in Table \ref{obs_setup}.

\textbf{APO}: We obtained $B$, $V$, $R$, and $I-$band images of 4 galaxies in our sample with the 3.5\,m Apache Point Observatory (APO) Astrophysical Research Consortium (ARC) telescope. The ARC Telescope Imaging Camera (ARCTIC) on the 3.5\,m is a 4096 x 4096 pixel CCD with an FOV of 7$\farcm$85 x 7$\farcm$85 and a pixel scale of 0$\farcs$114 pixel$^{-1}$. Because unbinned imaging tends to over sample the seeing, we employed 2 x 2 binning for all objects imaged by ARCTIC, which yields an effective pixel scale of 0$\farcs$228 pixel$^{-1}$.

\textbf{SMARTS}: Optical images of 4 galaxies were obtained with the Cerro Tololo Inter-American Observatory/Small and Moderate Aperture Research Telescope System (CTIO/SMARTS) 0.9\,m telescope. The 0.9\,m utilizes a 2048 x 2048 pixel CCD with a FOV of 13$\farcm$6 x 13$\farcm$6 and pixel scale of 0$\farcs$401 pixel$^{-1}$. Targets were observed through Tek 2K Set II Johnson-Cousins $B$, $V$, and $R$ filters.

\textbf{ARCSAT}: Using the 0.5\,m ARC Small Aperture Telescope (ARCSAT), we targeted galaxies which had not been observed with APO, MDM, or SMARTS or did not have $I-$band imaging. ARCSAT employs the SurveyCam CCD imager, which has 4096 x 4096 pixels, a FOV of 31$\farcm$1 x 31$\farcm$1, and pixel scale of 	0$\farcs$465 pixel$^{-1}$. For most observations, the large pixels and low spatial resolution caused the AGN and bulge to blend together in most images, eliminating the possibility of separation in the modeling process (see Sec.\ \ref{sec:modeling}). The images for only 2 galaxies were found to be useful for surface brightness modeling; NGC 4151 and NGC 6814 were observed with Cousins $I$ for the former and Johnson-Cousins $B$, $V$, $R$, and $I$ for the latter. Dates and exposure times for these 2 galaxies are listed in Table \ref{obs_setup}.

\begin{figure*}
\gridline{
\includegraphics[trim={8.1cm 4.95cm 6.02cm 1.72cm},clip,scale=0.37]{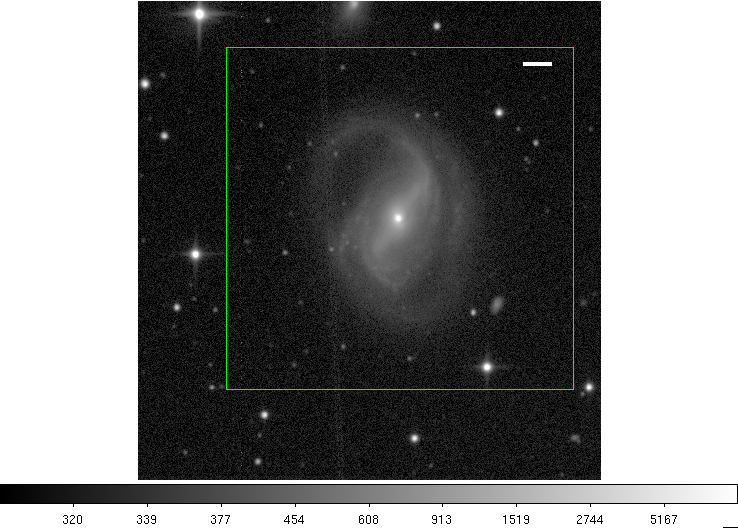}
\includegraphics[trim={5.1cm 2cm 5.1cm 0.9cm},clip,scale=0.281]{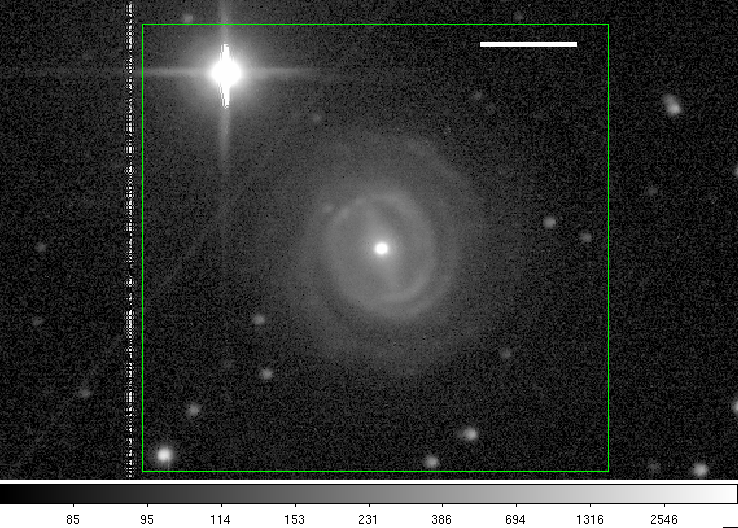}
\includegraphics[trim={5.615cm 5.935cm 8.185cm 0.49cm},clip,scale=0.3625]{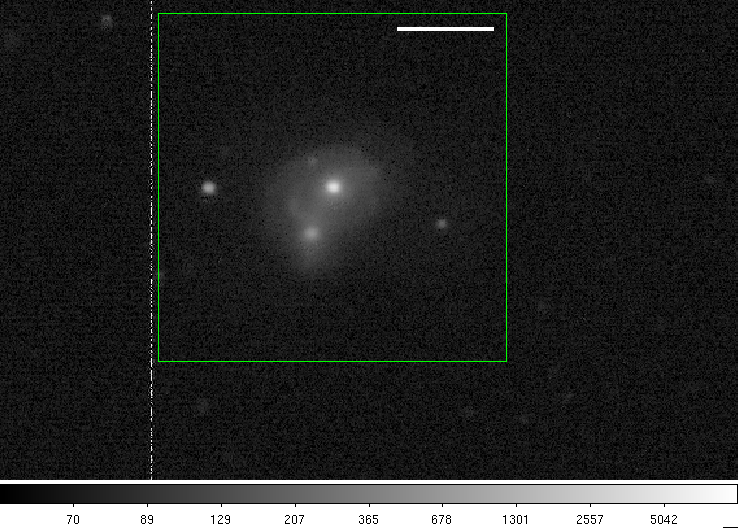}
\includegraphics[trim={11.55cm 3.05cm 6.20cm 7.3cm},clip,scale=0.534]{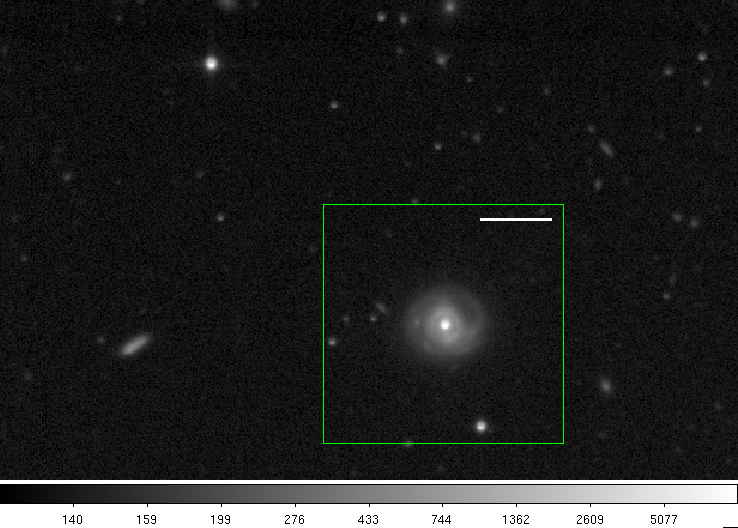}
}\vspace{-0.4cm}
\gridline{
\includegraphics[trim={8.1cm 4.95cm 6.02cm 1.72cm},clip,scale=0.37]{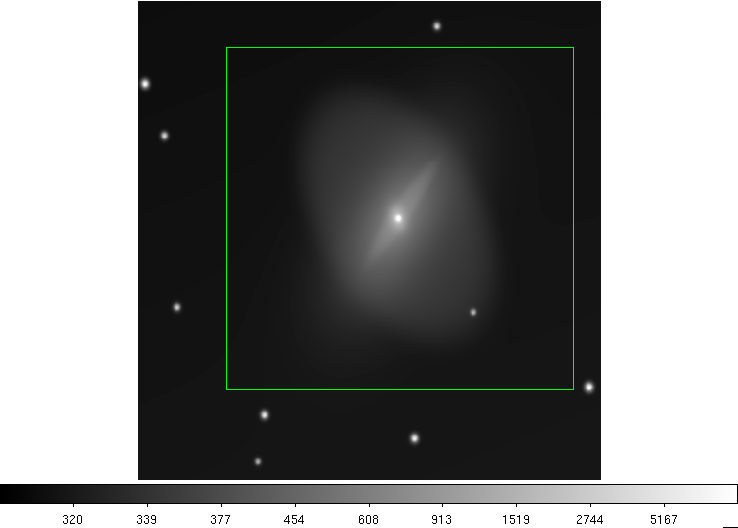}
\includegraphics[trim={5.1cm 2.05cm 5.1cm 0.9cm},clip,scale=0.282]{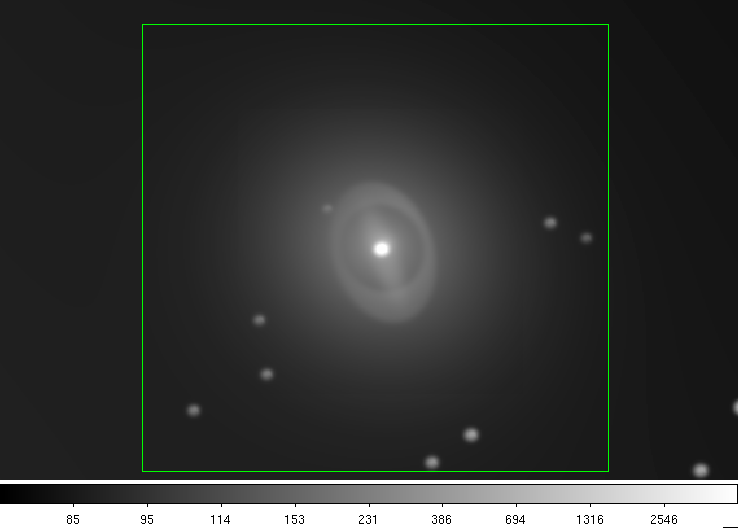}
\includegraphics[trim={5.615cm 5.935cm 8.185cm 0.49cm},clip,scale=0.3625]{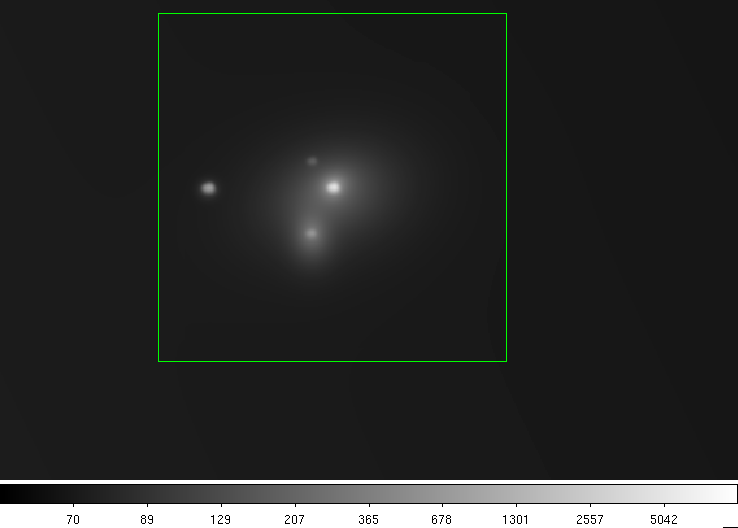}
\includegraphics[trim={11.55cm 3.05cm 6.20cm 7.3cm},clip,scale=0.534]{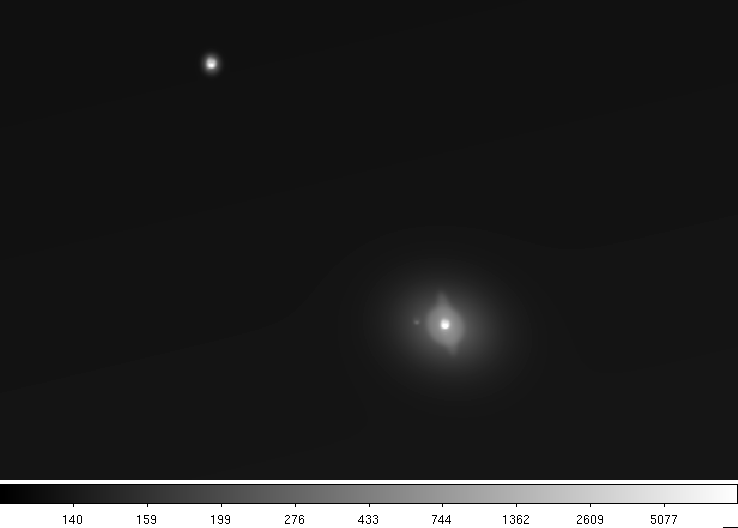}
}\vspace{-0.4cm}
\gridline{
\includegraphics[trim={8.1cm 4.95cm 6.02cm 1.72cm},clip,scale=0.37]{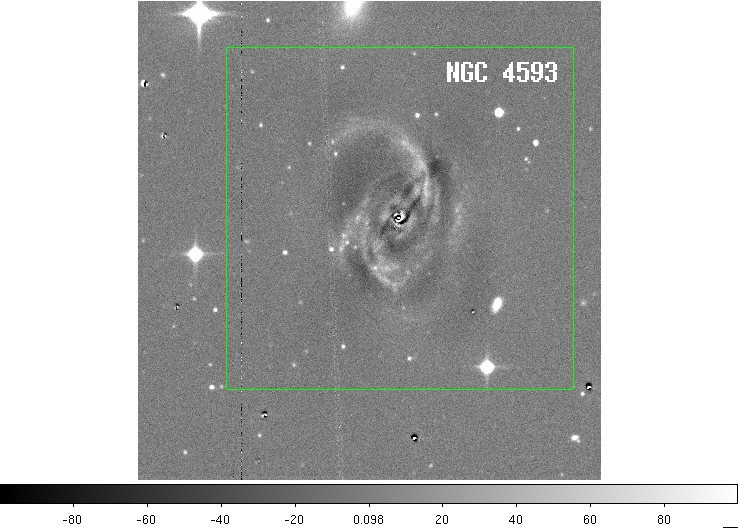}
\includegraphics[trim={9.1cm 6cm 9.11cm 4.9cm},clip,scale=0.572]{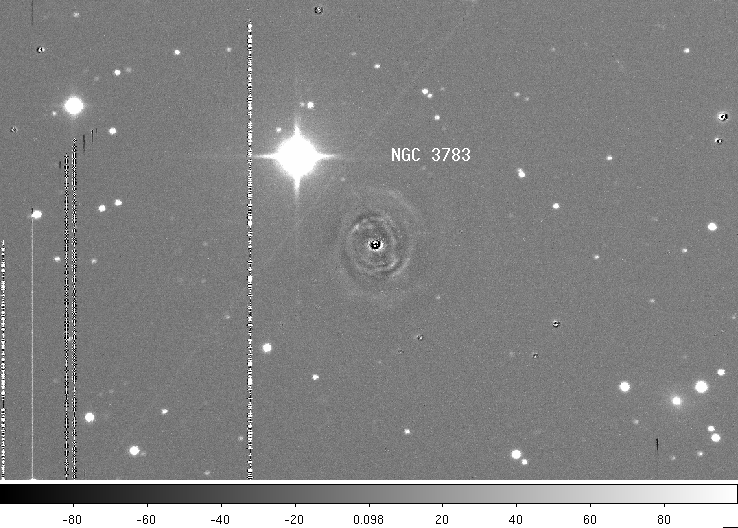}
\includegraphics[trim={5.615cm 5.935cm 8.19cm 0.49cm},clip,scale=0.3625]{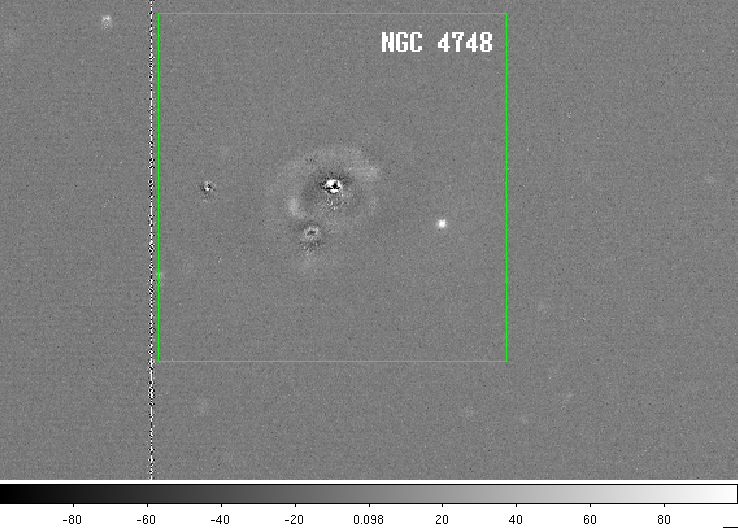}
\includegraphics[trim={11.55cm 3.05cm 6.20cm 7.3cm},clip,scale=0.534]{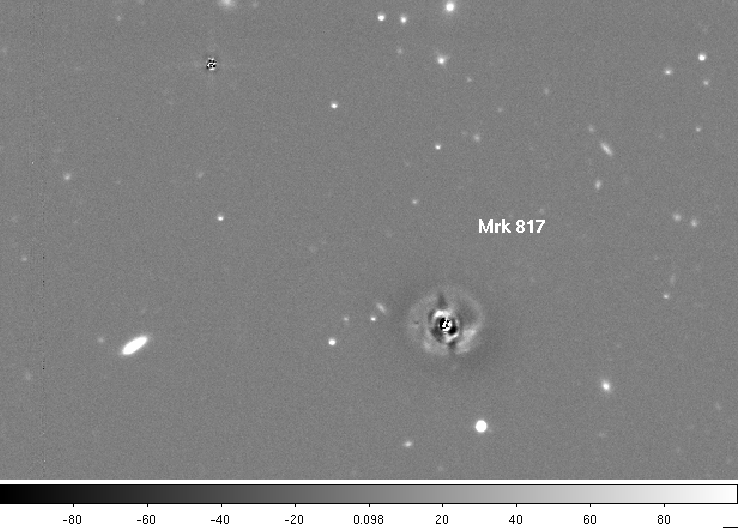}
}
\caption{Selected ground-based $B-$band AGN host galaxy images (top), $\textsc{Galfit}$ models (center), and residuals (bottom). From left to right: NGC 4593, NGC 3783, NGC 4748, Mrk 817. The images and models are displayed with a logarithmic stretch, and the residuals are displayed with a linear stretch centered around zero counts. The scale bars in each row are 30$''$ in length. Due to the varying levels of compactness of each galaxy, the selected fits correspond to the quality range of the separation of the AGN and galaxy light, from good (NGC 4593, NGC 3783), to moderate (NGC 4748), to poor (Mrk 817). Higher uncertainties were assigned to galaxies with poor fits (see Sec.\ \ref{quality_explanation} for a quantitative explanation). Magnitudes were calculated as $m = -2.5 \log (counts/s) + zpt$.}\label{galfit_models}
\end{figure*} 

\section{Surface Brightness Modeling}\label{sec:modeling}
As demonstrated by \cite{misty2013}, up to 30\% of the total galaxy brightness of nearby active galaxies may be contributed by a central AGN. The TF relation uses galaxy light as a mass tracer, thus accurate measurements of the galaxy starlight (which traces the stellar mass) are needed. In order to remove the AGN contamination from the host galaxy brightness, we conducted two-dimensional surface brightness decompositions using $\textsc{Galfit}$ \citep{peng2002, peng2010}. For the vast majority of our targets, the decompositions of the $V-$band $HST$ images have already been published \citep{misty2009a,misty2013,misty2018}. We follow a similar procedure here in the modeling of the new ground-based images. 

\begin{deluxetable*}{lrrrrrrrl}
\tablewidth{18cm} 
\tablecaption{Galaxy Surface Brightness Parameters}
\tablehead{
\colhead{Target} & \colhead{$V_{galaxy}$ (HST)} & \colhead{$B_{galaxy}$} & \colhead{$V_{galaxy}$} & \colhead{$R_{galaxy}$} & \colhead{$I_{galaxy}$} & \colhead{$H_{-0.5,galaxy}$}  & \colhead{\textit{Disk Radius}}  & \colhead{$q_{d}$}  \\
& \colhead{(mag)} & \colhead{(mag)} & \colhead{(mag)} & \colhead{(mag)} & \colhead{(mag)} & \colhead{(mag)} & \colhead{(arcsec)} &    \\
\colhead{(1)} & \colhead{(2)} &\colhead{(3)} & \colhead{(4)} & \colhead{(5)} & \colhead{(6)} & \colhead{(7)} & \colhead{(8)} & \colhead{(9)} \\
}
\startdata
Mrk 1044 & 14.20 $\pm$ 0.20 & \nodata & 13.96 $\pm$ 0.20 & 13.71 $\pm$ 0.20 & 12.76 $\pm$ 0.20 & \nodata & 12.09 $\pm$ 0.78 $^{[ a ]}$ & 0.87 \\
Ark 120 & 14.38 $\pm$ 0.20 & \nodata & \nodata & \nodata & \nodata & 11.85 $\pm$ 0.20 & 3.97 $\pm$ 0.56 $^{[ b ]}$ & 0.81 \\
MCG+08-11-011 & 11.80 $\pm$ 0.20 & \nodata & \nodata & \nodata & \nodata & \nodata & 56.84 $\pm$ 10.15 $^{[ a ]}$ & 0.62 \\
Mrk 6 & 13.97 $\pm$ 0.20 & \nodata & 13.74 $\pm$ 0.20 & \nodata & \nodata & 10.92 $\pm$ 0.20 & 29.45 $\pm$ 6.11 $^{[ a ]}$ & 0.61 \\
Mrk 374 & 14.34 $\pm$ 0.20 & \nodata & \nodata & \nodata & \nodata & \nodata & 8.47 $\pm$ 0.85 $^{[ b ]}$ & 0.58 \\
Mrk 79 & 14.04 $\pm$ 0.20 & 14.22 $\pm$ 0.30 & 13.75 $\pm$ 0.20 & 13.25 $\pm$ 0.20 & \nodata & 11.24 $\pm$ 0.20 & 31.68 $\pm$ 3.22 $^{[ a ]}$ & 0.79 \\
NGC 2617 & 12.66 $\pm$ 0.20 & 14.10 $\pm$ 0.30 & 12.80 $\pm$ 0.20 & 12.51 $\pm$ 0.20 & \nodata & \nodata & 23.80 $\pm$ 8.61 $^{[ a ]}$ & 0.97 \\
NGC 3227 & 11.00 $\pm$ 0.20 & 11.85 $\pm$ 0.20 & 10.87 $\pm$ 0.20 & 10.48 $\pm$ 0.20 & \nodata & 8.22 $\pm$ 0.20 & 91.57 $\pm$ 18.72 $^{[ a ]}$ & 0.42 \\
SBS1116+583A & 15.57 $\pm$ 0.20 & \nodata & \nodata & \nodata & \nodata & 13.65 $\pm$ 0.20 & 3.14 $\pm$ 0.44 $^{[ b ]}$ & 0.87 \\
NGC 3783 & 12.09 $\pm$ 0.20 & 12.89 $\pm$ 0.20 & 12.03 $\pm$ 0.20 & 11.50 $\pm$ 0.20 & \nodata & \nodata & 58.39 $\pm$ 6.58 $^{[ a ]}$ & 0.96 \\
Mrk 1310 & 14.93 $\pm$ 0.20 & \nodata & 15.02 $\pm$ 0.20 & 14.40 $\pm$ 0.20 & \nodata & 12.47 $\pm$ 0.20 & 11.67 $\pm$ 1.19 $^{[ a ]}$ & 0.73 \\
NGC 4051 & 10.11 $\pm$ 0.20 & 10.84 $\pm$ 0.20 & 10.21 $\pm$ 0.20 & 9.93 $\pm$ 0.20 & \nodata & 8.57 $\pm$ 0.20 & 127.67 $\pm$ 7.22 $^{[ a ]}$ & 0.58$^{[ c ]}$ \\
NGC 4151 & 10.80 $\pm$ 0.20 & 11.29 $\pm$ 0.20 & 10.63 $\pm$ 0.20 & 10.18 $\pm$ 0.20 & 9.73 $\pm$ 0.20 & 9.01 $\pm$ 0.20 & 64.12 $\pm$ 6.00 $^{[ a ]}$ & 0.94$^{[ d ]}$ \\
NGC 4593 & 11.21 $\pm$ 0.20 & 12.05 $\pm$ 0.20 & 11.21 $\pm$ 0.20 & 10.70 $\pm$ 0.20 & \nodata & 9.46 $\pm$ 0.20 & 89.44 $\pm$ 19.00 $^{[ a ]}$ & 0.69$^{[ c ]}$ \\
NGC 4748 & 13.56 $\pm$ 0.20 & 14.43 $\pm$ 0.30 & 13.19 $\pm$ 0.30 & 12.54 $\pm$ 0.20 & \nodata & 10.94 $\pm$ 0.20 & 24.46 $\pm$ 5.46 $^{[ a ]}$ & 0.69 \\
NGC 5548 & 12.72 $\pm$ 0.20 & 13.39 $\pm$ 0.20 & 12.58 $\pm$ 0.20 & 12.24 $\pm$ 0.20 & \nodata & \nodata & 36.03 $\pm$ 3.35 $^{[ a ]}$ & 0.83 \\
Mrk 817 & 14.30 $\pm$ 0.20 & 15.04 $\pm$ 0.30 & 13.99 $\pm$ 0.30 & 13.58 $\pm$ 0.30 & \nodata & 11.76 $\pm$ 0.20 & 13.03 $\pm$ 2.12 $^{[ a ]}$ & 0.81 \\
Mrk 478 & 15.54 $\pm$ 0.20 & \nodata & \nodata & \nodata & \nodata & \nodata & 3.39 $\pm$ 0.34 $^{[ b ]}$ & 0.85 \\
NGC 5940 & 13.35 $\pm$ 0.20 & \nodata & \nodata & \nodata & \nodata & \nodata & 9.21 $\pm$ 0.92 $^{[ b ]}$ & 0.86 \\
Mrk 290 & 15.31 $\pm$ 0.20 & \nodata & \nodata & \nodata & \nodata & \nodata & 4.12 $\pm$ 0.41 $^{[ b ]}$ & 0.81 \\
Zw 229-015 & 15.08 $\pm$ 0.20 & \nodata & \nodata & \nodata & \nodata & 12.77 $\pm$ 0.20 & 9.31 $\pm$ 1.33 $^{[ b ]}$ & 0.60 \\
1H1934-063 & 13.16 $\pm$ 0.20 & 14.47 $\pm$ 0.20 & 13.59 $\pm$ 0.20 & 13.08 $\pm$ 0.20 & \nodata & \nodata & 22.90 $\pm$ 3.12 $^{[ a ]}$ & 0.85 \\
NGC 6814 & 11.18 $\pm$ 0.20 & 12.16 $\pm$ 0.30 & 11.22 $\pm$ 0.20 & 10.62 $\pm$ 0.20 & 9.91 $\pm$ 0.20 & 9.80 $\pm$ 0.20 & 66.91 $\pm$ 15.75 $^{[ a ]}$ & 0.98 \\
NGC 7469 & 12.51 $\pm$ 0.20 & 13.05 $\pm$ 0.20 & 12.29 $\pm$ 0.20 & 12.10 $\pm$ 0.20 & 11.12 $\pm$ 0.20 & 9.84 $\pm$ 0.20 & 33.44 $\pm$ 2.03 $^{[ a ]}$ & 0.81 \\
\enddata
\tablecomments{Total calibrated galaxy magnitudes after removal of AGN contamination. Full decomposition information of the $HST$ galaxy images is available in \cite{misty2018}. The majority of the surface brightness fits to the ground-based images were guided by the $HST$-based solutions (see Sec.\ \ref{sec:modeling}). The $B$, $V$, $R$, and $I$ magnitudes are total magnitudes, the $H_{-0.5}$ magnitude is calibrated for the aperture system of \cite{aaronson1980}. Disk axis ratios ($q_{d}$) are adopted from the $HST$ decompositions of \cite{misty2018} unless otherwise noted. Disk radii are either measured at the $B-$band 25th mag arcsec$^{-2}$ isophote ($R_{25}$) or derived from the exponential disk radii ($R_{d}$) as noted. Uncertainties on the magnitudes were determined based on the quality of the surface brightness fits (see Sec.\ \ref{quality_explanation}).
\footnotetext{Radius at the $B-$band 25th mag arcsec$^{-2}$ isophote (see Sec.\ \ref{distances}).}
\footnotetext{Derived from the exponential disk radii (see Sec.\ \ref{sec:modeling}).}
\footnotetext{Averaged from ground-based decompositions.}
\footnotetext{Derived from the inclination of the resolved HI disk \citep{mundell1995}.}
}
\label{decomps}
\end{deluxetable*}

$\textsc{Galfit}$ allows a galaxy image to be modeled by a combination of analytical surface brightness components. These components are not always physically meaningful, although they may correspond to the morphological components of each spiral galaxy (i.e., disks, bulges, bars, rings, etc.). We utilized the general \cite{sersic} profile, which has the form

\begin{equation}
\sum\nolimits(r) = \sum\nolimits_{e}exp\Bigg[-k\bigg(\Big(\frac{r}{r_{e}}\Big)^{1/n} - 1 \bigg)\Bigg]
\end{equation}
where $\sum_{e}$ is the surface brightness of a pixel at an effective radius of $r_{e}$, $n$ is the S\'ersic index which dictates the profile's degree of curvature, and $k$ is defined such that $\Gamma(2n) = 2\gamma(2n,k)$, where $\Gamma$ and $\gamma$ are the complete and incomplete gamma functions, respectively. The analytical profiles of the disks are extrapolated out to 0 counts. The integrated magnitude of each component is calculated as $m = -2.5 \log (counts/s) + zpt$, where $m$ is the magnitude and $zpt$ is the zeropoint. A S\'ersic index of $n = 1$ yields an exponential disk profile. Indices of $n = 0.5$ and $n = 4$ correspond to Gaussian and \cite{dv1948} profiles, respectively. Disks were modeled by holding the S\'ersic index fixed at $n = 1$. Bulges are modeled with a typical index of  $n > 1$, while bars are usually modeled by a shallower curvature of $n < 1$. For the few galaxies in our sample which displayed rings, we employed an exponential disk profile with the index fixed at $n = 1$ and truncation functions to remove the inner and outer regions of the profile.

The AGN in each galaxy is an unresolved point source, therefore the brightness can be constrained with an accurate model of the point-spread-function (PSF) of each image. The PSF models were built by first selecting an isolated star in the field, and then modeling a small portion of the image centered on the star.  The star itself was modeled with a set of Gaussians whose widths, magnitudes, axis ratios, and position angles were left as free parameters, and the background sky was modeled as a gradient. A PSF model was considered suitable when the residuals, after subtracting the model from the image, contained only random noise. Evidence of regular patterns such as a `bulls-eye' signal the need for additional Gaussians. PSF models typically consisted of 3-4 Gaussian components. As we describe below, we fit several field stars in each image with the suitable PSF model in order to constrain the magnitude zeropoint. When fit, the residuals of the field stars were small, demonstrating that any changes in the PSF shape across the FOV were minimal for the images considered here.

Once the PSF model was constructed, the entire galaxy image was modeled. The background sky was again modeled as a gradient across the frame. The unresolved AGN was fit with the PSF model, and the galaxy was fit with bulge and disk components. The surface brightness decompositions of the $HST$ images, due to the higher spatial resolution, lower sky levels, and no seeing effects, were used to guide the number and type of components included in the decompositions of the ground-based images. Additional morphological components, such as a bar or ring, were included when present in the $HST$ decompositions. Most bulge characteristics were fixed to the parameters determined from the $HST$ image. This included radii, S\'ersic indices, and axis ratios, while the magnitudes were left as free parameters. For very compact galaxies (e.g., Mrk 79, 1H1934-063), bar and ring parameters were also held fixed to their $HST$ values. Field stars that were superimposed on or near the galaxy were also fit with the PSF model to fully isolate the galaxy brightness. Any saturated stars in the field were masked out before the fitting process began. 

\section{Measurements}\label{measurements}
With the surface brightness modeling complete and HI spectral data in hand, we discuss here the measurements conducted for use in TF distance determinations. These include the AGN-free, calibrated galaxy magnitudes and constraints on maximum rotational velocities from the HI emission lines.

\begin{figure*}
\gridline{
\includegraphics[trim={1.4cm 3.1cm 27cm 3.7cm},clip,scale=0.55]{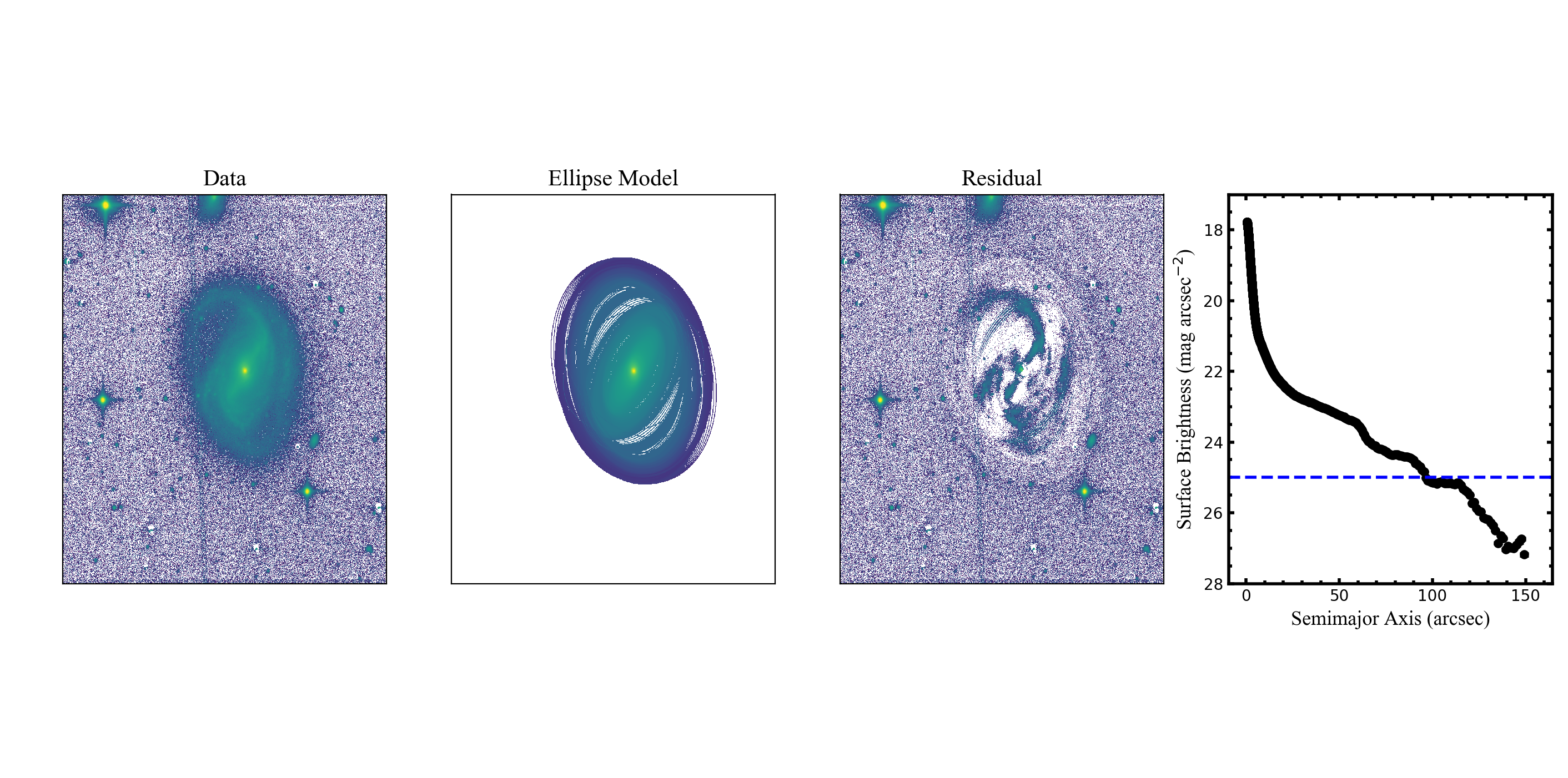}
\includegraphics[trim={10.2cm 3.1cm 18.1cm 3.7cm},clip,scale=0.55]{ellipse.png}
\includegraphics[trim={19.1cm 3.1cm 9.1cm 3.7cm},clip,scale=0.55]{ellipse.png}
\includegraphics[trim={26.8cm 3.1cm 0.3cm 3.7cm},clip,scale=0.55]{ellipse.png}
}
\caption{From left to right: original $B-$band image of NGC 4593, surface brightness isophotes from the \textsc{Ellipse} task in IRAF, data minus the model, and the surface brightness measurements as a function of semimajor axis in arcseconds. The 25 mag arcsec$^{-2}$ surface brightness is indicated with the dashed blue line. East is up and north is right in the above images, and the FOV is 7$\farcm$78 x 8$\farcm$49.} \label{ellipse}
\end{figure*}

\subsection{Optical Galaxy Magnitudes}\label{quality_explanation}
The TF relation for the optical bandpasses requires total, integrated galaxy magnitudes. Thus, with acceptable surface brightness models determined for each image, the image zeropoints were then constrained in order to properly calibrate the model magnitudes. We achieved this by first modeling stars in the field which matched those in optical and near-infrared catalogs. The number of stars modeled was mainly dependent on how many were within the FOV, but was typically between $5-10$. For all of our $B$ and $V-$band images, we drew stellar magnitudes from the AAVSO Photometric All-Sky Survey (APASS; \citealt{hm2014,henden2016}), assuring none of the selected stars were flagged as variable. For NGC 6814 and Mrk 817, we utilized the $R$ and $I-$band field star magnitudes determined by Crimean Astrophysical Observatory imaging \citep{crimea}.

For the remaining 12 galaxies in which data from the Crimean Observatory were not available, we collected $r$ and $i-$band stellar magnitudes first from the Sloan Digital Sky Survey (SDSS) data release 16 \citep{ahumada2019}, or we collected the $r$ and $i-$band magnitudes from APASS. To transform the $r$ and $i$ magnitudes to $R$ and $I$, we calculated synthetic photometry with the IRAF task  $\textsc{synphot}$. We first estimated the spectral type of each star using the spectral classifications as a function of SDSS $g$$-$$i$ (using $g$ stellar magnitudes from either SDSS or APASS) color from Table 4 of \cite{covey2007}. Once the spectral type was assigned, we employed the corresponding stellar template from the Kurucz 1993 Atlas of Model Atmospheres \citep{kurucz1993}, and used $\textsc{synphot}$ to calculate the difference between magnitudes of the template through the SDSS and Johnson-Cousins throughputs. The color differences were small for $R$ and $r$, $-$0.05 $<$ $m_{R}$ $-$ $m_{r}$ $<$ $-$0.26, and slightly larger for $I$ and $i$, $-$0.06 $<$ $m_{I}$ $-$ $m_{i}$ $<$ $-$0.76. 

We adjusted the zeropoint in \textsc{Galfit} to minimize the difference between the measured and expected magnitudes of the field stars, thus calibrating the photometry of the galaxy components as well. Lastly, we combined all the host-galaxy surface brightness components to determine total galaxy magnitudes, which are listed in Table \ref{decomps}.

We determine a typical uncertainty of 0.2\,mag for the integrated galaxy magnitudes, consistent with \cite{misty2018} based on our previous experience using \textsc{Galfit} as well as the level of agreement between fitting results to $HST$ images of compact PG quasars \citep{Veilleux2009}. In some cases, poor seeing conditions or bright sky background induced higher uncertainty in the separation of AGN light from bulge light, or disk light from the sky contribution. For these cases, we assigned a slightly larger uncertainty 0.3\,mag to the final galaxy magnitudes (Mrk 79, NGC 2617 NGC 4748, Mrk 817, NGC 6814). We were unable to separate the disk light from sky contribution in the $B-$band images of Mrk 1044 and Mrk 6. Additionally for Mrk 1310, the seeing conditions coupled with focusing offsets in the $B-$band image caused substantial blending of the AGN and bulge light, thus we were unable to remove the AGN contamination. For these three galaxies, we omit the $B-$band data from our analysis.

Fig.\ \ref{galfit_models} displays selected $B-$band galaxy images, surface brightness models, and residuals which show the range of quality in our surface brightness models of the ground-based images. For the most extended galaxies, like NGC 4593 and NGC 3783 (first and second column, respectively), we have good surface brightness models due to the larger size of the galaxy on the detector and, hence, easier separation of each surface brightness component in the modeling process. More compact galaxies like NGC 4748 (third column) had surface brightness models of moderate quality, and the most compact galaxies, such as Mrk 817 (fourth column), had relatively poor quality surface brightness models. The quality of our models was mainly determined by comparing our galaxy $V-$band magnitudes to their $HST$ $V-$band magnitudes. Good models had excellent agreement, usually within $\sim0.01-0.02$\,mag. While moderate and poor models had larger discrepancies ($\sim0.1-0.3$\,mag), they are still in agreement within the larger uncertainties attributed to the compactness of the galaxy and the seeing conditions that complicated the modeling process of the ground-based images.

\begin{deluxetable}{lrcrrc}
\tablecaption{21\,cm Spectral Characteristics}
\tablehead{
\colhead{Target}&\colhead{V$_{R}$}&\colhead{W$_{m50}$}&\colhead{S/N}&\colhead{Flux}&\colhead{Resolution}\\
\colhead{}&\colhead{(km s$^{-1}$)}&\colhead{(km s$^{-1}$)}&\colhead{}&\colhead{(Jy km s$^{-1}$)}&\colhead{(km s$^{-1}$)}\\
\colhead{(1)}&\colhead{(2)}&\colhead{(3)}&\colhead{(4)}&\colhead{(5)}
&\colhead{(6)}}
\startdata
Mrk 1044 & 4912.0 & 178.0 $\pm$ 8.0 & 20.4 & 4.98 & 1.1 \\
Ark 120 & 9810.0 & 350.0 $\pm$ 13.0 & 10.1 & 3.63 & 2.4 \\
MCG+08- & 6132.0 & 309.0 $\pm$ 8.0 & 28.7 & 14.97 & 0.8 \\
11-011 \\
Mrk 6 & 5632.0 & 477.0 $\pm$ 19.0 & 2.6 & 3.31 & 3.3 \\
Mrk 374 & 13246.0 & 271.0 $\pm$ 18.0 & 4.6 & 0.60 & 9.0 \\
Mrk 79 & 6661.0 & 178.0 $\pm$ 12.0 & 11.4 & 5.71 & 3.4 \\
NGC 2617 & 4265.0 & 119.0 $\pm$ 8.0 & 41.7 & 18.62 & 0.3 \\
NGC 3227 & 1148.0 & 428.0 $\pm$ 9.0 & 14.8 & 29.87 & 3.2 \\
SBS1116 & 8373.0 & 168.0 $\pm$ 18.0 & 3.8 & 0.43 & 6.8 \\
+583A\\
NGC 3783 & 2916.0 & 147.0 $\pm$ 8.0 & 31.0 & 21.03 & 0.7 \\
Mrk 1310 & 5838.0 & 258.0 $\pm$ 15.0 & 7.5 & 2.38 & 3.4 \\
NGC 4051 & 703.0 & 245.0 $\pm$ 8.0 & 147.2 & 61.59 & 0.6 \\
NGC 4151 & 999.0 & 131.0 $\pm$ 8.0 & 45.4 & 74.4 & 0.6 \\
NGC 4593 & 2502.0 & 370.0 $\pm$ 9.0 & 15.4 & 13.52 & 0.7 \\
NGC 4748 & 4184.0 & 332.0 $\pm$ 15.0 & 8.0 & 2.79 & 2.0 \\
NGC 5548 & 5150.0 & 212.0 $\pm$ 15.0 & 8.2 & 2.12 & 5.3 \\
Mrk 817 & 9438.0 & 353.0 $\pm$ 19.0 & 3.1 & 0.74 & 8.2 \\
Mrk 478 & 23881.0 & 314.0 $\pm$ 23.0 & 1.9 & 0.73 & 8.8 \\
NGC 5940 & 10209.0 & 186.0 $\pm$ 13.0 & 10.6 & 3.33 & 1.3 \\
Mrk 290 & 9104.0 & 245.0 $\pm$ 18.0 & 3.8 & 0.76 & 6.8 \\
Zw 229-015 & 8317.0 & 209.0 $\pm$ 17.0 & 5.2 & 0.94 & 6.4 \\
1H1934-063 & 3191.0 & 165.0 $\pm$ 10.0 & 13.6 & 4.54 & 2.0 \\
NGC 6814 & 1562.0 & 84.0 $\pm$ 8.0 & 60.4 & 55.92 & 0.2 \\
NGC 7469 & 4930.0 & 215.0 $\pm$ 16.0 & 6.1 & 2.00 & 3.3 \\
\enddata
\tablecomments{Columns (2) and (3) list the measured recessional velocity and W$_{m50}$ values, respectively, with the updated HI line width measurement method described in Section \ref{distances}. Columns (4) and (5) list the S/N and flux after spectral smoothing to the final spectral resolution, reported in column (6).}
\label{characteristics}
\end{deluxetable}

\subsection{$H-$band Aperture Photometry}
The TF relation for the $H-$band utilizes the $H_{-0.5}$ magnitude, which is calibrated for the aperture system ${\log(A/D_{25}^{i}) = -0.5}$, where $A$ is the aperture through which the galaxy intensity is measured and $D_{25}^{i}$ is the galaxy diameter at the $B-$band 25 mag arcsec$^{-2}$ isophote, corrected for inclination \citep{aaronson1980}. We employed the $\textsc{Ellipse}$ task in IRAF to measure isophotes from our sky-subtracted, ground-based $B-$band images, with Galactic extinction corrections applied to the $B-$band magnitudes. We then fit an exponential disk function to the outer surface brightness profile to arrive at the semimajor axis in arcseconds at which the surface brightness reached 25 mag arcsec$^{-2}$. An example is shown in Fig.\ \ref{ellipse}, where the blue dashed line indicates the 25th mag arcsec$^{-2}$ surface brightness. For galaxies that had $H-$band imaging but did not have $B-$band images (Ark 120, SBS1116+583A, Zw 229-015), we utilized the relation between the radius at the 25 mag arcsec$^{-2}$ isophote ($R_{25}$) and the exponential disk scale length ($R_{d}$) of $R_{25} = 3.2R_{d}$ (\citealt{Catinella2006,deBlok2014}) to estimate $D_{25}$ from our exponential disk profile fits. Lastly, our inclination corrections to the diameters follow the formula from the Second Reference Catalog of Bright Galaxies (\citealt{dv1976}, hereafter RC2):

\begin{figure}
    \centering
    \includegraphics[trim={2.2cm 13cm 0cm 2.5cm},clip,scale=0.5]{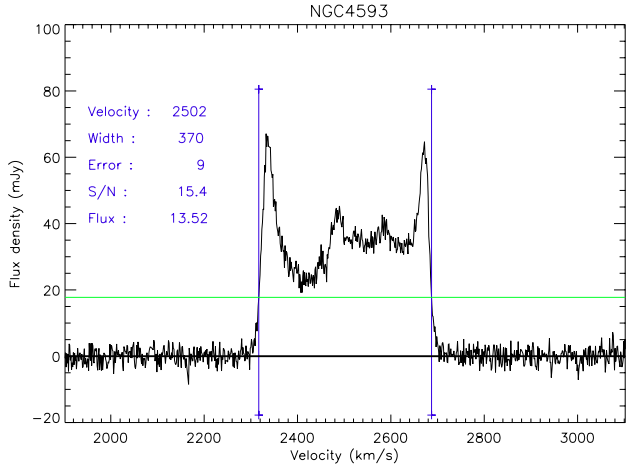}
    \caption{Example of the W$_{m50}$ line width measurement method for the HI emission spectrum of NGC 4593. The blue vertical lines indicate the range of spectral channels which contain 90\% of the HI flux, while the green horizontal line indicates 50\% of the mean flux in this range. The integrated flux measurement is in units of Jy km s$^{-1}$, and the width and velocity measurements are in units of km s$^{-1}$. The error is the uncertainty on the width and in units of km s$^{-1}$.}
    \label{wm50}
\end{figure}

\begin{equation}
    D_{25}^{i} = D_{25}(a/b)^{-C}
\end{equation}
where $a/b$ is the ratio of major to minor axes, and \cite{tf1985} determined $C = 0.22 \pm 0.03$ based on their best-fit to the deviations in $H-$band surface brightness as a function of galaxy inclination. $H_{-0.5}-$band magnitudes in addition to the galaxy radii and corresponding measurement method are listed in Table \ref{decomps}.

\cite{aaronson1980} originally assumed a typical uncertainty of 0.1\,mag for their $H_{-0.5}-$band values. However, the aperture photometry for our sample was conducted on galaxy images in which we removed the AGN contamination in the surface brightness modeling process, which induces additional uncertainty in the total galaxy magnitude. Therefore, we conservatively adopt 0.2\,mag uncertainty on all $H_{-0.5}$ magnitudes, consistent with our typical uncertainties on the AGN-free galaxy magnitudes for the optical bands. 

\subsection{HI 21cm Line Widths}\label{line_widths}
The TF relation utilizes the width of the unresolved, rotationally-broadened HI 21\,cm emission line from late-type galaxies, which is directly related to the maximum rotation rate \citep{epstein1964,roberts1969}. We follow the method originally described by \cite{tf1985}, with the updated definition of the HI line width \citep{courtois2009} which includes corrections for instrumental and redshift broadening:
\begin{equation}
    W_{m50}^{c} = \frac{W_{m50}}{(1+z)} - 2\Delta v\lambda
\end{equation}
where $z$ is the redshift of the HI line, $\Delta v$ is the smoothed resolution of the spectrum, and $\lambda$ is an empirically determined constant term given as $\lambda = 0.25$. We use the redshifts of the HI lines of our targets reported in Paper I. $W_{m50}$ is defined as the width of the HI profile at 50\% of the mean flux over the range of spectral channels which contain 90\% of the HI flux. This new definition by \cite{courtois2009} is preferred as it employs the mean flux rather than the peak, which makes the width measurement independent of the strengths of the flanks. Excluding 5\% of the flux on either side of the profile also aids in separation of the profile wings from the noise. The line widths reported in Paper I are widths calculated at 50\% and 20\% over 100\% of the flux. Therefore, we have remeasured the widths of our HI profiles using the updated definition and we list them in Table \ref{characteristics}. An example of this measurement for the HI emission from NGC 4593 is shown in Fig.\ \ref{wm50}.  

The corrected, observed width is then adjusted to agree statistically with twice the maximum rotational velocity, as the width of the HI line includes both redshifted and blueshifted gas motions. The translation is given as
\begin{equation}
\begin{split}
W_{mx}^{2} = (W_{m50}^{c})^{2} + (W_{t,m50})^{2}[1 - 2e^{-(W_{m50}^{c}/W_{c,m50})^{2}}]\\
& \hspace{-4cm} - 2W_{m50}^{c}W_{t,m50}[1 - e^{-(W_{m50}^{c}/W_{c,m50})^{2}}]
\end{split}
\end{equation}
where $W_{c,m50}$ = 100 km s$^{-1}$ and
$W_{t,m50}$ = 9 km s$^{-1}$ are the values found by \cite{courtois2009} to produce the best match between maximum rotation rate and adjusted HI line width. The width is then deprojected to edge-on orientation by
\begin{equation}
    W_{mx}^{i} = W_{mx} / sin(i)
\end{equation}
where $i$ is the inclination of the galaxy disk. 

The inclinations were generally derived from the axis ratios of the galaxy disk, as listed in Table \ref{decomps}.  For most of the galaxies, we adopted the axis ratios reported by \cite{misty2018}. For NGC 4051 and NGC 4593, where the disk extended beyond the FOV of the $HST$ image, we adopted the axis ratios from our $\textsc{Galfit}$ models of the ground-based images. For NGC 4151, the spatially resolved HI study by \cite{mundell1999} reveals the inclination of the HI disk to be 21$^\circ$, much more face-on than the disk axis ratio that has been typically found in the optical based on the high surface brightness stellar distribution ($\sim$0.6; \citealt{dv1976, dv1991, misty2018}; see Sec.\ \ref{previous_meas}). We therefore adopt 21$^\circ$ as the true inclination of the HI disk for NGC 4151. We follow the standard prescription from the photovisual analysis of \cite{holmberg1958} adopted by the main TF works in the literature \citep{tf1977,tp2000,tully2008,tully2013}:

\begin{deluxetable*}{lcrrrrrrr}
\tablewidth{18cm} 
\tablecaption{Corrected HI line widths and Magnitudes}
\tablehead{
\colhead{Target} & \colhead{Morphology} & \colhead{W$_{mx}^{i}$} & \colhead{$m_{V}^{b,i,k}$ ($HST$)} & \colhead{$m_{B}^{b,i,k}$} & \colhead{$m_{V}^{b,i,k}$} & \colhead{$m_{R}^{b,i,k}$} & \colhead{$m_{I}^{b,i,k}$} & \colhead{$m_{H_{-0.5}}^{b,i,k}$} \\ \colhead{(1)} & \colhead{(2)} & \colhead{(3)} & \colhead{(4)} & \colhead{5} & \colhead{(6)} & \colhead{(7)} & \colhead{(8)} & \colhead{(9)}
}
\startdata
Mrk 1044 &  SB(s)c & 329.89 $\pm$ 28.33 & 14.02 $\pm$ 0.20 & \nodata & 13.79 $\pm$ 0.20 & 13.56 $\pm$ 0.20 & 12.65 $\pm$ 0.20 & \nodata \\
Ark 120 & Sb pec & 549.21 $\pm$ 33.78 & 13.82 $\pm$ 0.20 & \nodata & \nodata & \nodata & \nodata & 11.71 $\pm$ 0.20 \\
MCG+08-11-011 &  SBc & 360.16 $\pm$ 11.95 & 10.93 $\pm$ 0.20 & \nodata & \nodata & \nodata & \nodata & \nodata \\
Mrk 6 &  Sb & 565.76 $\pm$ 25.94 & 13.19 $\pm$ 0.20 & \nodata & 12.96 $\pm$ 0.20 & \nodata & \nodata & 10.78 $\pm$ 0.20 \\
Mrk 374 & SBc & 295.93 $\pm$ 22.24 & 13.93 $\pm$ 0.20 & \nodata & \nodata & \nodata & \nodata & \nodata \\
Mrk 79 & SBb & 261.91 $\pm$ 21.85 & 13.74 $\pm$ 0.20 & 13.77 $\pm$ 0.30 & 13.45 $\pm$ 0.20 & 12.95 $\pm$ 0.20 & \nodata & 11.15 $\pm$ 0.20 \\
NGC 2617 &  Sc & 445.10 $\pm$ 149.27 & 12.53 $\pm$ 0.20 & 13.93 $\pm$ 0.30 & 12.68 $\pm$ 0.20 & 12.38 $\pm$ 0.20 & \nodata & \nodata \\
NGC 3227 & SAB(s) pec & 448.88 $\pm$ 10.74 & 10.38 $\pm$ 0.20 & 11.00 $\pm$ 0.20 & 10.25 $\pm$ 0.20 & 9.85 $\pm$ 0.20 & \nodata & 8.13 $\pm$ 0.20 \\
SBS1116+583A & SBc & 301.51 $\pm$ 41.10 & 15.45 $\pm$ 0.20 & \nodata & \nodata & \nodata & \nodata & 13.58 $\pm$ 0.20 \\
NGC 3783 & (R')SB(r)a & 480.53 $\pm$ 120.76 & 11.71 $\pm$ 0.20 & 12.38 $\pm$ 0.20 & 11.65 $\pm$ 0.20 & 11.17 $\pm$ 0.20 & \nodata & \nodata \\
Mrk 1310 &  Sbc & 347.49 $\pm$ 24.08 & 14.67 $\pm$ 0.20 & \nodata & 14.76 $\pm$ 0.20 & 14.13 $\pm$ 0.20 & \nodata & 12.40 $\pm$ 0.20 \\
NGC 4051 &  SAB(rs)bc & 282.83 $\pm$ 15.66 & 9.87 $\pm$ 0.20 & 10.45 $\pm$ 0.20 & 9.97 $\pm$ 0.20 & 9.61 $\pm$ 0.20 & \nodata & 8.52 $\pm$ 0.20 \\
NGC 4151 & (R')SAB(rs)ab & 342.95 $\pm$ 56.31 & 10.69 $\pm$ 0.20 & 11.14 $\pm$ 0.20 & 10.52 $\pm$ 0.20 & 10.09 $\pm$ 0.20 & 9.66 $\pm$ 0.20 & 8.98 $\pm$ 0.20 \\
NGC 4593 & (R)SB(rs)b & 437.95 $\pm$ 23.30 & 10.81 $\pm$ 0.21 & 11.51 $\pm$ 0.20 & 10.81 $\pm$ 0.20 & 10.29 $\pm$ 0.20 & \nodata & 9.39 $\pm$ 0.20 \\
NGC 4748 &  Sab & 429.70 $\pm$ 23.25 & 13.17 $\pm$ 0.20 & 13.89 $\pm$ 0.30 & 12.80 $\pm$ 0.30 & 12.17 $\pm$ 0.20 & \nodata & 10.86 $\pm$ 0.20 \\
NGC 5548 & (R')SA(s)0/a & 534.08 $\pm$ 30.90 & 12.49 $\pm$ 0.20 & 13.08 $\pm$ 0.20 & 12.35 $\pm$ 0.20 & 12.03 $\pm$ 0.20 & \nodata & \nodata \\
Mrk 817 &  SBc & 549.93 $\pm$ 40.97 & 14.09 $\pm$ 0.20 & 14.76 $\pm$ 0.30 & 13.78 $\pm$ 0.30 & 13.38 $\pm$ 0.30 & \nodata & 11.67 $\pm$ 0.20 \\
Mrk 478 &  Sab & 516.02 $\pm$ 163.75 & 15.24 $\pm$ 0.22 & \nodata & \nodata & \nodata & \nodata & \nodata \\
NGC 5940 &  SBc & 327.54 $\pm$ 32.79 & 13.13 $\pm$ 0.20 & \nodata & \nodata & \nodata & \nodata & \nodata \\
Mrk 290 &  S0 & 376.64 $\pm$ 34.87 & 15.08 $\pm$ 0.20 & \nodata & \nodata & \nodata & \nodata & \nodata \\
Zw 229-015 & (R)SBc & 234.32 $\pm$ 21.15 & 14.71 $\pm$ 0.20 & \nodata & \nodata & \nodata & \nodata & 12.66 $\pm$ 0.20 \\
1H1934-063 &  Sbc & 286.26 $\pm$ 25.28 & 12.29 $\pm$ 0.20 & 13.30 $\pm$ 0.20 & 12.72 $\pm$ 0.20 & 12.33 $\pm$ 0.20 & \nodata & \nodata \\
NGC 6814 &  SAB(rs)bc & 388.09 $\pm$ 195.55 & 10.66 $\pm$ 0.20 & 11.46 $\pm$ 0.30 & 10.69 $\pm$ 0.20 & 10.20 $\pm$ 0.21 & 9.62 $\pm$ 0.20 & 9.71 $\pm$ 0.20 \\
NGC 7469 & (R')SAB(rs)a & 335.80 $\pm$ 30.96 & 12.19 $\pm$ 0.20 & 12.60 $\pm$ 0.20 & 11.97 $\pm$ 0.20 & 11.83 $\pm$ 0.20 & 10.92 $\pm$ 0.20 & 9.76 $\pm$ 0.20 \\
\enddata
\tablecomments{The morphological classifications listed in Column (2) are consistent with those reported from Paper I. Column (3) lists the corrected HI line width in km s$^{-1}$, which is statistically equal to twice the maximum rotation rate, deprojected to edge-on orientation (see Sec.\ \ref{line_widths}). Columns (4-8) list the observed magnitudes in each band corrected for Galactic extinction, inclination-dependent extinction, and redshift (see Sec.\ \ref{current_calibrations}, \ref{vband_cal}).}
\label{line_widths_mags}
\end{deluxetable*}

\begin{equation}
    cos(i) = [(q_{d}^{2} - q_{0,d}^{2})/(1 - q_{0,d}^{2})]^{1/2}
\end{equation}
where $q_{d}=b/a$ is the disk axis ratio and $q_{0,d}$ is the intrinsic axial ratio of a disk galaxy viewed edge-on. Following \cite{tp2000}, we adopt $q_{0,d} = 0.2$ as the single, global value for the flattening. The uncertainties in the deprojected line widths increase as galaxy inclinations become more face-on. Consequently, the galaxies in this sample with the lowest inclination (NGC 2617, NGC 3783, NGC 6814) have the highest uncertainties in $W_{mx}^{i}$.

\section{TF Distances}\label{distances}
With HI widths in hand and the AGN contamination removed from the galaxy brightness, we employed the TF method to constrain the distance to each AGN host.

\subsection{Current TF Calibrations}\label{current_calibrations}
The current calibrations for the $B$, $R$, $H_{-0.5}$ \citep{tully2008}, and $I-$band \citep{tc2012} TF relations are as follows:

\begin{equation}
    M_{B}^{b,i,k} = -19.99 - 7.27(\log W_{mx}^{i} - 2.5)
\end{equation}
\begin{equation}
    M_{R}^{b,i,k} = -21.00 - 7.65(\log W_{mx}^{i} - 2.5)
\end{equation}
\begin{equation}
    M_{I}^{b,i,k} = -21.39 - 8.81(\log W_{mx}^{i} - 2.5)
\end{equation}
\begin{equation}
    M_{H_{-0.5}}^{b,i,k} = -22.17  - 9.55(\log  W_{mx}^{i} - 2.5)
\end{equation}
where $b$, $i$, and $k$ are Galactic extinction, inclination, and redshift corrections, respectively, and the superscripts on the magnitudes indicate that the corresponding corrections have been applied. We estimate the extinction along the line of sight in each bandpass using the \cite{schlafly2011} recalibration of the Milky Way dust map of \cite{schlegel1998}.

The inclination correction is given by the expression ${A_{i}^{\lambda} = \gamma_{\lambda}\log(a/b)}$, originally formulated by \cite{tully1998} and subsequently used by \cite{tully2008} and \cite{tc2012}, where $\lambda$ is the passband, $a/b$ is the ratio of major to minor axes of the galaxy disk, and $\gamma$ is defined as
\begin{equation}\label{b_inc}
    \gamma_{B} = 1.57 + 2.75(\log W_{mx}^{i} - 2.5)
\end{equation}
\begin{equation}
    \gamma_{R} = 1.15 + 1.88(\log W_{mx}^{i} - 2.5)
\end{equation}
\begin{equation}
    \gamma_{I} =0.92 + 1.63(\log W_{mx}^{i} - 2.5)
\end{equation}

However, \cite{tully1998} do not include a prescription for the $H_{-0.5}$ magnitudes. \cite{Sakai2000} approximated the correction as $A_{i}^{H_{-0.5}} = 0.5A_{i}^{I}$, however we adopt the original correction from \cite{tf1985} of $A_{i}^{H_{-0.5}} = 0.1A_{i}^{B}$ as it was derived from measurements in the $H_{-0.5}$ band.

Finally, the $k-$corrections for the $B$, $R$, and $I$ bands utilized in the TF calibrations of \cite{tully2008} and \cite{tc2012} are described in \cite{tp2000} and \cite{Chilingarian2010} for the optical bands and near-infrared band, respectively, and are as follows:
\begin{equation}\label{b_k}
    A_{k}^{B} = (3.6 - 0.36T)z
\end{equation}
\begin{equation}
    A_{k}^{R} = [4.24(R - I) - 1.10]z
\end{equation}
\begin{equation}
    A_{k}^{I} = 0.302z + 8.768z^{2} - 68.680z^{3} + 181.904z^{4}
\end{equation}
where $T$ is the galaxy morphological type (1, 3, 5, and 7 corresponding to Sa, Sb, Sc, and Sd) and $z$ is the redshift. The $k-$corrections in this work utilize the morphological classifications reported in Paper I and are listed in Table \ref{line_widths_mags}. Once more the $H_{-0.5}$ band lacks a prescription from \cite{tp2000}, thus we adopt the original $k-$correction from \cite{aaronson1980} as $A_{k}^{H_{-0.5}} = 1.9z$. 

\vspace{2cm}
 \begin{longtable}[c]{@{\extracolsep{\fill}}lcccc@{}}
\caption{\\ TF Distance and $V_{\textsc{PEC}}$ Measurements}\\
\hline\hline
\multicolumn{1}{l}{Target} & \multicolumn{1}{c}{Band}  & \multicolumn{1}{c}{D} & \multicolumn{1}{c}{$V_{\textsc{PEC}}$} &
\multicolumn{1}{c}{Flag}\\
& & \multicolumn{1}{c}{(Mpc)} &  \multicolumn{1}{c}{(km s$^{-1}$)} & \\ 
\midrule
\endhead
\hline
\endfoot 
\bottomrule
\endlastfoot
Mrk 1044 & $V$ ($HST$) & 81.3 $\pm$ 13.0 & \nodata & \nodata \\
& $V$ & 73.0 $\pm$ 11.7 & \nodata & \nodata \\
& $R$ & 86.9 $\pm$ 14.0 & \nodata & \nodata \\
& $I$ & 69.1 $\pm$ 12.2 & \nodata & \nodata \\
& Best Estimate & 81.3 $\pm$ 16.3 & -1275 $\pm$ 1212 & a \\
Ark 120 & $V$ ($HST$) & 161.2 $\pm$ 21.2 & \nodata & \nodata \\
& $H_{-0.5}$ & 171.4 $\pm$ 25.6 & \nodata & \nodata \\
& Best Estimate & 161.2 $\pm$ 32.2 & -1855 $\pm$ 2394 & a \\
MCG+08-11-011 & $V$ ($HST$) & 22.4 $\pm$ 4.5 & 4586 $\pm$ 335 & c \\
Mrk 6 & $V$ ($HST$) & 126.2 $\pm$ 14.7 & \nodata & \nodata \\
& $V$ & 113.6 $\pm$ 13.2 & \nodata & \nodata \\
& $H_{-0.5}$ & 117.9 $\pm$ 15.0 & \nodata & \nodata \\
& Best Estimate & 126.2 $\pm$ 25.2 & -3625 $\pm$ 1873 & b \\
Mrk 374 & $V$ ($HST$) & 66.1 $\pm$ 13.2 & 8878 $\pm$ 981 & c \\
Mrk 79 & $V$ ($HST$) & 50.2 $\pm$ 7.9 & \nodata & \nodata \\
& $B$ & 42.9 $\pm$ 7.9 & \nodata & \nodata \\
& $V$ & 43.9 $\pm$ 6.9 & \nodata & \nodata \\
& $R$ & 46.3 $\pm$ 7.3 & \nodata & \nodata \\
& $H_{-0.5}$ & 32.2 $\pm$ 5.9 & \nodata & \nodata \\
& Best Estimate & 50.2 $\pm$ 10.0 & 3181 $\pm$ 743 & b \\
NGC 2617 & $V$ ($HST$) & 64.7 $\pm$ 33.6 & \nodata & \nodata \\
& $B$ & 100.0 $\pm$ 50.7 & \nodata & \nodata \\
& $V$ & 69.1 $\pm$ 35.9 & \nodata & \nodata \\
& $R$ & 79.9 $\pm$ 41.7 & \nodata & \nodata \\
& Best Estimate & 64.7 $\pm$ 19.5 & -183 $\pm$ 1446 & a \\
NGC 3227 & $V$ ($HST$) & 24.3 $\pm$ 2.4 & \nodata & \nodata \\
& $B$ & 26.3 $\pm$ 2.6 & \nodata & \nodata \\
& $V$ & 22.9 $\pm$ 2.3 & \nodata & \nodata \\
& $R$ & 25.3 $\pm$ 2.5 & \nodata & \nodata \\
& $H_{-0.5}$ & 22.4 $\pm$ 2.3 & \nodata & \nodata \\
& Best Estimate & 24.3 $\pm$ 4.9 & -323 $\pm$ 364 & a \\
SBS1116+583A & $V$ ($HST$) & 136.7 $\pm$ 31.1 & \nodata & \nodata \\
& $H_{-0.5}$ & 129.0 $\pm$ 35.6 & \nodata & \nodata \\
& Best Estimate & 136.7 $\pm$ 27.3 & -1393 $\pm$ 2030 & a \\
NGC 3783 & $V$ ($HST$) & 49.8 $\pm$ 19.6 & \nodata & \nodata \\
& $B$ & 54.8 $\pm$ 20.7 & \nodata & \nodata \\
& $V$ & 48.5 $\pm$ 19.1 & \nodata & \nodata \\
& $R$ & 51.6 $\pm$ 20.4 & \nodata & \nodata \\
& Best Estimate & 49.8 $\pm$ 10.0 & -427 $\pm$ 743 & a \\
Mrk 1310 & $V$ ($HST$) & 118.7 $\pm$ 16.7 & \nodata & \nodata \\
& $V$ & 123.4 $\pm$ 17.3 & \nodata & \nodata \\
& $R$ & 122.7 $\pm$ 17.3 & \nodata & \nodata \\
& $H_{-0.5}$ & 98.3 $\pm$ 15.8 & \nodata & \nodata \\
& Best Estimate & 118.7 $\pm$ 23.7 & -2480 $\pm$ 1762 & b \\
NGC 4051 & $V$ ($HST$) & 9.5 $\pm$ 1.2 & \nodata & \nodata \\
& $B$ & 10.4 $\pm$ 1.1 & \nodata & \nodata \\
& $V$ & 10.0 $\pm$ 1.1 & \nodata & \nodata \\
& $R$ & 11.2 $\pm$ 1.2 & \nodata & \nodata \\
& $H_{-0.5}$ & 11.1 $\pm$ 1.3 & \nodata & \nodata \\
& Best Estimate & 9.5 $\pm$ 1.9 & 227 $\pm$ 141 & a \\
NGC 4151 & $V$ ($HST$) & 18.6 $\pm$ 5.0 & \nodata & \nodata \\
& $B$ & 18.9 $\pm$ 4.8 & \nodata & \nodata \\
& $V$ & 17.2 $\pm$ 4.6 & \nodata & \nodata \\
& $R$ & 18.7 $\pm$ 5.0 & \nodata & \nodata \\
& $I$ & 18.7 $\pm$ 5.7 & \nodata & \nodata \\
& $H_{-0.5}$ & 19.9 $\pm$ 6.5 & \nodata & \nodata \\
& Best Estimate & 18.6 $\pm$ 3.7 & -127 $\pm$ 275 & a \\
NGC 4593 & $V$ ($HST$) & 28.5 $\pm$ 3.6 & \nodata & \nodata \\
& $B$ & 32.0 $\pm$ 3.3 & \nodata & \nodata \\
& $V$ & 28.6 $\pm$ 3.0 & \nodata & \nodata \\
& $R$ & 29.9 $\pm$ 3.1 & \nodata & \nodata \\
& $H_{-0.5}$ & 38.2 $\pm$ 4.2 & \nodata & \nodata \\
& Best Estimate & 28.5 $\pm$ 5.7 & 756 $\pm$ 424 & a \\
NGC 4748 & $V$ ($HST$) & 82.2 $\pm$ 10.2 & \nodata & \nodata \\
& $B$ & 93.2 $\pm$ 14.9 & \nodata & \nodata \\
& $V$ & 69.3 $\pm$ 11.2 & \nodata & \nodata \\
& $R$ & 68.7 $\pm$ 8.5 & \nodata & \nodata \\
& $H_{-0.5}$ & 72.3 $\pm$ 10.0 & \nodata & \nodata \\
& Best Estimate & 82.2 $\pm$ 16.4 & -1513 $\pm$ 1219 & b \\
NGC 5548$^{1}$ & $V$ ($HST$) & 83.6 $\pm$ 10.7 & \nodata & \nodata \\
& $B$ & 88.0 $\pm$ 11.0 & \nodata & \nodata \\
& $V$ & 78.4 $\pm$ 10.0 & \nodata & \nodata \\
& $R$ & 90.0 $\pm$ 11.5 & \nodata & \nodata \\
& Best Estimate & 83.6 $\pm$ 16.7 & -753 $\pm$ 1242 & a \\
Mrk 817 & $V$ ($HST$) & 182.8 $\pm$ 26.8 & \nodata & \nodata \\
& $B$ & 198.9 $\pm$ 35.0 & \nodata & \nodata \\
& $V$ & 158.7 $\pm$ 28.4 & \nodata & \nodata \\
& $R$ & 175.0 $\pm$ 31.4 & \nodata & \nodata \\
& $H_{-0.5}$ & 168.8 $\pm$ 28.6 & \nodata & \nodata \\
& Best Estimate & 182.8 $\pm$ 36.6 & -3770 $\pm$ 2721 & c \\
Mrk 478 & $V$ ($HST$) & 282.2 $\pm$ 139.4 & 4655 $\pm$ 10323 & c \\
NGC 5940 & $V$ ($HST$) & 53.4 $\pm$ 10.7 & 6687 $\pm$ 795 & c \\
Mrk 290 & $V$ ($HST$) & 162.3 $\pm$ 32.5 & -2644 $\pm$ 2416 & a \\
Zw 229-015 & $V$ ($HST$) & 66.3 $\pm$ 11.0 & \nodata & \nodata \\
& $H_{-0.5}$ & 52.2 $\pm$ 10.2 & \nodata & \nodata \\
& Best Estimate & 66.3 $\pm$ 13.3 & 3424 $\pm$ 989 & c \\
1H1934-063 & $V$ ($HST$) & 29.5 $\pm$ 4.8 & \nodata & \nodata \\
& $B$ & 39.3 $\pm$ 6.2 & \nodata & \nodata \\
& $V$ & 36.0 $\pm$ 5.9 & \nodata & \nodata \\
& $R$ & 39.3 $\pm$ 6.4 & \nodata & \nodata \\
& Best Estimate & 29.5 $\pm$ 5.9 & 818 $\pm$ 439 & a \\
NGC 6814 & $V$ ($HST$) & 22.1 $\pm$ 17.1 & \nodata & \nodata \\
& $B$ & 26.3 $\pm$ 19.6 & \nodata & \nodata \\
& $V$ & 22.5 $\pm$ 17.4 & \nodata & \nodata \\
& $R$ & 23.8 $\pm$ 18.5 & \nodata & \nodata \\
& $I$ & 22.9 $\pm$ 20.4 & \nodata & \nodata \\
& $H_{-0.5}$ & 35.1 $\pm$ 33.9 & \nodata & \nodata \\
& Best Estimate & 22.1 $\pm$ 8.0 & -283 $\pm$ 593 & a \\
NGC 7469 & $V$ ($HST$) & 36.0 $\pm$ 6.1 & \nodata & \nodata \\
& $B$ & 36.0 $\pm$ 5.9 & \nodata & \nodata \\
& $V$ & 32.5 $\pm$ 5.5 & \nodata & \nodata \\
& $R$ & 40.4 $\pm$ 6.8 & \nodata & \nodata \\
& $I$ & 32.3 $\pm$ 6.0 & \nodata & \nodata \\
& $H_{-0.5}$ & 27.3 $\pm$ 5.4 & \nodata & \nodata \\
& Best Estimate & 36.0 $\pm$ 7.2 & 1949 $\pm$ 535 & c \\

\label{distances_allbands}
\end{longtable}
\begin{figure}
\end{figure}
\makeatletter
\def\blfootnote{\xdef\@thefnmark{}\@footnotetext}
\makeatother
\blfootnote{\parbox[3in]{3.3in}{\vspace{-21cm}\textbf{Note.} $--$ TF Distance and $V_{\textsc{PEC}}$ measurements for all bands. We quote the distance predicted from our $V-$band TF calibration of the $HST$ decompositions as our adopted distance. For galaxies in which we were limited to $V(HST)$, we adopt the values that band predicts. We list the uncertainty on the Best Estimate distance as 20\%, which is the typical uncertainty noted by CF1 and CF2 for TF-based distances. In cases where the uncertainty of the mean of all available TF distances for each galaxy is $>20$\%, we list that value as the uncertainty. We have marked each galaxy with a quality flag for the distance based on comparison with literature, the mass distributions in the CF3 DVC, and peculiar velocities (see Sec.\ \ref{measurements}), a = best quality, b = moderate quality, c = poor quality. \\
$^{1}$Adopted using the HI line width from \cite{ho2008_data}}}\vspace{3.3cm}

\subsection{$V-$band TF Calibration}\label{vband_cal}
The TF relation has been calibrated for most optical and near-infrared bands \citep{tully2008,tc2012}, and most recently for SDSS and WISE filters \citep{Kourkchi2020}. However, the Johnson $V$ band has so far been neglected. We have therefore conducted the first  calibration of the $V-$band TF relation.

\begin{figure*}
\includegraphics[trim={0.3cm 0cm 0cm 1.2cm},clip,scale=0.5]{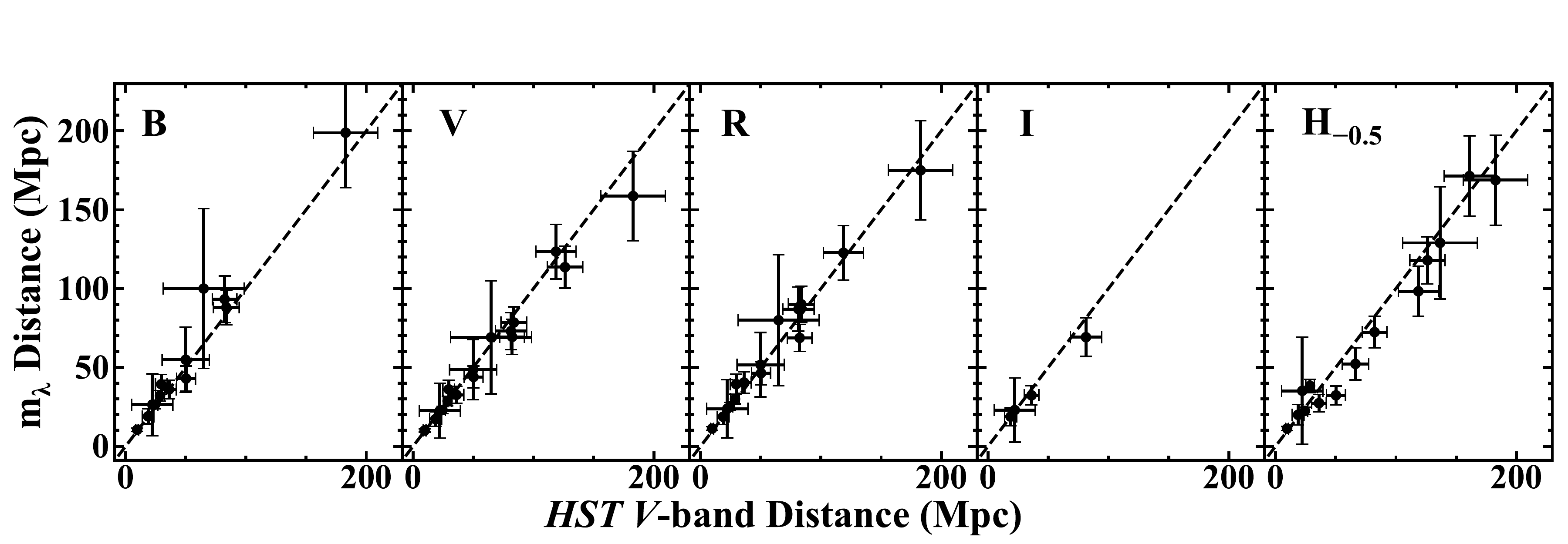}
\caption{Comparisons between the TF distances determined by our calibration of the $V-$band relationship using our $HST$ images to (from left to right) ground-based $B$, $V$, $R$, $I$, and $H_{-0.5}-$band distances. A line of unity is drawn in all panels. The error bars are the individual uncertainties of each distance calculation (see Table \ref{distances_allbands}). We find good agreement for all calibrations within the uncertainties.} \label{band_comp}
\end{figure*}

We began by identifying the galaxies which were used to calibrate the most recent definitions of the optical TF relations \citep{tully2008}. These included galaxies with distances determined from either Cepheid variable stars (26 galaxies), TRGB stars (13 galaxies), or SBF (7 galaxies). We then retrieved the $V-$band magnitudes for these galaxies from the Third Reference Catalogue of Bright Galaxies (\citealt{dv1991}, hereafter RC3), the same source for the $B-$band magnitudes from the calibrations of \cite{tully2008}. These are purely observed magnitudes which have not been corrected for Galactic extinction, inclination-dependent extinction, or redshift. We followed \cite{tully2008} and adopted the \cite{schlegel1998} extinction corrections. We derived the inclination-dependent and redshift corrections in $V$ following the same methods used to define them in $B$, $R$, and $I$. In-depth descriptions of the corrections and final calibration are available in the \nameref{Appendix}.

\cite{tully1998} detailed the extinction corrections due to inclination in the $B$, $R$, and $I$ bands. Following the same procedure and adopting the same formalism for the extinction parameter, ${A_{i}^{\lambda} = \gamma_{\lambda}\log(a/b)}$, where $a/b$ is inverse of the disk axis ratio, we find
\begin{equation}\label{v_inc}
    \gamma_{V} = (1.01 \pm 4.06) + (2.94 \pm 1.09)(\log W_{\textsc{R}}^{i} - 2.5)
\end{equation}

The method for deriving the $k-$corrections adopted by \cite{tp2000} is not described, however they are quite similar to the $k-$corrections based on the analysis of \cite{frei1994}. We therefore adopt the \cite{frei1994} methodology and find a $V-$band $k-$correction of
\begin{equation}\label{v_k}
    A_{k}^{V} = (2.23 - 0.22T)z
\end{equation}
where $T$ is the morphological type (1, 3, 5, and 7 again corresponding to Hubble types Sa, Sb, Sc, and Sd) and $z$ is the redshift. 

The extinction, inclination, and $k-$corrections were then applied to the apparent $V$ magnitudes from RC3 of the \cite{tully2008} calibrating sample. Using the accurate distances to these galaxies, which are based on Cepheids, TRGB, or SBF, we derived their absolute magnitudes. Finally, we fit a linear relationship between the absolute magnitudes and the HI line widths, following the formalism adopted for the other bandpasses. Our best-fit result is:
\begin{equation}\label{vband_eq}
    M_{V}^{b,i,k} = (-20.39 \pm 0.03) - (7.62 \pm 0.15)(\log W_{mx}^{i} - 2.5)
\end{equation}
We find a negligible change to the final result if we instead employ the updated Galactic extinction values from \cite{schlafly2011}. When substituted, the slope and intercept shift slightly to $-7.59$ and $-20.36$, respectively. The calibrated relationship for the $V$ band sits between the existing calibrated relationships for the $B$ and $R$ bands, and also agrees well with the recent TF calibrations of \cite{Kourkchi2020} for SDSS bands, especially when compared to $g$ and $r$. 

\subsection{Final Distances and Peculiar Velocities}\label{final_d_vpec}
To constrain the TF distances, we utilized the deprojected HI line widths and calibrated TF relationships to derive absolute magnitudes for each galaxy in each available bandpass. We then calculated the distance moduli between our corrected apparent magnitudes and derived absolute magnitudes to constrain each distance. All corrected HI line widths and apparent magnitudes used in distance calculations are listed in Table \ref{line_widths_mags}, and the distance measurements for all bands are tabulated in Table \ref{distances_allbands}. 

In Fig.\ \ref{band_comp}, we compare the ground-based distances to the distances based on the $V-$band $HST$ apparent magnitudes due to their superior spatial resolution and lack of seeing effects. Within the uncertainties, we generally find close agreement between the distances derived from the different photometric bands. In addition to the superior image quality of the $HST$ data, the models of the ground-based images were guided by, and in some cases held fixed to, the parameters determined from the $HST$ images. The axis ratios from the models of the $HST$ images were used to derive the deprojected HI line widths in most cases, except for NGC 4051 and NGC 4593 as we described in Sec.\ \ref{line_widths}. Therefore we prefer the distances based on the photometry derived from the $HST$ images and quote them as our adopted TF distances in Table \ref{distances_allbands}. We adopt a typical uncertainty of 20\%, as used by CF1 and CF2 for TF-based distances. However, the ground-based photometry, especially when multiple bandpasses were available, can provide some additional insight into the uncertainties, so we list the uncertainty of the weighted mean as the final adopted uncertainty in cases where it was larger than 20\% of the distance (3/24 galaxies). Though we employ the updated Galactic extinction values of \cite{schlafly2011}, the calibrations of \cite{tully2008} and \cite{tc2012} utilize the previous values of \cite{schlegel1998}. We find a negligible change to our final distances if we instead employ the \cite{schlegel1998} values, with a median fractional change of  0.4\% for all galaxies in our sample.

Peculiar velocities relative to the Hubble Flow, or $V_{\textsc{PEC}}$, can be calculated as a check on the reliability of the TF distances. Galaxies in the local universe are generally observed to have $V_{\textsc{PEC}}$ values $\lesssim$ 500 km s$^{-1}$ \citep{tully2008}, therefore larger values require additional scrutiny. We calculated the modified, cosmologically adjusted galaxy velocity $V_{\textsc{MOD}}$, described in \citet{tully2013,tully2016} as
\begin{equation}
    V_{MOD} = cz[1 + 0.5(1-q_{0})z - (1/6)\left(1 - q_{0} - 3q_{0}^{2} + 1\right)z^{2}]
\end{equation}
where $z$ is the redshift with respect to the Cosmic Microwave Background rest frame, $q_{0}$ = 0.5($\Omega_{\textsc{M}}$ - 2$\Omega_{\Lambda}$), $\Omega_{\textsc{M}}$ = 0.27, and $\Omega_{\Lambda}$ = 0.73. This velocity includes relativistic corrections to the observed velocity assuming $\Lambda$CDM cosmology, which are small for galaxies with $z < 0.1$, like our sample. 

$V_{\textsc{PEC}}$ is then calculated as
\begin{equation}
    V_{PEC} = V_{MOD} - H_{0}D
\end{equation}
where we adopt $H_{0} = 74$ km s$^{-1}$ Mpc$^{-1}$ \citep{riess2019}, and $D$ is the adopted distance to the galaxy in Mpc. 

\subsection{Notes on Individual Objects}\label{previous_meas}
The TF distances we have determined are the first redshift independent distances for many of the galaxies in our sample. However, in a select few cases there are previously measured distances with which we can compare our results, primarily by Cepheid and SBF methods, and secondly the Cosmicflows programs (CF1, CF2, CF3; \citealt{tully2008,tully2013,tully2016}). Previous TF measurements have been reported for 9 galaxies in our sample, mostly in the $B$ band, yet none have taken into account the contamination of the predominantly blue AGN in the nucleus. As discussed in Sec.\ \ref{sec:modeling}, the brightness contribution of an AGN can be significant and will bias the distance modulus towards smaller values, as we have found with the majority of TF distances discussed below. We have tabulated previous distance measurements with their respective methods in Table \ref{previous}. 

Mrk 1044: There are previously published distances for Mrk 1044 from the $J$, $H$, and $K-$band TF calibrations of \cite{theureau2007} of 86.8 $\pm$ 18.4, 78.5 $\pm$ 17.0, and 68.5 $\pm$ 14.2 Mpc, respectively. We find a distance to Mrk 1044 of 81.3 $\pm$ 16.3 Mpc, which lies within the estimates of \cite{theureau2007}. Our surface brightness decomposition of the $HST$ $V-$band image is mostly consistent with the decomposition of the same image from \cite{wang2014}, however we find a larger exponential disk radius of 21\farcs9 compared to their value of 21\farcs2.

NGC 3227: NGC 3227 is interacting with its neighboring elliptical galaxy NGC 3226, which has a SBF distance measurement of $23.7 \pm 2.6$ Mpc from \cite{tonry2001}, with a slight correction from \cite{Blakeslee2010}. NGC 3227 also has two previously reported $B-$band TF distance determinations: \cite{bottinelli1984} report a $B-$band distance of 15.2 Mpc, and \cite{tully1988} report an updated $B-$band distance of $20.6 \pm 3.8$. We report a distance of $24.3 \pm 4.9$ Mpc, which shows good agreement with the SBF measurement to its companion. The removal of the AGN contamination decreases the galaxy's apparent magnitude and results in the determination of a larger distance than both of the previous $B-$band TF determinations. 

\begin{deluxetable}{lclr}
\tablewidth{8.5cm} 
\tablecaption{Previous Measurements}
\tablehead{
\colhead{Target}&\colhead{Distance}&\colhead{Method}&\colhead{Ref} \\
& \colhead{(Mpc)} & &}
\startdata
Mrk 1044 & 68.5 $\pm$ 14.2 & TF $K-$band & 1\\
& 78.5 $\pm$ 17.0 & TF $H-$band & 1\\
& 86.8 $\pm$ 18.4 & TF $J$-band & 1\\
NGC 3227 & 15.2 & TF $B-$band  & 2\\
& 20.6 $\pm$ 3.8 & TF $B-$band  & 3\\
& 23.7 $\pm$ 2.6$^{[a]}$ & SBF  & 4 \\
NGC 3783 & 38.5 $\pm$ 14.2 & TF $B-$band$^{[b]}$ & 2\\
NGC 4051 & 8.8 $\pm$ 1.8 & TF 3.6$\mu$m & 5\\
& 11.0 $\pm$ 1.0 & TF $B-$band &  6\\
& 11.0 $\pm$ 2.0 & TF $I-$band &  7\\
& 11.7 $\pm$ 1.6 & TF $B-$band &  8\\
& 12.2 $\pm$ 2.0 & CF2  & 9\\
& 12.4 $\pm$ 2.3 & TF $B-$band  & 10\\
& 12.6 $\pm$ 1.5 & TF $B-$band  & 3\\
& 16.6 $\pm$ 0.3 & Cepheids & 11\\
& 17.0 $\pm$ 3.1 & TF $B-$band & 12\\
NGC 4151 & 3.9 $\pm$ 0.6 & CF1  & 13\\
& 4.5 $\pm$ 0.8 & TF $B-$band &  10\\ 
& 4.8 $\pm$ 1.0 & TF $B-$band  & 8\\
& 5.0 $\pm$ 0.7 & TF $B-$band & 3\\
& 5.7 $\pm$ 1.2 & TF $K-$band  & 1\\
& 6.4 $\pm$ 1.4 & TF $H-$band &  1\\
& 6.9 $\pm$ 1.5 & TF $J$-band &  1\\
& 15.8 $\pm$ 0.4 & Cepheids &  14\\
& 16.6 $\pm$ 1.1 & SN 2018aoq &  15 \\
& 20.0 $\pm$ 1.6 & SN 2018aoq &  15\\
& 20.3 $\pm$ 3.7 & TF $B-$band &  11\\
NGC 4593 & 25.6 $\pm$ 5.30 & TF $K-$band & 1 \\
& 26.2 $\pm$ 5.7 & TF $H-$band &  1\\
& 26.6 $\pm$ 5.6 & TF $J$-band &  1\\
& 39.5 $\pm$ 14.5 & TF $B-$band$^{[b]}$ &  11\\
NGC 5548 & 34.0 $\pm$ 8.8 & TF $B-$band  & 3\\
NGC 6814 & 8.6 $\pm$ 2.7 & TF $B-$band  & 3\\
& 21.65 $\pm$ 0.41 & Cepheids &  16\\
& 22.8 $\pm$ 4.2 & TF $B-$band &  11\\
NGC 7469 & 35.6 & TF $B-$band &  3\\
& 50.0 $\pm$ 10.4 & TF $K-$band & 1\\
& 55.3 $\pm$ 12.0 & TF $H-$band & 1\\
& 57.30 $\pm$ 2.11 & SN 2008ec &  17\\
& 59.6 $\pm$ 12.7 & TF $J$-band  & 1\\
& 59.70 $\pm$ 4.57 & SN 2008ec &  17\\
& 66.40 $\pm$ 6.23 & SN 2008ec &  18\\
\enddata
\tablecomments{References are as follows: 1. \cite{theureau2007}, 
2. \cite{tully1988}, 3. \cite{bottinelli1984}, 4. \cite{tonry2001}, 
5. \cite{sorce2014}, 6. \cite{tully2013}, 7. \cite{tully2016}, 8. \cite{bottinelli1985}, 9. \cite{tully2009}, 10. \cite{dv1981}, 11. \cite{yuan4051}, 12. \cite{tully1988}, 13. \cite{tully2008}, 14. \cite{yuan2020}, 15. \cite{yu2019}, 16. \cite{misty2019}, 17. \cite{koshida2017}, 18. \cite{Ganeshalingam2013}
\footnotetext{Distance to interacting companion, with correction from \cite{Blakeslee2010}}
\footnotetext{Estimated distance using the diameter-HI line width method (See Sec.\ \ref{previous_meas})}
}
\label{previous}
\end{deluxetable}

\begin{deluxetable*}{lcrrrrrrr}
\tablewidth{18cm} 
\tablecaption{Final Adopted Distances and Mass Estimates}
\tablehead{
\colhead{Target}&\colhead{Distance}&\colhead{Ref}&\colhead{$R_{\textsc{HI}}$}&\colhead{$M_{\textsc{BH}}$}&\colhead{$M_{\star}$}&\colhead{$M_{\textsc{BARY}}$}&\colhead{$M_{\textsc{DYN}}$}&\colhead{$M_{\textsc{DM}}$} \\
& \colhead{(Mpc)} & & \colhead{(Kpc)}&\colhead{(M$_{\odot}$)} &\colhead{(M$_{\odot}$)}&\colhead{(M$_{\odot}$)} & \colhead{(M$_{\odot}$)} & \colhead{(M$_{\odot}$)}\\
\colhead{(1)}&\colhead{(2)}&\colhead{(3)}&\colhead{(4)}&\colhead{(5)}&\colhead{(6)}&\colhead{(7)}&\colhead{(8)}&\colhead{(9)}}
\startdata
Mrk 1044 & 81.3 $\pm$ 16.3 & 1 & 8.10 $\pm$ 1.87 & 6.71 $^{+ 0.12 }_{- 0.10 }$ & 10.02 $\pm$ 0.43 & 10.20 $\pm$ 0.32 & 10.71 $\pm$ 0.11 & 10.55 $\pm$ 0.22 \\
Ark 120 & 161.2 $\pm$ 32.2 & 1 & 16.74 $\pm$ 4.39 & 8.07 $^{+ 0.05 }_{- 0.06 }$ & 11.13 $\pm$ 0.22 & 11.18 $\pm$ 0.21 & 11.47 $\pm$ 0.11 & 11.15 $\pm$ 0.27 \\
MCG+08-11-011 & 84.2 $\pm$ 6.9 & 2 & 39.45 $\pm$ 8.60 & 7.43 $^{+ 0.15 }_{- 0.15 }$ & 11.10 $\pm$ 0.42 & 11.15 $\pm$ 0.39 & 11.47 $\pm$ 0.09 & 11.19 $\pm$ 0.38 \\
Mrk 6 & 126.2 $\pm$ 25.2 & 1 & 30.63 $\pm$ 9.29 & 8.10 $^{+ 0.04 }_{- 0.04 }$ & 11.05 $\pm$ 0.22 & 11.08 $\pm$ 0.21 & 11.76 $\pm$ 0.12 & 11.65 $\pm$ 0.16 \\
Mrk 374 & 185.3 $\pm$ 7.0 & 2 & 41.08 $\pm$ 5.81 & 7.30 $^{+ 0.31 }_{- 0.31 }$ & 10.84 $\pm$ 0.42 & 10.86 $\pm$ 0.41 & 11.32 $\pm$ 0.08 & 11.13 $\pm$ 0.28 \\
Mrk 79 & 50.2 $\pm$ 10.0 & 1 & 13.10 $\pm$ 3.19 & 7.61 $^{+ 0.11 }_{- 0.14 }$ & 10.13 $\pm$ 0.22 & 10.19 $\pm$ 0.20 & 10.72 $\pm$ 0.11 & 10.56 $\pm$ 0.17 \\
NGC 2617 & 64.7 $\pm$ 19.5 & 1 & 12.70 $\pm$ 6.10 & 7.49 $^{+ 0.14 }_{- 0.14 }$ & 10.54 $\pm$ 0.44 & 10.68 $\pm$ 0.36 & 11.17 $\pm$ 0.26 & 10.99 $\pm$ 0.37 \\
NGC 3227 & 23.7 $\pm$ 2.6 & 3 & 17.89 $\pm$ 4.48 & 6.77 $^{+ 0.08 }_{- 0.11 }$ & 11.04 $\pm$ 0.20 & 11.05 $\pm$ 0.20 & 11.32 $\pm$ 0.10 & 10.99 $\pm$ 0.27 \\
SBS1116+583A & 136.7 $\pm$ 27.3 & 1 & 11.23 $\pm$ 2.94 & 6.56 $^{+ 0.08 }_{- 0.09 }$ & 10.48 $\pm$ 0.22 & 10.50 $\pm$ 0.22 & 10.77 $\pm$ 0.14 & 10.44 $\pm$ 0.32 \\
NGC 3783 & 49.8 $\pm$ 10.0 & 1 & 23.98 $\pm$ 5.95 & 7.37 $^{+ 0.08 }_{- 0.08 }$ & 11.08 $\pm$ 0.21 & 11.11 $\pm$ 0.20 & 11.51 $\pm$ 0.19 & 11.29 $\pm$ 0.30 \\
Mrk 1310 & 118.7 $\pm$ 23.7 & 1 & 11.41 $\pm$ 2.78 & 6.21 $^{+ 0.07 }_{- 0.09 }$ & 10.28 $\pm$ 0.22 & 10.38 $\pm$ 0.18 & 10.90 $\pm$ 0.11 & 10.75 $\pm$ 0.17 \\
NGC 4051 & 16.6 $\pm$ 0.3 & 4 & 17.47 $\pm$ 1.94 & 6.13 $^{+ 0.12 }_{- 0.16 }$ & 10.10 $\pm$ 0.19 & 10.19 $\pm$ 0.16 & 10.91 $\pm$ 0.06 & 10.82 $\pm$ 0.09 \\
NGC 4151 & 15.8 $\pm$ 0.4 & 5 & 8.35 $\pm$ 1.13 & 7.56 $^{+ 0.05 }_{- 0.05 }$ & 10.35 $\pm$ 0.19 & 10.41 $\pm$ 0.17 & 10.76 $\pm$ 0.13 & 10.50 $\pm$ 0.24 \\
NGC 4593 & 28.5 $\pm$ 5.7 & 1 & 21.04 $\pm$ 6.45 & 6.88 $^{+ 0.08 }_{- 0.10 }$ & 10.58 $\pm$ 0.22 & 10.60 $\pm$ 0.22 & 11.37 $\pm$ 0.12 & 11.29 $\pm$ 0.15 \\
NGC 4748 & 82.2 $\pm$ 16.4 & 1 & 16.58 $\pm$ 5.21 & 6.41 $^{+ 0.11 }_{- 0.18 }$ & 10.70 $\pm$ 0.22 & 10.73 $\pm$ 0.21 & 11.25 $\pm$ 0.12 & 11.09 $\pm$ 0.19 \\
NGC 5548 & 83.6 $\pm$ 16.7 & 1 & 24.84 $\pm$ 5.96 & 7.72 $^{+ 0.02 }_{- 0.02 }$ & 11.19 $\pm$ 0.22 & 11.20 $\pm$ 0.22 & 11.61 $\pm$ 0.10 & 11.40 $\pm$ 0.20 \\
Mrk 817 & 130.8 $\pm$ 6.9 & 2 & 14.04 $\pm$ 2.74 & 7.59 $^{+ 0.06 }_{- 0.07 }$ & 10.92 $\pm$ 0.19 & 10.93 $\pm$ 0.19 & 11.39 $\pm$ 0.10 & 11.21 $\pm$ 0.17 \\
Mrk 478 & 342.7 $\pm$ 7.2 & 2 & 30.39 $\pm$ 4.19 & 7.40 $^{+ 0.18 }_{- 0.18 }$ & 11.13 $\pm$ 0.42 & 11.17 $\pm$ 0.40 & 11.67 $\pm$ 0.22 & 11.51 $\pm$ 0.34 \\
NGC 5940 & 141.6 $\pm$ 6.9 & 2 & 34.15 $\pm$ 4.94 & 7.04 $^{+ 0.07 }_{- 0.06 }$ & 11.03 $\pm$ 0.29 & 11.07 $\pm$ 0.27 & 11.33 $\pm$ 0.10 & 10.98 $\pm$ 0.34 \\
Mrk 290 & 162.3 $\pm$ 32.5 & 1 & 17.52 $\pm$ 4.24 & 7.28 $^{+ 0.06 }_{- 0.06 }$ & 10.79 $\pm$ 0.30 & 10.81 $\pm$ 0.29 & 11.16 $\pm$ 0.12 & 10.90 $\pm$ 0.29 \\
Zw 229-015 & 115.5 $\pm$ 6.9 & 2 & 28.16 $\pm$ 5.07 & 6.91 $^{+ 0.07 }_{- 0.12 }$ & 10.26 $\pm$ 0.19 & 10.30 $\pm$ 0.18 & 10.95 $\pm$ 0.10 & 10.84 $\pm$ 0.13 \\
1H1934-063 & 29.5 $\pm$ 5.9 & 1 & 5.57 $\pm$ 1.45 & 6.40 $^{+ 0.17 }_{- 0.20 }$ & 10.15 $\pm$ 0.21 & 10.17 $\pm$ 0.21 & 10.42 $\pm$ 0.12 & 10.07 $\pm$ 0.31 \\
NGC 6814 & 21.6 $\pm$ 0.4 & 6 & 11.91 $\pm$ 3.03 & 7.04 $^{+ 0.06 }_{- 0.06 }$ & 10.33 $\pm$ 0.19 & 10.40 $\pm$ 0.16 & 11.02 $\pm$ 0.31 & 10.90 $\pm$ 0.38 \\
NGC 7469 & 61.9 $\pm$ 3.3 & 7 & 17.06 $\pm$ 2.12 & 6.96 $^{+ 0.05 }_{- 0.05 }$ & 10.78 $\pm$ 0.19 & 10.81 $\pm$ 0.18 & 11.05 $\pm$ 0.09 & 10.67 $\pm$ 0.28 \\
\enddata
\tablecomments{Final adopted distances and mass estimates for the AGN hosts in this study. The reference for each adopted distance in Column (3) is as follows: 1. TF distance; this work, 2. Redshift-based distances consistent with those reported in Paper I, adjusted for $H_{0}$ $=$ 74 km s$^{-1}$, 3. SBF estimate to interacting companion NGC 3226 \citep{tonry2001}, 4. Cepheids measurement \citep{yuan4051},
5. Cepheids measurement \citep{yuan2020}, 6. Cepheids measurement \citep{misty2019}, 7. Average SN1a distance \citep{koshida2017,Ganeshalingam2013}. The majority of black hole masses are from the reverberation-mapping database of \cite{bhdatabase} (see Sec.\ \ref{bhmass}). The calculations for $M_{\textsc{BARY}}$ are detailed in Paper I, $M_{\textsc{DYN}}$ and $M_{\textsc{DM}}$ are described in Sec.\ \ref{sec:masses}.
}
\label{final_values}
\end{deluxetable*}

NGC 3783: NGC 3783 has a previous $B-$band TF estimate of 38.5 $\pm$ 14.2 Mpc \citep{tully1988} based on the diameter-HI line width relation. We report a slightly larger distance of 49.8 $\pm$ 10.0 Mpc. However, NGC 3783 is one of the most highly inclined systems in our sample, with an axis ratio of 0.96. Near face-on systems cause large uncertainties in $W_{mx}^{i}$, and consequently a large uncertainty in the distance.

NGC 4051: There are numerous TF distance estimates for NGC 4051 with a large span of values, the most accurate of which is the recent Cepheids measurement by \cite{yuan4051} of $16.6 \pm 0.3$ Mpc. \cite{sorce2014} report a 3.6$\mu$m TF distance of $8.8 \pm 1.8$ Mpc. $B-$band TF determinations span the range of $11.0 - 17.0$ Mpc \citep{dv1981,bottinelli1984,bottinelli1985,tully1988,tully2009}. Finally, CF3 reports an $I-$band measurement of $11.0 \pm 2.0$ Mpc. Our distance is $9.5 \pm 1.9$ Mpc, surprisingly smaller than the previous TF distances given the removal of AGN contamination. The original axis ratio used by CF1 (and subsequently by CF2 and CF3) for NGC 4051 is 0.66, which is slightly more face-on than our constrained axis ratio of 0.58 from the ground-based surface brightness modeling (see Sec.\ \ref{line_widths}). The higher inclination used by CF1 would produce a larger deprojected HI line width and subsequently brighter absolute magnitude predicted by the TF relation, thus resulting in a slightly larger distance of 10.9 Mpc.

NGC 4151: NGC 4151 has been studied by numerous groups in an attempt to constrain its distance, finding values that range from 4.5 to 20.3 Mpc \citep{dv1981,bottinelli1984,bottinelli1985,tully1988}. The most accurate distance comes from a recent Cepheid study, which found $15.8 \pm 0.4$ Mpc \citep{yuan2020}. Almost all of the TF studies underpredict the distance, which seems to be caused by the adoption of an axis ratio of 0.6 when constraining the galaxy inclination. Resolved HI imaging of NGC 4151 \citep{mundell1999} suggests a much more face-on orientation of $\sim21^{\circ}$. Adopting this value constrains our TF estimate of the distance to $18.6 \pm 3.7$ Mpc, slightly larger than but consistent with the Cepheids distance.

NGC 4593: \cite{theureau2007} measured $J$, $H$, and $K-$band TF distances to NGC 4593 of $\sim$26 Mpc, which agrees fairly well with our finding of $28.5 \pm 5.7$ Mpc. \cite{tully1988} estimate a much larger distance of $39.5 \pm 14.5$ Mpc based on the HI line width-diameter TF relation. However, as shown originally by \cite{tf1977} and noted by \cite{bottinelli1983}, the diameter relation is much less accurate than the luminosity-HI line width relationship.

NGC 5548: The previous $B-$band TF measurement from \cite{bottinelli1984} places NGC 5548 at a distance of $34.0 \pm 8.8$ Mpc. We find a distance of 45.0 $\pm$ 3.8 Mpc with a large $V_{\textsc{PEC}}$ of 2104 $\pm$ 288 km s$^{-1}$. However, the large predicted $V_{\textsc{PEC}}$, in addition to the turbulent and low S/N HI profile, suggest this may not be a reliable distance. \cite{ho2008_data} collected a higher S/N HI spectrum with $W_{20} = 321.1 \pm 6.8$ km s$^{-1}$. Using this measurement predicts $D = 83.6 \pm 16.7$ Mpc with a more reasonable $V_{\textsc{PEC}} = -753 \pm 1242$ km s$^{-1}$. We therefore adopt this distance for NGC 5548 and list it in Table \ref{distances_allbands}.
 
NGC 6814: \cite{misty2019} recently reported a Cepheid-based distance to NGC 6814 of $21.6 \pm 0.4$ Mpc. There are also $B-$band TF estimates which range from $8.6 - 22.8$ Mpc \citep{bottinelli1984,tully1988}. Even though NGC 6814 is almost perfectly face on (with an axis ratio of 0.98) and therefore has a large uncertainty, the TF distance we predict of $22.1 \pm 8.0$ Mpc is in good agreement with the Cepheids value.

NGC 7469: NGC 7469 was host to SN 2008ec, a type Ia supernova.  Analysis of the supernova light curve by \cite{koshida2017} and \cite{Ganeshalingam2013} constrained distances of $57.30 - 66.40$ Mpc. There are also multiple TF distance determinations to NGC 7469, including a $B-$band measurement of 35.6 Mpc \citep{bottinelli1984} and $JHK$ measurements of $50.0 - 59.6$ Mpc \citep{theureau2007}. However, in our analysis of the HI spectrum in Paper I, we commented on possible flux contribution to the blueshifted flank of NGC 7469 from companion galaxy IC 5283. Higher S/N emission line detections in the literature most likely include the flux contribution of the companion \citep{mw1984,ms1988,ho2008_data}, while our lower S/N profile does not share the same signature. We have tested distances predicted using the $W_{20}$ measurements (with the $W_{\textsc{R}}^{i}$ definition) from the literature to compare to our result. If 525.1 km s$^{-1}$ from \cite{ho2008_data} is used, we calculate 140 Mpc. If we use 395 km s$^{-1}$ from \cite{mw1984}, the resulting distance is 91 Mpc. Using our width, our distance is $36.0 \pm 7.2$ Mpc with a $V_{\textsc{PEC}}$ of $1949 \pm 535$ km s$^{-1}$. Due to the interaction of IC 5283 and uncertainty in the width of the emission line, resolved HI observations are necessary to both separate the interacting galaxies and improve on the current distance estimates which rely on the HI line width.

\subsection{Selection of Final Adopted Distances}\label{final_selection}
For all galaxies except Mrk 478, we are able to compare our distances to the distances predicted by the CF3 Distance-Velocity Calculator (\citealt{Kourkchi2020}, hereafter CF3 DVC) based on the velocity field from the Numerical Action Methods model (\citealt{shaya2017}, ${D < 38}$\,Mpc) and the Velocity and Density Field Model (\citealt{graziani2019}, ${D < 200}$\,Mpc). The CF3 DVC predicts a distance based on the Cosmicflows model of the local velocity field in a specific region of the sky. It also displays distances and velocities of known galaxy groups and clusters within the search region that define the local model, allowing us to analyze the density of matter in a particular region. The local gravitational interactions between a galaxy and its environment cause individual $V_{\textsc{PEC}}$ values.

For Mrk 1044 ($V_{\textsc{PEC}} = -1275 \pm 1212$), Ark 120 ($V_{\textsc{PEC}} = -1855 \pm 2394$), SBS1116+583A ($V_{\textsc{PEC}} = -1393 \pm 2030$), NGC 4748 ($V_{\textsc{PEC}} = -1513 \pm 1219$), and Mrk 290 ($V_{\textsc{PEC}} = -2644 \pm 2416$), the $V_{\textsc{PEC}}$ values we calculate agree with the range of peculiar velocities observed by CF1 within the large uncertainties. The peculiar velocities of Mrk 6 ($V_{\textsc{PEC}} = -3625 \pm 1873$), Mrk 79 ($V_{\textsc{PEC}} = 3181 \pm 743$), and Mrk 1310 ($V_{\textsc{PEC}} = -2480 \pm 1762$) are large, but still within those observed by the larger CF2 and CF3 catalogs (maximum observed $V_{\textsc{PEC}}$ of $\sim$4000 km s$^{-1}$), and could be caused by the mass distributions near each galaxy's position on the sky present in the CF3 catalog. We confirmed that the CF3 DVC shows known, localized mass concentrations occupying distributions of either smaller or larger distances than those predicted by the DVC. These suggest local gravitational wells, and the resultant blueshifts or redshifts would cause each galaxy to appear closer or farther, assuming the recessional velocities are equivalent to the Hubble flow. 

For MCG+08-11-011, Mrk 374, Mrk 817, Mrk 478, NGC 5940, and Zw 229-015, however, we are unable to further check our TF distances with the CF3 DVC results (mostly due to the lack of clusters present in the CF3 catalog near the position of each galaxy) or any literature results. Within the uncertainties, these $V_{\textsc{PEC}}$ constraints are all $>$1000 km s$^{-1}$, which we set as the cutoff for galaxies in which we were unable to analyze the local mass distribution. The same is true even if we adopt the uncertainty typically considered by Tully et al.\ of 20\%. Our estimated distance for Mrk 478 is 282.2 Mpc, which is beyond the 200 Mpc limit of the CF3 DVC. Additionally, as previously discussed, any distance measurement which relies on current HI line width measurements of NGC 7469 is suspect. Thus, we have deemed the TF distances to these 7 galaxies as uncertain, and have assigned each with the poor quality flag `c' in Table \ref{distances_allbands}.

\begin{figure}
\includegraphics[trim={0.16cm 0cm 0.8cm 1.8cm},clip,scale=0.53]{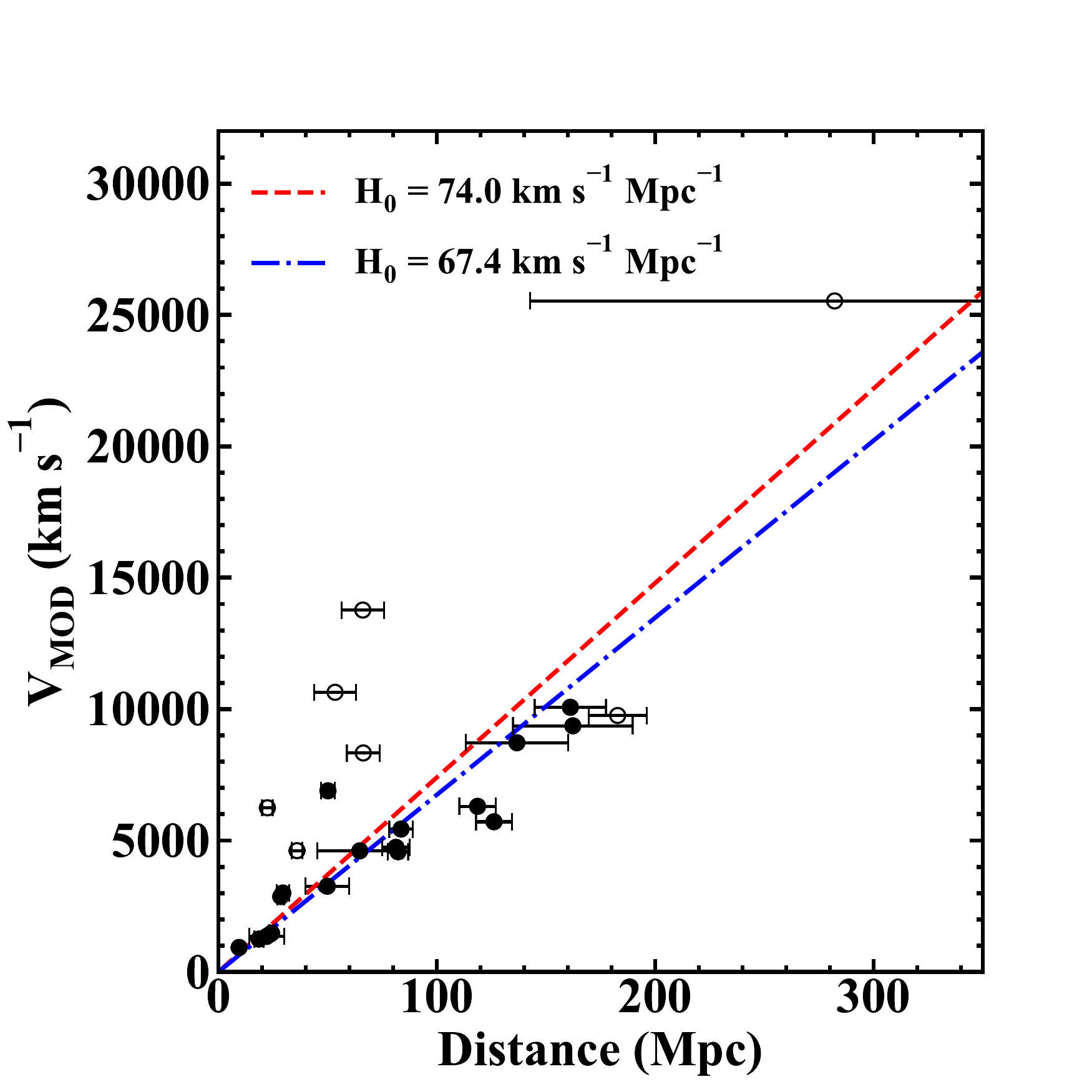}
\caption{Hubble diagram exhibiting our TF distances vs their cosmologically-adjusted velocity $V_{\textsc{MOD}}$. Points shown in open circles represent the 7 TF distances we have deemed as uncertain (see sec.\ \ref{previous_meas}), MCG+08-11-011, Mrk 374, Mrk 817, Mrk 478, NGC 5940, Zw 229-015, and NGC 7469. The closed points are our remaining, certain distance measurements. We compare these to the Hubble-Lema\^itre Law, with H$_{0}$ values of 74.0 km s$^{-1}$ Mpc$^{-1}$ determined by \cite{riess2019} and 67.4 km s$^{-1}$ Mpc$^{-1}$ determined by \cite{planck2018}.} \label{hub_plot}
\end{figure}

We note that the galaxy inclinations of the majority of this sample lie below the usual limit of 45$^{\circ}$ for TF studies in the literature, namely the Cosmicflows programs. Systems with tendencies towards face-on orientations cause large uncertainties in the deprojected HI line widths and, consequently, high uncertainties on distance and $V_{\textsc{PEC}}$ constraints. We also note that the majority of the outliers in Fig.\ \ref{hub_plot} have TF distance measurements far too small for their $V_{\textsc{MOD}}$ given the Hubble flow for either value of H$_{0}$ we have displayed. The TF relation is calibrated with inactive galaxies, thus it is of interest to explore potential differences between active and inactive galaxies that could cause discrepancies in the predicted distances for this sample of AGN hosts.

The analysis of color-magnitude diagrams for mass-matched samples of AGN and non-AGN hosts from the Chandra Deep Field North and South surveys by \cite{Xue2010} found that the star formation rates in active galaxies are typically a factor of $\sim2-3$ higher than quiescent galaxies for $0 < z < 1$. Increased star formation has been shown to lead to an increase in surface brightness (e.g., \citealt{Graves2010,mould2020}), and a higher surface brightness would lead to a brighter apparent magnitude, thus causing a galaxy to appear closer than it is. Additionally, in their study of the $R-$band TF relation for close galaxy pairs, \cite{Barton2001} found that triggered star formation is a significant contributor to the difference in slope from the TF relation. Whether higher star formation rates are caused by AGN activity or interactions, the observational effect would be consistent in that the galaxy would appear brighter than a comparable inactive or isolated galaxy, leading to a distance estimate that is biased low. Active galaxies might therefore be reasonably expected to show a larger scatter about the canonical TF relation.

All final TF distances are shown in the Hubble diagram in Fig.\ \ref{hub_plot}, where the aforementioned uncertain distances are open circles, and the rest are closed circles. For the remainder of this work, we apply the redshift-based distances for MCG+08-11-011, Mrk 374, Mrk 817, Mrk 478, NGC 5940, and Zw 229-015, consistent with those reported in Paper I. We employ the SBF distance to NGC 3226 as the adopted distance to NGC 3227 \citep{tonry2001,Blakeslee2010}, the Cepheid distance measurements to NGC 4051, NGC 4151, and NGC 6814 \citep{yuan4051,yuan2020,misty2019}, and the average of the SN1a distances to NGC 7469 \citep{Ganeshalingam2013,koshida2017}. Final adopted distances are tabulated in Table \ref{final_values}.

\section{Masses}\label{sec:masses}
With distances and physical surface brightness details constrained, we are able to estimate the masses of the AGN hosts in our sample and explore their relationships to the central SMBH masses. Here we detail the methods of measurement for dynamical mass ($M_{\textsc{DYN}}$) and dark matter mass ($M_{\textsc{DM}}$) and adopted measurements of $M_{\textsc{BH}}$.

\subsection{Dynamical and Dark Matter Mass}\label{ddmass}
Measurement of the maximum rotation rate ($V_{mx}$) of a disk galaxy allows the total enclosed mass of the system, or $M_{\textsc{DYN}}$, to be measured. HI is one of the best tracers of galaxy rotational velocity at the outer extents of the disk, as its distribution usually extends much farther than the high surface brightness stellar component (i.e., \citealt{walter2008,ott2012,koribalski2018}). $V_{mx}$ is most precisely measured from the flat portion of HI rotation curves (i.e., \citealt{deBlok2008}), however the unresolved HI emission line is more commonly used as it requires far fewer observational resources to acquire. 

\begin{figure}
\includegraphics[scale=0.55,trim={1cm 0cm 0cm 1.6cm},clip]{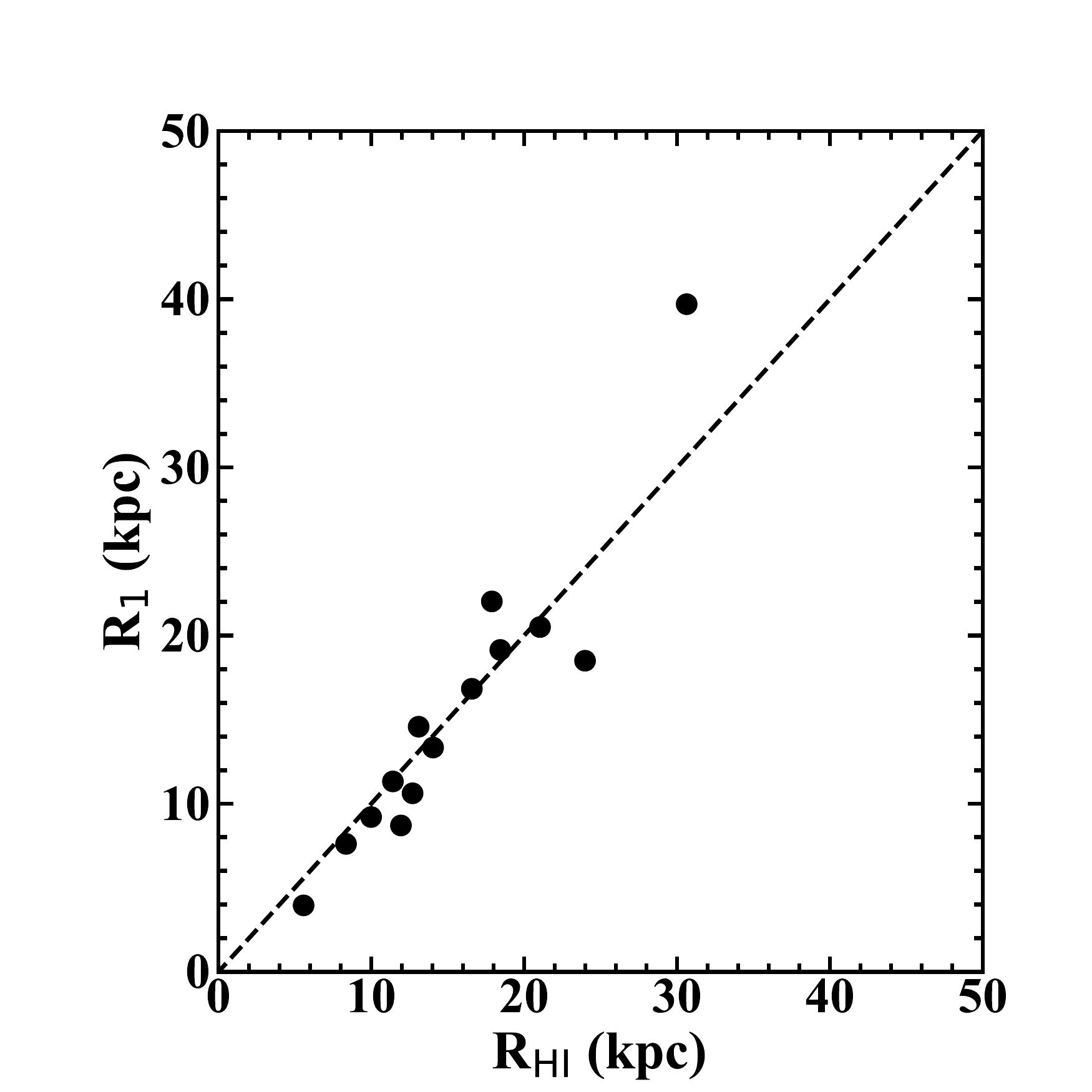}
\caption{Comparison between $R_{\textsc{HI}}$ and the radius at which the stellar mass density reaches 1 M$_{\odot}$ pc$^{-2}$, $R_{1}$, a proxy for the star formation threshold. We have followed the prescriptions from \cite{trujillo2020} to convert the $B-$band surface brightness profiles of the galaxies in our sample to mass density profiles (see Sec.\ \ref{sec:masses}), after which $R_{1}$ was able to be measured. A line of unity is drawn, and we find good agreement between the two radii, with an average fraction of $R_{\textsc{HI}}/R_{1} \sim 1.1$.} \label{r1_comp}
\end{figure}

The large scale velocity dispersion of HI is negligible, $\sim$10 km s$^{-1}$ \citep{Tamburro2009,Ianjamasimanana2012} relative to the rotational velocity. The broadening of the emission line is thus dominated by virial rotation, and the virial theorem describes the mass enclosed in the system as $M_{\textsc{DYN}} = (RV_{mx}^{2})/\textrm{G}$, where $R$ is a characteristic radius and G is the gravitational constant. In our case, $R$ is the radial extent of the HI distribution, and $V_{mx}$ is equivalent to $W_{mx}^{i}$/2, as \cite{tf1985} have shown $W_{mx}$ to be statistically equal to twice the maximum rotation rate. 

There are multiple ways of estimating the HI radius in the literature. In their 21\,cm study of 108 spiral galaxies, \cite{br1997} found a relation between the HI radius and $R_{25}$. For galaxies where we have $B-$band images, we therefore adopt their relation of $R_{\textsc{HI}} = (1.70 \pm 0.16)R_{25}$ with our $R_{25}$ measurements from $B-$band isophote analyses (see Sec.\ \ref{distances}). In the cases where $B-$band data are unavailable for the sample (Ark 120, Mrk 374, SBS1116+583A, Mrk 478, NGC 5940, Mrk 290, Zw 229-015), we look to the collection of template disk galaxy rotation curves by \cite{Catinella2006} and \cite{deBlok2014}, and use the relation $R_{25} = 3.2R_{d}$, with the $R_{d}$ measurements listed in Table \ref{decomps}. Combining these definitions yields $R_{\textsc{HI}} = (5.4 \pm 0.5)R_{d}$, which agrees with the HI rotation curve analysis of \cite{deBlok2014} that shows the maximum extent of all curves to be $\sim$5R$_{d}$. Therefore, for the remainder of the sample that do not have $B-$band data, we adopt $R_{\textsc{HI}} = (5.4 \pm 0.5)R_{d}$.

Additionally, \cite{wang2016} recalibrated the relation between the diameter of the HI disk and HI mass, resulting in an extremely tight relationship over 5 orders of magnitude in mass. With this calibrated relationship, the integrated 21\,cm HI flux measurement may be employed to estimate $R_{\textsc{HI}}$, as opposed to relying on the assumption of uniform scaling between the HI and optical sizes for all morphological types. We find a median fractional decrease of $\sim$18\% in $R_{\textsc{HI}}$ with this method compared to our adopted method of estimating $R_{\textsc{HI}}$, which is within our typical uncertainty of $\sim$24\%.

A recent study by \cite{trujillo2020} sought to derive a physically motivated galaxy radius definition. Such a radius would correspond to a clearly measurable galaxy property, and they suggest the radius at which the star formation threshold is reached ($R_{1}$). The gas density for this threshold is usually estimated to be $\sim$3-10 M$_{\odot}$ pc$^{-2}$ for gas-to-star transformation efficiencies of $\sim$100\% \citep{Schaye2004}. However, \cite{trujillo2020} argue that if the efficiency is less than 100\%, a more reasonable estimation is 1 M$_{\odot}$ pc$^{-2}$, which corresponds to an efficiency of $\sim$10\%. HI has been observed to condense to molecular hydrogen at a threshold of $\sim$10 M$_{\odot}$ pc$^{-2}$ \citep{mk2001,wb2002,bigiel2008}, and molecular clouds are the typical locations of star formation (e.g., \citealt{leroy2008}). HI surface density should therefore be linked to star formation, and Trujillo et al.\ suggest that $R_{1}$ could be closely related to $R_{\textsc{HI}}$. To investigate this, we follow the prescriptions of \cite{trujillo2020}, which were derived from SDSS colors and a Chabrier initial mass function \citep{Chabrier2003}, to transform the surface brightness profiles of the galaxies in our sample to stellar mass densities, and measure radii at 1 M$_{\odot}$ pc$^{-2}$. When compared to our $R_{\textsc{HI}}$ estimates, we find an average ratio of $R_{\textsc{HI}}$ to $R_{1}$ of $\sim$1.1, supporting the similarity between the two. We display the comparison between the two measurements in Fig.\ \ref{r1_comp}. While we do not employ this method to estimate $R_{\textsc{HI}}$, and while our sample is somewhat small and we employ estimates rather than measurements of $R_{\textsc{HI}}$, the agreement between $R_{1}$ and $R_{\textsc{HI}}$ may suggest a promising new avenue for constraining $R_{\textsc{HI}}$ without resolved HI studies.

With $R_{\textsc{HI}}$ estimates for all galaxies in our sample, the total enclosed mass $M_{\textsc{DYN}}$ is calculated by
\begin{equation}
    M_{DYN} = \bigg(\frac{R_{HI}(W_{mx}^{i}/2)^{2}}{G}\bigg)
\end{equation}
with $R_{\textsc{HI}}$ in units of kpc translated from the angular disk sizes by our adopted distances. The constraint on the amount of dark matter, then, is simply the difference between the total enclosed mass and luminous mass, $M_{\textsc{BARY}}$, calculated as
\begin{equation}
    M_{DM} = M_{DYN} - M_{BARY}
\end{equation}
where our $M_{\textsc{BARY}}$ values are the sum of the gas mass and the stellar mass ($M_{\star}$), or $M_{\textsc{BARY}}$ = 1.4$M_{\textsc{HI}}$ + $M_{\star}$. The factor of 1.4 on $M_{\textsc{HI}}$ accounts for the contribution of molecular gas. HI masses are adopted from Paper I, and the stellar masses are adopted from \cite{misty2018} with a few additions in Paper I, both of which have been updated with our final adopted distances reported in this work. $M_{\star}$, $M_{\textsc{BARY}}$, $M_{\textsc{DYN}}$, and $M_{\textsc{DM}}$ are reported in Table \ref{final_values}.

\subsection{Black Hole Mass}\label{bhmass}
All of the galaxies in this work belong to the sample of AGNs with direct black hole mass measurements from RM \citep{bm1982,peterson1993}. RM measures the echo between the continuum variations of the nucleus, likely arising from the accretion disk, and the response of optically thick gas in the broad line region (BLR) moving at Doppler velocities. The time delay ($\tau$) in the BLR variations is due to the extra path length travelled by the ionizing photons, and provides a measurement of the radius of the BLR ($R_{\textsc{BLR}}$). When $R_{\textsc{BLR}}$ is combined with the Doppler-broadened emission line width via the virial theorem, a constraint on the enclosed mass is obtained, the majority of which is due to the SMBH. The mass is given by
\begin{equation}
    M_{BH} = f\frac{c\tau V^{2}}{G}
\end{equation}
where $c\tau$ is the effective radius, $V$ is the width of the broad emission line, $G$ is the gravitational constant, and $f$ is an order-unity scale factor accounting for the unknown geometry and kinematics of the unresolved BLR. We adopt $\langle f \rangle$ = 4.3 \citep{grier2013}.

The majority of $M_{\textsc{BH}}$ values are adopted from the AGN Black Hole Mass Database \citep{bhdatabase}, and are the same as those used in our analysis in Paper I. For MCG+08-11-011, Mrk 374, and NGC 2617, we utilized the virial $M_{\textsc{BH}}$ from
\cite{Fausnaugh2017} and scaled them with $\langle f \rangle$ = 4.3. For Mrk 1044, we used the H$\beta$ time delay from \cite{hu2015} and the rms H$\beta$ line width from \cite{du2016} with our adopted $\langle f \rangle$ to arrive at an $M_{\textsc{BH}}$ estimate. For NGC 5940, we adopt the rms line width from \cite{barth2015} and time delay from \cite{barth2013} to estimate $M_{\textsc{BH}}$. Lastly, the black hole masses for Mrk 478 and 1H1934-063 are based on current work on in-hand RM data (G.\ de Rosa 2020, private communication; M.\ C.\ Bentz et al. 2020, in preparation). All $M_{\textsc{BH}}$ values are listed in Table \ref{final_values}.

\section{Discussion}\label{sec:discussion}
With distances and derived masses in hand, we are able to place both in the context of other established and well-studied relationships. We first explore the most recent calibration of the Baryonic Tully-Fisher (BTF) relation \citep{mcgaugh2000}, which can also be inverted to predict distance. Several scaling relations between $M_{\textsc{BH}}$ and host galaxy characteristics have also been explored in the literature, and in the past two decades have strongly suggested a co-evolving relationship between black holes and their host galaxies. Such examples include the $M_{\textsc{BH}}$ $-$ $\sigma_\star$ relation \citep{fm2000,gebhardt2000,kh2013}, the $M_{\textsc{BH}}$ $-$ $L_{\textsc{BULGE}}$ relation \citep{kr1995,kh2013}, and the $M_{\textsc{BH}}$ $-$ $M_{\star}$ relation \citep{misty2018}, many of which are utilized as inputs to large cosmological simulations of galaxy evolution (e.g., \citealt{steinborn2015}, \citealt{volonteri2016}, \citealt{mutlu2018}). Here, we investigate relationships between $M_{\textsc{BH}}$ and $M_{\textsc{DYN}}$, $M_{\textsc{DM}}$, and galaxy halo mass ($M_{\textsc{HALO}}$).

Linear regressions were carried out with $\textsc{linmix\textunderscore err}$ \citep{linmix_err}, which is a Bayesian-based fitting algorithm that includes uncertainties in both variables in addition to a random scatter component. The slopes, intercepts, and scatter we report for each relationship are the medians of the respective large, random draws from the posterior probability distributions. The uncertainties are the 1-$\sigma$ deviations of each distribution.

\begin{figure}
\includegraphics[scale=0.58,trim={1.5cm 0cm 0cm 1.6cm},clip]{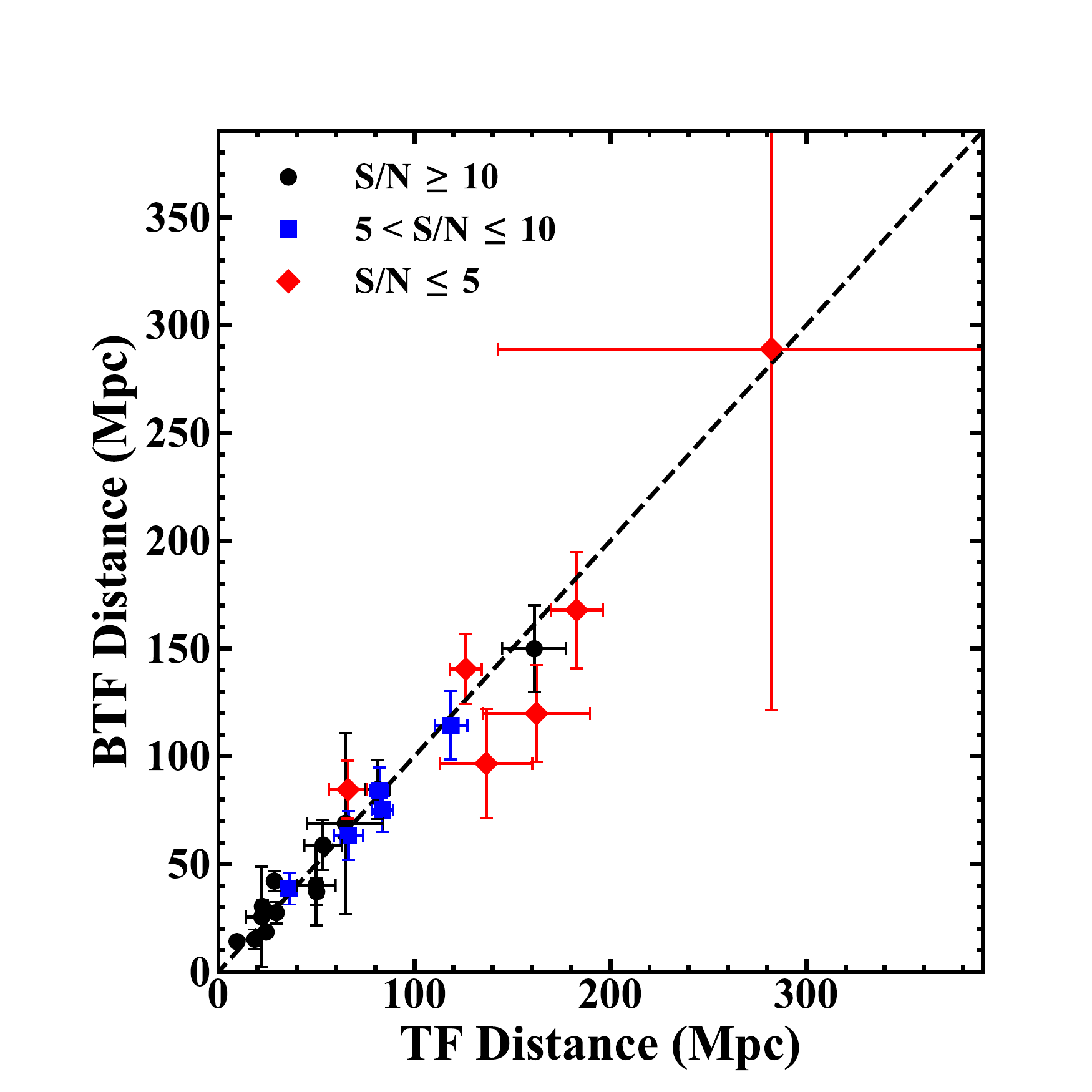}
\caption{Comparison of our adopted distances (see Sec.\ \ref{previous_meas}) to those predicted by the BTF relation, using the $W_{m50}$ calibration from \cite{lelli2019}. A line of unity is drawn. We display 3 HI emission line S/N thresholds (see Sec.\ \ref{btfr_discussion}), where black circles are $\geq$10, blue squares are $>$5 and $\leq$10, and red diamonds are $\leq$5, which may relate to the source of discrepancy for the few outliers.} \label{btfr_comp}
\end{figure}

\begin{figure*}
\centering
\includegraphics[scale=0.65,trim={0cm 0cm 0cm 1cm},clip]{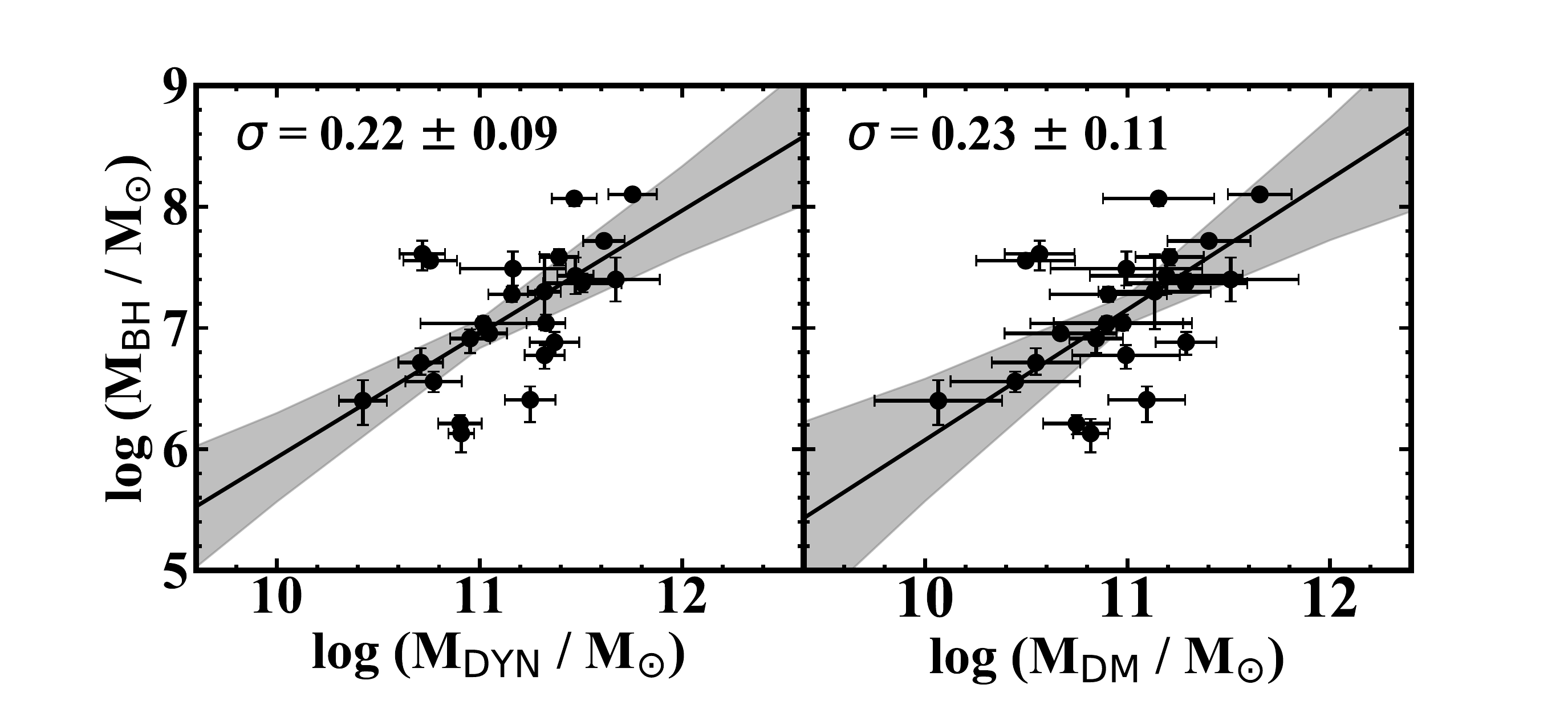}
\caption{$M_{\textsc{BH}}$ as a function of $M_{\textsc{DYN}}$ (left) and $M_{\textsc{DM}}$ (right). The best fits are displayed as solid, black lines in both panels. Uncertainties on the fits are displayed with gray shading around the best-fit lines. $M_{\textsc{DYN}}$ values are calculated using $W_{mx}^{i}/2$ as the maximum rotation rate and $R_{\textsc{HI}}$ as the enclosing radius. $M_{\textsc{DM}}$ is calculated as $M_{\textsc{DYN}} - M_{\textsc{BARY}}$. $M_{\textsc{BARY}}$ values were updated from those calculated in Paper I with the adopted distances in this work (see Sec.\ \ref{sec:masses}).} \label{dyn_dm}
\end{figure*}

\subsection{The Baryonic Tully-Fisher Relation}\label{btfr_discussion}
The past two decades have revealed that the most fundamental form of the TF relation is the BTF relation, which shows a tightly-correlated linear relation between rotational velocity and total baryonic mass over 5 decades of mass \citep{mcgaugh2000,mcgaugh2005,lelli2015,iorio2017}. The mass contribution of gas in massive galaxies is small, therefore the BTF relation is equivalent to the classic TF relation on the high mass end. As calculations of both gas and stellar masses, which constitute $M_{\textsc{BARY}}$, rely on distance, it is therefore of interest to compare our distances to those predicted by the BTF relation. Though some deviations from the relation have been observed, such as HI massive galaxies ($M_{\textsc{HI}} \gtrsim 10^{10} M_{\odot}$; \citealt{courtois2015}) and HI-rich ultra diffuse galaxies \citep{mancera2019}, the distances predicted by the BTF relation here are expected to be largely similar to those predicted by the TF relation, as the typical gas-to-stellar mass fraction for this sample is $\sim$13\%. However, discrepancies may surface for a few galaxies with higher gas-to-stellar mass fractions (e.g., Mrk 1044, NGC 2617, Mrk 1310, Ark 120).
 
The BTF relation has been recently calibrated for several different definitions of rotational velocity measurement. We employ the calibration for our adopted velocity definition of W$_{m50}$ from \cite{lelli2019}, which has a slope of $3.62 \pm 0.09$ and an intercept of $2.33 \pm 0.20$. This agrees well with the examination of the BTF relation by \cite{Zaritsky2014} using the Spitzer Survey of Stellar Structure in Galaxies \citep{Sheth2010}, which found a slope of $3.5 \pm 0.2$. We employ the HI fluxes given in Table \ref{characteristics} for the HI mass estimates, with the scaling factor to convert HI mass to total gas mass of 1.33 for consistency with the BTF definition. We note that we use a scale factor of 1.33 here, rather than the value of 1.4 that we employ throughout the rest of this work, only to ensure that we calculate values in the same way as they were calculated in the calibration of the relation. Additionally, we note that the BTF relation employs the smaller scale factor of 1.33 as the helium contribution (which the scale factor accounts for) is assumed to be lower for the HI-rich calibrating sample of the BTF relation (\citealt{mcgaugh2012} and references therein). We use the stellar mass-to-light ratios of \cite{bd2001} to estimate $M_{\star}$. Comparison between the BTF distances and TF distances is shown in Fig.\ \ref{btfr_comp}. A line of unity is drawn, and we find generally good agreement between the TF distances and those predicted by the BTF relation.

The points in Fig.\ \ref{btfr_comp} are labeled by HI emission line S/N, where profiles of S/N $\geq$ 10 are black circles, 5 $<$ S/N $\leq$ 10 are blue squares, and S/N $\leq$ 5 are red diamonds. Mrk 290 and SBS1116+583A lie below the unity line, along with Ark 120 and Mrk 817, which lie below the line but agree within the uncertainties. The HI emission lines of these outliers all have a S/N $<$ 5, which could result in underpredicting the flux and may be related to the cause of the discrepancies. 

\subsection{Dynamical Mass - Black Hole Mass and Dark Matter Mass - Black Hole Mass Relations}\label{dyn_dm_fits}
As described in Sec.\ \ref{sec:masses}, we utilize $R_{\textsc{HI}}$ as the enclosing radius to estimate $M_{\textsc{DYN}}$ for the galaxies in this sample. Here, we have examined the relationship between black hole mass and the total mass enclosed within the HI radius. The left panel of Fig.\ \ref{dyn_dm} displays a clear trend between $M_{\textsc{BH}}$ and $M_{\textsc{DYN}}$, with the best fit given by
\begin{equation}
\text{log}\frac{M_{BH}}{M_{\odot}} = ( 1.02 \pm 0.35 )\text{log}\bigg(\frac{M_{DYN}}{10^{12}M_{\odot}}\bigg) + ( 6.95 \pm 0.12 )
\end{equation}
with a scatter of ($0.22 \pm 0.09$).

Using the difference between the total enclosed mass and the luminous mass, we have also explored the relationship between $M_{\textsc{BH}}$ and dark matter mass. We plot $M_{\textsc{BH}}$ vs $M_{\textsc{DM}}$ in the right panel of Fig.\ \ref{dyn_dm} and find a weaker, but still significant, correlation.  The best fit to the relation is given by
\begin{equation}
\text{log}\frac{M_{BH}}{M_{\odot}} = ( 1.08 \pm 0.49 )\text{log}\bigg(\frac{M_{DM}}{10^{12}M_{\odot}}\bigg) + ( 7.15 \pm 0.12 )
\end{equation}
with a scatter of ($0.23 \pm 0.11$). 

The average $M_{\textsc{DM}}$/$M_{\textsc{DYN}}$ fraction is $62 \pm 12\%$. However, we note that we are relying on the extent of the HI disk to measure $M_{\textsc{DYN}}$ (and consequently $M_{\textsc{DM}}$). The dark matter halo (DMH) is known to extend far beyond the visible radius, and thus we are probing only a fraction of the mass associated with each galaxy.

\subsection{Halo Mass - Black Hole Mass Relation}\label{halo_relation}

We have also attempted to estimate the total enclosed mass within the halo radius, or $M_{\textsc{HALO}}$. This includes estimates of the halo radius in relation to the HI radius and assumptions of the disk velocity at the halo radius ($V_{\textsc{HALO}}$).

\cite{kravtsov2013} conducted a study relating $R_{200}$ to several galaxy radius definitions. $R_{200}$ is commonly treated as the radius of the DMH, and is the radius which encloses 200 times the critical density of the universe ($\rho_{\textsc{cr}}(z)$). Halo radius hereafter is assumed to be equivalent to $R_{200}$. Through abundance matching of halo mass functions \citep{Tinker2008,Klypin2011} and stellar mass functions \citep{bernardi2010,Papastergis2012}, they first defined a relationship between stellar mass and the halo mass enclosed within $R_{200}$, or $M_{200}$. $M_{200}$ was then estimated from $M_{\star} - M_{200}$ for a sample of galaxies with measured stellar masses that span 8 decades in $M_{\star}$ and all morphological types \citep{mh2011,leroy2008,zhang2012}. $R_{200}$ was then estimated from $M_{200}$, where $M_{200} = (4\pi/3)200\rho_{\textsc{cr}}(z)R^{3}_{200}$. \cite{kravtsov2013} found $R_{25}$ = $0.048R_{200}$, which when combined with $R_{\textsc{HI}}$ = ($1.70 \pm 0.16)R_{25}$ \citep{br1997} yields ${R_{200} \sim (12.3 \pm 1.2)R_{\textsc{HI}}}$. 

Additionally, \cite{lapi2018} derived global galaxy properties such as $M_{200}$ and $R_{200}$ for a sample of 546 nearby late-type galaxies \citep{ps1995} by constructing templates of the rotation curve compilations of \cite{Persic1996}, \cite{Catinella2006}, and \cite{Yegorova2011}. Templates were derived as a function of $I-$band luminosity. By modeling a DMH profile \citep{Burkert1995}, $R_{200}$ was derived by extrapolating the rotation curve to where the halo density reached $\rho_{\textsc{cr}}(z)$. They then explored the relation between $R_{200}$ and the effective galaxy stellar radius $R_{e}$ (assumed to be equivalent to 1.68$R_{d}$). While they quote a polynomial form for the relationship, it is nearly linear within the sizeable scatter, except for a break in the trend towards smaller $R_{200}$ at $R_{e} \sim 4$ kpc. Their best fit (approximated using solely the linear component) is log$R_{e} = 0.73\log R_{200} - 0.91$, which yields $R_{200} \sim (31.1 \pm 1.6)R_{e}$. Using their formula ${R_{e} \sim 1.68R_{d}}$ and the relation $R_{\textsc{HI}} \sim (5.4 \pm 0.5)R_{d}$ \citep{Catinella2006,deBlok2014} yields ${R_{200} \sim (9.7 \pm 1.0)R_{\textsc{HI}}}$. Given the numerous approximations in both studies, we adopt ${R_{200} \sim (11 \pm 1) R_{\textsc{HI}}}$, the average of the results of \cite{kravtsov2013} and \cite{lapi2018}. 

In regards to $V_{\textsc{HALO}}$, if the DMH is assumed to have a constant density profile, it follows that the rotation curve would be flat out to $R_{200}$, thus $V_{\textsc{HALO}} \sim W_{mx}^{i}/2$. This assumption is also adopted in the first derivation of a $M_{\textsc{BH}} - M_{\textsc{HALO}}$ relation by \cite{ferrarese2002}. Though for any density profile other than a constant, the disk circular velocity would decrease out to $R_{200}$, in which case $W_{mx}^{i}/2$ would be an upper limit to $V_{\textsc{HALO}}$ (as noted by \citealt{ferrarese2002}). The effect of the concentration parameter (the ratio of $R_{200}$ to a characteristic inner radius) of the DMH on disk circular velocity was explored in the $\Lambda$CDM simulation of \cite{Bullock2001}. The median of concentration parameters in the simulation showed an approximately flat rotation curve out to $\sim$40 kpc. Additionally, The HI Nearby Galaxy Survey \citep{walter2008} has measured flat rotation curves out to a maximum of $\sim$50 kpc, compared to the median $R_{200}$ estimate of \cite{lapi2018} of $\sim150$ kpc. Therefore, we assume a flat rotation curve out to $R_{200}$, and as follows $V_{\textsc{HALO}} \sim W_{mx}^{i}/2$

\begin{figure}
\includegraphics[scale=0.57,trim={1.6cm 0.5cm 0cm 1.7cm},clip]{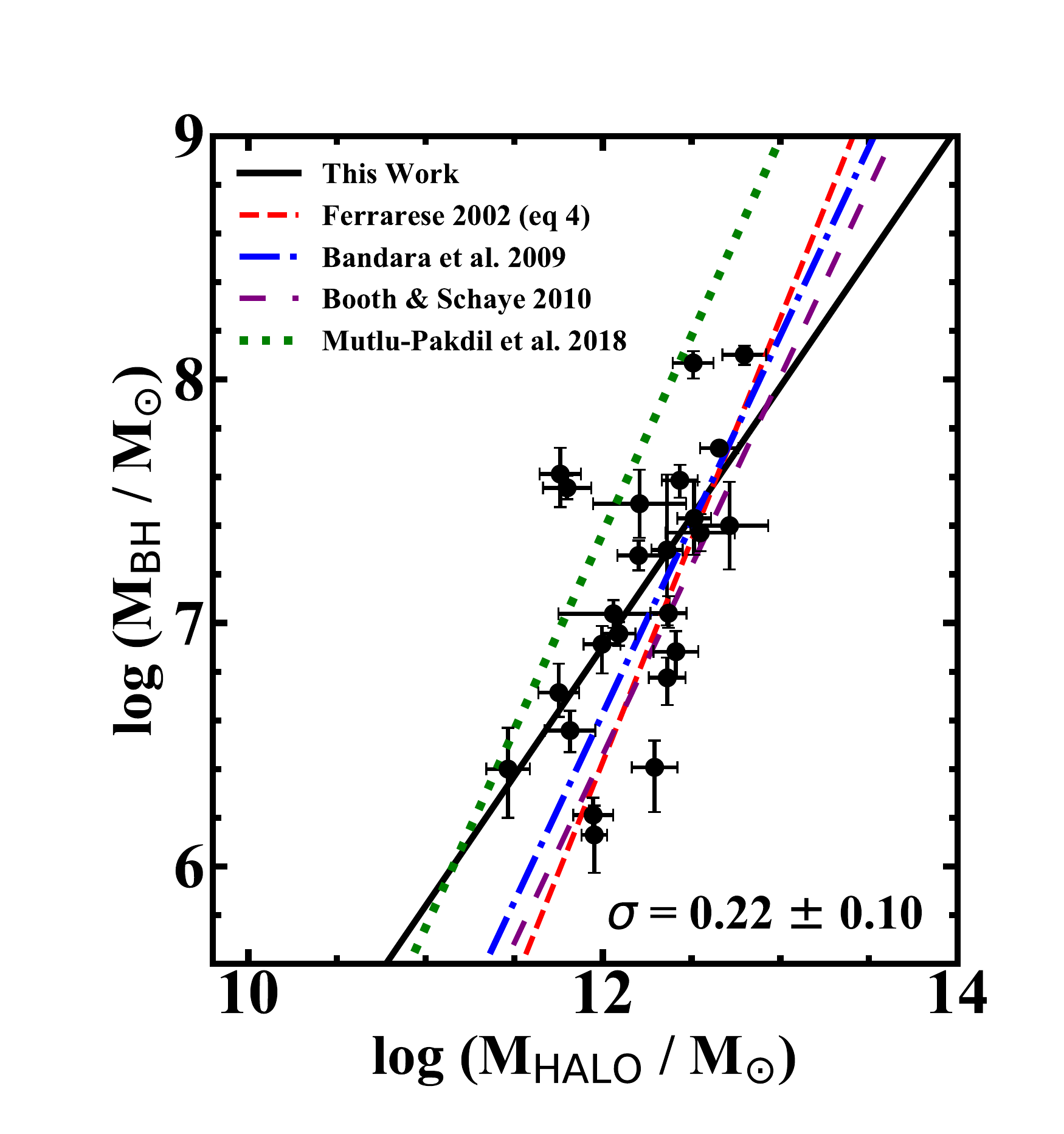}
\caption{$M_{\textsc{BH}}$ vs estimates of $M_{\textsc{HALO}}$. We first use the average results of the relationships between observed radii and halo radii from \cite{kravtsov2013} and \cite{lapi2018} to scale the $R_{\textsc{HI}}$ values by ($11 \pm 1$) to estimate $R_{200}$. We then assume a flat rotation curve out to $R_{\textsc{200}}$ ($V_{\textsc{HALO}} \sim W_{mx}^{i}/2$) to arrive at an approximation of $M_{\textsc{HALO}}$. The solid black line is the formal fit to the data. The red dashed line is the first result from \cite{ferrarese2002}, using the same assumption of a flat rotation curve. The blue dot-dashed line is the relationship of $M_{\textsc{BH}}$ to total gravitational mass from \cite{bandara2009}, which utilized gravitational lens modeling to constrain $M_{\textsc{HALO}}$. The purple dash-spaced line and green dotted line are the $M_{\textsc{BH}}$ $-$ $M_{\textsc{HALO}}$ relations from the simulations of \cite{BS2010} and \cite{mutlu2018}, respectively.} \label{dm}
\end{figure}

We thus calculate the total enclosed mass $M_{\textsc{HALO}}$ with the same formalism used to estimate $M_{\textsc{DYN}}$. In Fig.\ \ref{dm} we explore the relationship between $M_{\textsc{BH}}$ and $M_{\textsc{HALO}}$. The best fit, which is displayed with the solid line, is 
\begin{equation}
\text{log}\frac{M_{BH}}{M_{\odot}} = ( 1.07 \pm 0.37 )\text{log}\bigg(\frac{M_{HALO}}{10^{13}M_{\odot}}\bigg) + ( 7.97 \pm 0.31 )
\end{equation}
with a scatter of ($0.22 \pm 0.10$)\,dex. The average fraction of dark matter within $R_{200}$ is $97 \pm 1$\%. The typical $M_{\textsc{BH}}/M_{\textsc{HALO}}$ fraction is 10$^{-5}$, but shows a trend with less massive black holes making up a smaller fraction of the total mass of the system, similar to what was found for $M_{\textsc{BH}}/M_{\star}$ by \cite{misty2018}.

To place the $M_{\textsc{HALO}}$ estimates in context with other methods used to derive total enclosed galaxy mass, we first compare to the $M_{\textsc{HALO}}$ estimates used to construct the globular cluster system mass $-$ galaxy halo mass relation \citep{sf2009}. In that study, halo masses were estimated using $M_{\star} - M_{\textsc{HALO}}$ relations from weak gravitational lensing results \citep{Hoekstra2005,Mandelbaum2006}. The majority of the $M_{\textsc{HALO}}$ estimates found by \cite{sf2009} fell within the range of $\sim 10^{10} - 10^{13}$ M$_{\odot}$ for a sample consisting of dE, E, S0, and S-type galaxies. Additionally, the E-MOSAICS simulation \citep{Bastian2020} of the globular cluster system mass $-$ galaxy halo mass relation found that the majority of $M_{\textsc{HALO}}$ constraints ranged from $\sim 10^{11} - 10^{13}$ M$_{\odot}$. Comparatively, our sample of halo estimates span the right orders of magnitude for galaxies of similar morphological type.

We have compared our fit to several other estimates of the $M_{\textsc{BH}} - M_{\textsc{HALO}}$ relation. \cite{ferrarese2002} utilized $\sigma_{\star}$ measurements with the $M_{\textsc{BH}} - \sigma_{\star}$ relation to constrain SMBH masses which spanned $\sim$10$^{6} - 10^{9}$\,M$_{\odot}$, along with several methods of estimating total gravitational mass, which fell within the range of $\sim$10$^{11} - 10^{14}$\,M$_{\odot}$. We plot the first derivation (equation 4 in \citealt{ferrarese2002}), which assumes $V_{\textsc{HALO}} \sim W_{mx}^{i}/2$, as the red dashed line in Fig.\ \ref{dm}. We find a shallower slope than \cite{ferrarese2002}, even if we refit their relationship with their sample restricted to the same mass ranges we find. However, we employ direct measurements of $M_{\textsc{BH}}$ while those in  \cite{ferrarese2002} were estimated from the $M_{\textsc{BH}} - \sigma_\star$ relation. Additionally, we constrain $W_{mx}^{i}$ from unresolved HI line widths, while \cite{ferrarese2002} utilized $V_{mx}$ measurements from rotation curves. The differences in both measurements are potential sources of discrepancy in the $M_{\textsc{BH}} - M_{\textsc{HALO}}$ relations. \cite{ferrarese2002} describes two other methods of estimating $M_{\textsc{HALO}}$, however both assume nonconstant halo density profiles, and as such predict less massive DMHs and intercepts $\sim$0.5\,dex and $\sim$1.5\,dex smaller than what we find, respectively.

We also compare to the observationally-constrained relation from \cite{bandara2009}. Their $M_{\textsc{BH}}$ values were derived using $\sigma_{\star}$ measurements and the $M_{\textsc{BH}} - \sigma_{\star}$ relation, and total galaxy mass was estimated from strong gravitational lens modeling \citep{bolton2008}. We plot their relation as the blue dot-dashed line in Fig.\ \ref{dm}. We again find a slightly shallower slope in comparison (though the fits of our relation and \citealt{bandara2009} are statistically equivalent within the uncertainties). However, their study probes only the high mass end of both black hole and halo mass, with their sample spanning $\sim$10$^{8} - 10^{9}$ M$_{\odot}$ in $M_{\textsc{BH}}$ and $\sim$10$^{13} - 10^{14}$ M$_{\odot}$ in $M_{\textsc{HALO}}$.

Lastly, we have compared our results to those of large, hydrodynamical simulations. \cite{BS2010} explored correlations between $M_{\textsc{BH}} -  M_{\star}$ and $M_{\textsc{BH}} - M_{\textsc{HALO}}$, which they define as the mass enclosed within a sphere of a mean density of 200$\rho_{\textsc{cr}}(z)$. Their result is plotted as the purple dash-spaced line in Fig.\ \ref{dm}. The Illustris simulation also explored $M_{\textsc{BH}} - M_{\textsc{HALO}}$ (where the definition of $M_{\textsc{HALO}}$ is the same as \citealt{BS2010}), and we plot the result of \cite{mutlu2018} as the green dotted line. The intercepts of their relations differ by a decade, most likely due to the difference in DMH density profiles between the simulations. \cite{Pillepich2014} reported that the halo density profile for the Illustris galaxies is well characterized by a negative power law, which would result in less massive haloes due to the quicker drop-off in density. Interestingly, galaxies in the the upgraded IllustrisTNG simulation (\citealt{Lovell2018}, which incorporated a larger volume, higher resolution, and new physics such as black hole-driven winds) are a much better match to observations, with flat rotation curves out to large radii ($\sim$60 kpc for $M_{200} = 10^{13}$ M$_{\odot}$). While the $M_{\textsc{BH}} - M_{\textsc{HALO}}$ relationship has not yet been reexamined for IllustrisTNG, the flatter rotation curves will result in a larger enclosed mass within $R_{200}$ and may provide a better match to observationally-constrained relationships such as the one we present here.

\section{Summary} \label{summary}
We presented TF distance measurements for 24 AGN host galaxies with direct $M_{\textsc{BH}}$ measurements from reverberation-mapping. We also presented the first calibration of the $V-$band TF relation. These are the first redshift independent distance determinations for 14 active galaxies. We utilized $HST$ $V-$band and ground-based $B$, $V$, $R$, $I$, and $H-$band images to constrain multiple distance measurements for 19 galaxies. Removal of the bright nucleus via surface brightness modeling has allowed measurements of the distance moduli free of AGN contamination for the first time. Generally close agreement was found between distances derived from different photometric bandpasses within the uncertainties, and we found good agreement between the TF distances and those predicted by the BTF relation.

We collected available distance measurements from Cepheids (NGC 4051, NGC 4151, NGC 6814), SBF (NGC 3227), and SN1a (NGC 7469), $z$-based distances for 6 galaxies for which we deem the TF distances as uncertain (MCG+08-11-011, Mrk 374, Mrk 817, Mrk 478, NGC 5940, Zw 229-015), and the remaining 13 TF distances for estimates of galaxy dynamical and dark matter masses. We combined these with W$_{mx}^{i}/2$ and estimates of the HI radius to derive $M_{\textsc{DYN}}$, which when compared to $M_{\textsc{BARY}}$ allowed constraints on $M_{\textsc{DM}}$. The typical fraction for our sample is $M_{\textsc{DM}}$/$M_{\textsc{DYN}}$ = 62\%.

We also explored the relationships between $M_{\textsc{BH}}$ and constraints on $M_{\textsc{DYN}}$, $M_{\textsc{DM}}$, and $M_{\textsc{HALO}}$. We found significant correlations between $M_{\textsc{BH}} - M_{\textsc{DYN}}$ and $M_{\textsc{BH}} - M_{\textsc{DM}}$. To approximate $M_{\textsc{HALO}}$, we assumed $R_{200} \sim (11 \pm 1)R_{\textsc{HI}}$ and a flat rotation curve out to $R_{200}$. $M_{\textsc{HALO}}$ estimates for this sample showed good agreement with the range of halo masses found by other observationally based methods and simulations \citep{sf2009,Bastian2020}. Over the range of $M_{\textsc{BH}}$ and $M_{\textsc{HALO}}$ probed by this sample, we found good agreement between our $M_{\textsc{BH}} - M_{\textsc{HALO}}$ relation, the observationally-constrained relations of \cite{ferrarese2002} and \cite{bandara2009}, and the hydrodynamical simulations of \cite{BS2010} and \cite{mutlu2018}.

\acknowledgements
We thank the referee for helpful suggestions that improved the clarity of this paper. M.C.B gratefully acknowledges support from the NSF through CAREER grant AST-1253702 and grant AST-2009230. H.M.C. acknowledges support by the CNES and Institut Universitaire de France. We thank Mitchell Revalski for advisement of drizzling techniques for archival $HST$ data. We thank Wesley Peters for helpful observing strategies with the APO ARC telescope. We also thank Joanne Hughes for obtaining sky flat observations for the ARCTIC instrument.

This work is based on observations obtained at the MDM Observatory, operated by Dartmouth College, Columbia University, Ohio State University, Ohio University, and the University of Michigan.

This work is based on observations obtained with the Apache Point Observatory 3.5\,m telescope, which is owned and operated by the Astrophysical Research Consortium, and with Apache Point Observatory's 0.5\,m Astrophysical Research Consortium Small Aperture Telescope.

This research has been supported by RECONS (\url{www.recons.org}) members Michele Silverstein, Todd Henry, and Wei-Chun Jao, who provided data as part of the long-term astrometry/photometry program at the CTIO/SMARTS 0.9\,m, which is operated as part of the SMARTS Consortium.

This research has made use of the NASA/IPAC Extragalactic Database (NED), which is operated by the Jet Propulsion Laboratory, California Institute of Technology, under contract with the National Aeronautics and Space Administration.

Funding for SDSS-III has been provided by the Alfred P. Sloan Foundation, the Participating Institutions, the National Science Foundation, and the U.S. Department of Energy Office of Science. The SDSS-III web site is http://www.sdss3.org/.

\section*{Appendix}\label{Appendix}
In Sec.\ \ref{vband_cal}, we presented the derivations of the inclination and redshift corrections for the $V-$band TF calibration. Here, we describe those derivations in detail.
\appendices
\begin{figure}
\includegraphics[trim={2.1cm 13.5cm 0cm 3.5cm},clip,scale=0.73]{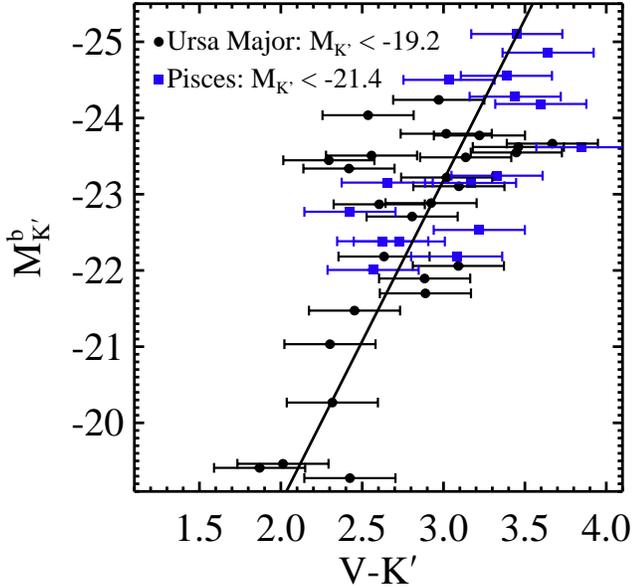}
\caption{Color-magnitude diagram for the Ursa Major and Pisces clusters
used for the analysis of the inclination-dependent extinction correction for our TF V-band calibration. Ursa major galaxies are displayed as black circles,
Pisces galaxies as blue squares.The Ursa Major cluster data is available in \cite{tully1996}, and the Pisces cluster in \cite{tp2000}. The $V-$band magnitudes were retrieved from RC3. The M$_{\textsc{K}'}$ values were derived using the same distance modulii in the original analysis of \cite{tully1998}, 31.33 for Ursa Major and 33.88 for Pisces. The cuts to M$_{\textsc{K}'}$ for each cluster are also consistent with their analysis. $V$ and $K'-$band magnitude uncertainties are assumed to be 0.2\,mag (see \nameref{Appendix}). The line is a linear regression with uncertainties in the color.} \label{cm_diagram}
\end{figure}

\begin{figure*}
\includegraphics[trim={3cm 16cm 0cm 6.5cm},clip,scale=1.1]{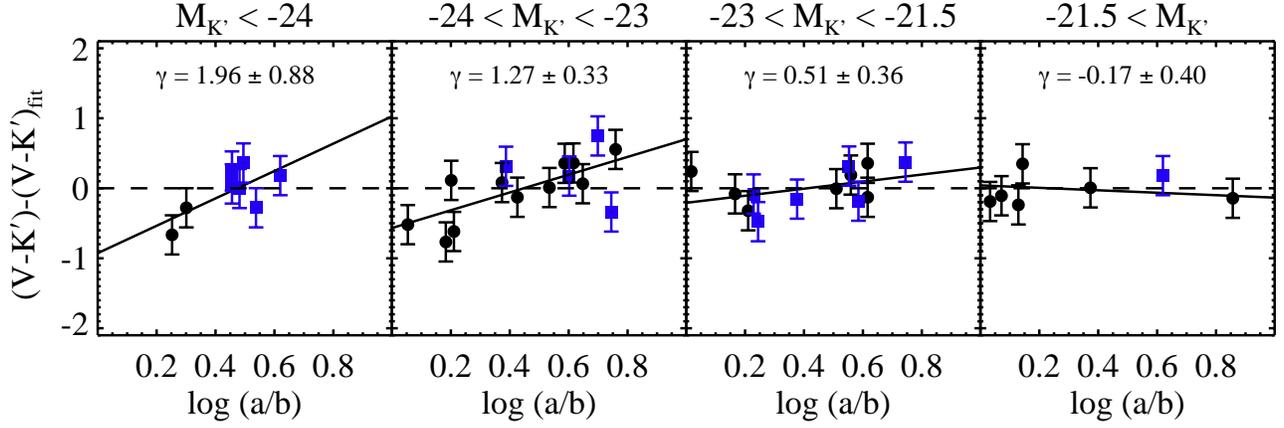}
\caption{Deviations of $V-K'$ from the mean (best-fit line) of the color-magnitude relation of Fig.\ \ref{cm_diagram} as a function of the log of the axis ratio of each galaxy. The centers of the 4 luminosity bins are adjusted from those used by \cite{tully1998} to better evenly separate the data. Black circles are members of the Ursa Major cluster, and blue squares are members of the Pisces cluster. The solid black lines are linear regression solutions to $A_{i}^{\lambda} = \gamma_{\lambda}\log(a/b)$. The lack of data in the lowest luminosity bin resulted in a non-physical negative slope, and we adopt $0.00 \pm 0.40$ for this bin for the remainder of our analysis.}  \label{axrat}
\end{figure*}
\begin{figure}
\includegraphics[trim={2.2cm 13cm 0cm 3cm},clip,scale=0.75]{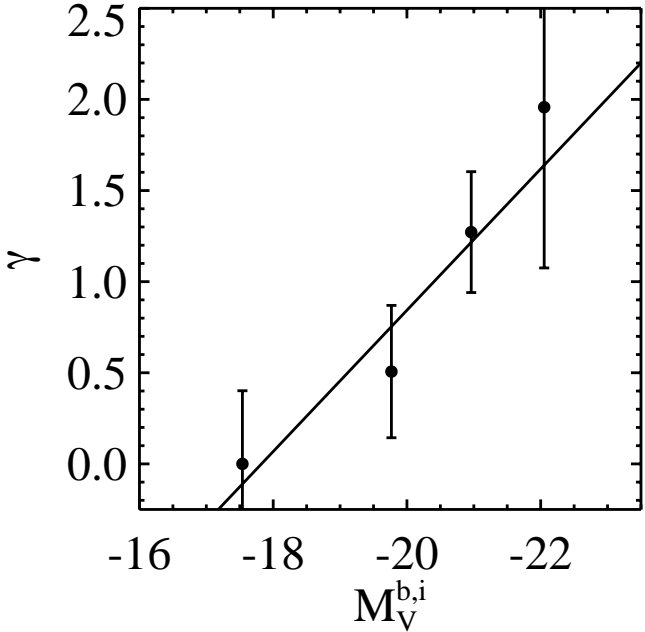}
\caption{Dependency of the $\gamma$ values from the fits displayed in Fig.\ \ref{axrat} of each $K'$ luminosity bin. The 4 points are the median $M_{V}^{b,i}$ values of each bin, and were derived from the distance moduli to each cluster used in the original analysis of \cite{tully1998}. The magnitudes were first corrected for galactic extinction and second for inclination-dependent extinction from each corresponding $K'$ luminosity bin. The solid line is a linear regression with uncertainties in $\gamma$ from the fits in each bin.} \label{gamma}
\end{figure}

\begin{figure} 
\includegraphics[trim={1.8cm 13cm 0cm 3cm},clip,scale=0.7]{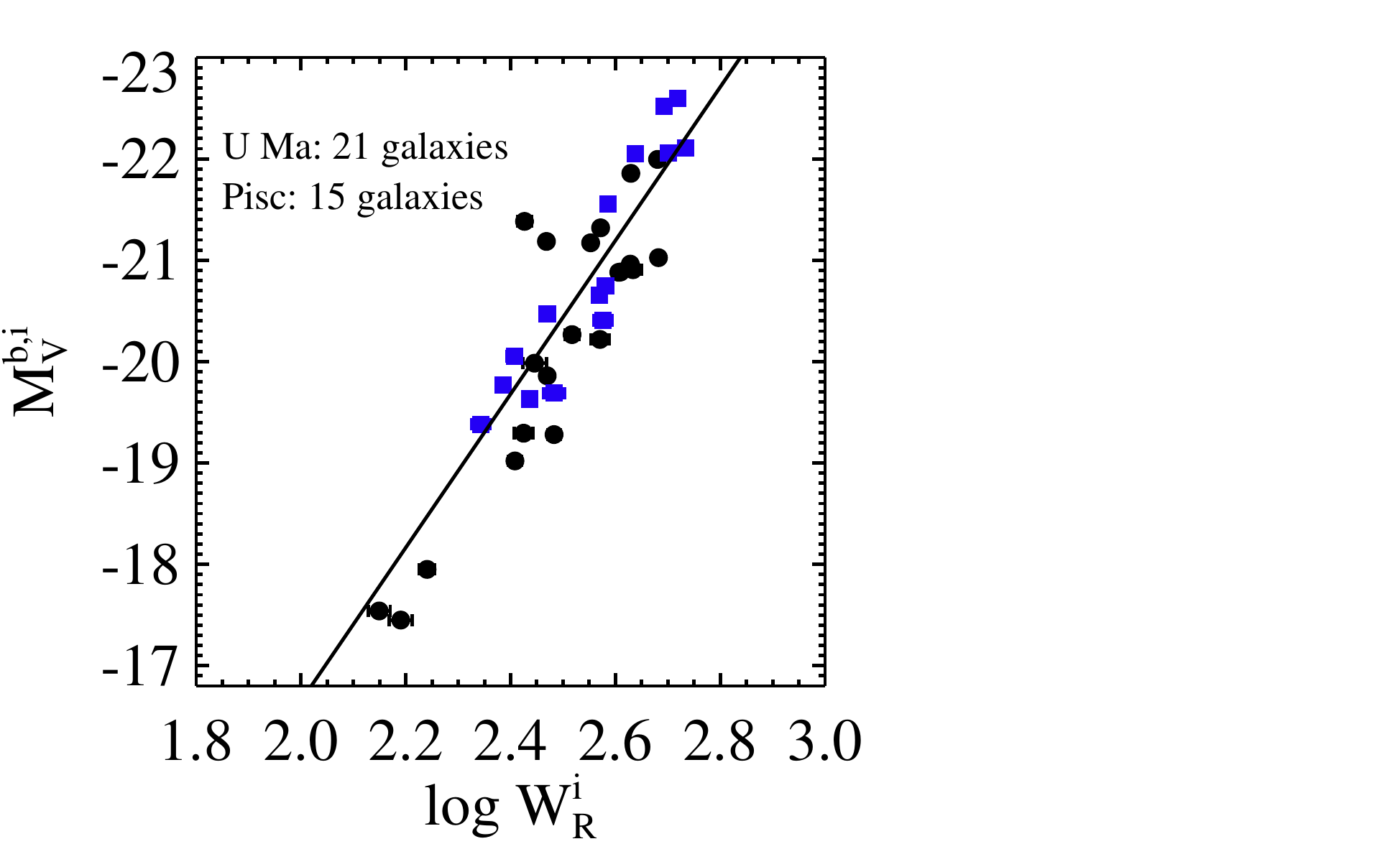}
\caption{The TF relation of the calibrating sample of \cite{tully1998} for the $V-$band inclination-dependent extinction correction. Ursa major galaxies are displayed as black circles, Pisces galaxies as blue squares. The absolute $V-$band magnitudes are corrected for galactic extinction and inclination-dependent extinction using the best-fit solution for $A_{i}^{\lambda} = \gamma_{\lambda}\log(a/b)$ in Fig.\ \ref{gamma}. The former width parameter $W_{\textsc{R}}$ is used here. The solid line is the best fit to the data, with uncertainties in the HI line widths from \cite{tully1998}.}\label{final_v}
\end{figure}
To constrain the inclination-dependent correction for the $V$-band TF calibration, we have followed the same procedures described by \cite{tully1998}. Their analysis includes $B$, $R$, $I$, and $K'$ photometric data from 2 nearby galaxy clusters. The first is the Ursa Major cluster, whose data are available from \cite{tully1996}. The second is the Pisces Cluster, whose data were used in the analysis of \cite{tully1998}, but were not made publicly available until the work of \cite{tp2000}. Two tests of extinction as a function of galaxy inclination were conducted by Tully et al., the first and most sensitive of which describes constructing color-magnitude diagrams of $B$, $R$, and $I$ passbands relative to the $K'$ passband. Reddening effects in $K'$ are small, therefore the extinction effects on the galaxies are dominant in the color terms. We identified all galaxies in both clusters used in the analysis of \cite{tully1998}, and collected available $V-$band magnitudes from RC3, as the adopted $B-$band magnitudes of Tully et al.\ show an almost exact match to those listed in RC3.

\begin{figure*}
\includegraphics[trim={3cm 15.7cm 0cm 5.5cm},clip,scale=1.2]{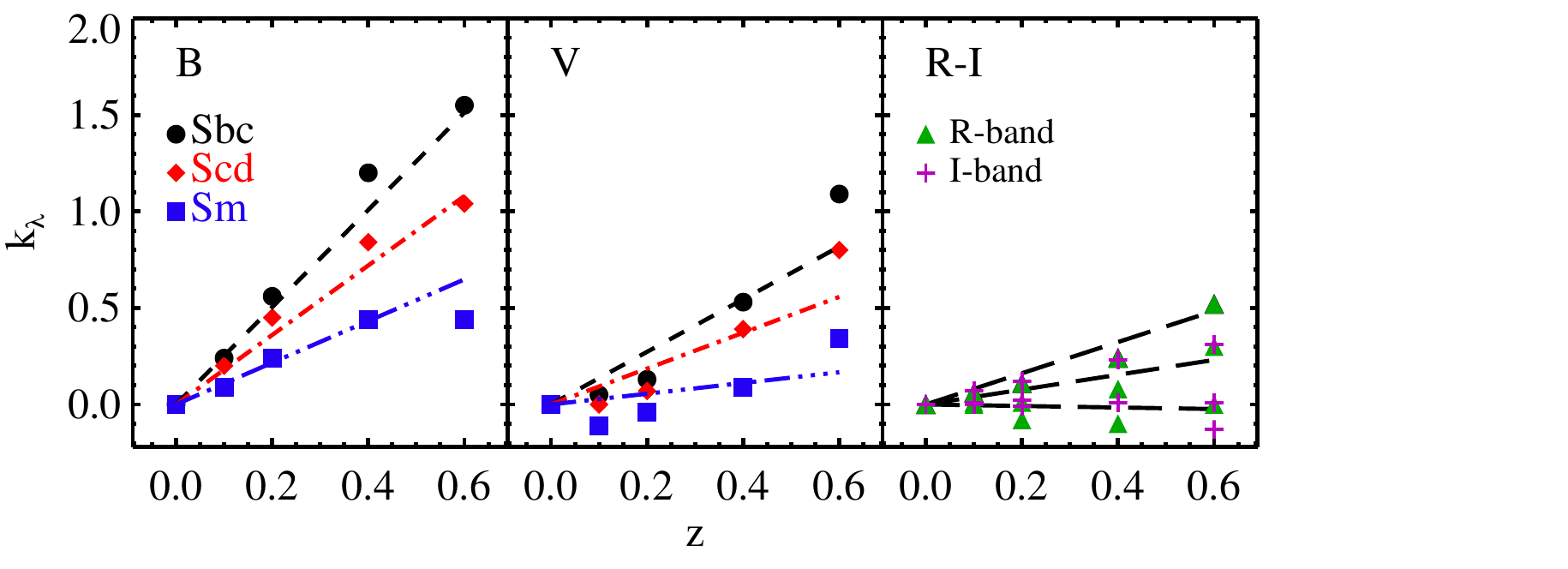}
\caption{$k-$corrections for $B$, $V$, $R$, and $I$ bands. Data were retrieved from \cite{frei1994}. The black (dashed), red (dot-dashed), and blue (dot-dot-dot-dashed) lines in the left panel, and the black double-dashed lines in the right panel, are the $B$ and $R-I$ $k-$correction prescriptions from \cite{tp2000}, respectively. Excellent agreement is found when Hubble type Sbc corresponds to $T$=3, Scd to $T$=5, and Sm to $T$=7. The fits in the right panel assume $R-I$ values of 0.45, 0.35, and 0.25 for the top, middle, and bottom lines. The middle panel displays our best-fits to the $V-$band $k-$corrections (where Sbc, Scd, and Sm correspond to $T$ values of 4, 6, and 9, respectively), constrained such that $k_{V}=0$ at $z=0$.} \label{kb}
\end{figure*}

We adopt the same $K'$ apparent magnitudes and distance moduli reported by \cite{tully1998} for each cluster (31.33 for Ursa Major, 33.88 for Pisces) to derive absolute $K'$ magnitudes. The apparent magnitudes in their analysis include $k-$corrections and Galactic extinction corrections. They detail that the $k-$corrections were extremely small given the proximity of each cluster; specifically, they report $k-$corrections of $\sim$0 mag for the Ursa Major galaxies and $\sim0.03 \pm 0.03$\,mag for the Pisces galaxies at $B$, and even less significant ($\lesssim 0.01$\,mag) in the redder bands. Therefore, we have not considered $k-$corrections for the $V-$band for this portion of our analysis. The Galactic extinction corrections for the $B-$band were drawn from \cite{bh1984}, and Tully et al.\ report extinction ratios between the separate bands, yet do not detail how they arrived at the given ratios. In order to estimate the extinction ratio for the $V$ band, we assumed an average extinction curve of $R = A_{V}/E(B-V) = 3.1$  (\citealt{sm1979}; which also approximates the $R/B$ and $I/B$ extinction ratios listed by \citealt{tully1998} to within $\sim$0.05), and arrived at an extinction ratio of $V/B = 0.76$. 

In Fig.\ \ref{cm_diagram}, we display the absolute $K'$ magnitudes vs $V - K'$ color. Ursa Major galaxies are plotted in black circles, Pisces galaxies in blue squares. We make consistent magnitude cuts ($M_{K'}$ < -19.2 for Ursa Major, $M_{K'}$ < -21.4 for Pisces), which Tully et al.\ detail as the cutoffs for their $K'-$band completeness limits. Several of the lower-luminosity galaxies that were originally included in the \cite{tully1998} analysis do not have $V-$band magnitudes available in RC3, so the low-luminosity end of the color-magnitude diagram is only sparsely populated. Tully et al.\ describe their linear regression to the color-magnitude diagram with uncertainties in the color, yet no uncertainties are provided for the magnitudes. We find that we can reasonably reproduce the results of Tully et al.\ for the $B$ band when a typical uncertainty of 0.2\,mag is assumed for the photometry. Thus, we adopt 0.2\,mag uncertainty for the $V$ and $K'$ magnitudes, and find a best fit of $M_{K'}^{b} = (-4.23 \pm 0.56)(V-K') - (10.50 \pm 1.67)$.

The next portion of the analysis assumes that deviations from the mean of the color-magnitude relation (given by the best-fit line in Fig.\ \ref{cm_diagram}) are dominated by inclination-dependent obscuration. Tully et al.\ found that the extinction dependence on inclination is also a function of luminosity, where brighter galaxies have a much higher dependence than fainter galaxies. To quantify the luminosity dependencies for the $B-K'$, $R-K'$, and $I-K'$ colors, they separated the galaxies into 4 luminosity bins and plotted the deviations as a function of disk axis ratio. In Fig.\ \ref{axrat}, we show the deviations from the $V-K'$ fit as a function of axis ratio.  Following Tully et al., we have split the sample into four bins, but with slightly different bin centers that account for the small number of galaxies at the lowest luminosities.

The extinction parameter is often described as ${A_{i}^{\lambda} = \gamma_{\lambda}\log(a/b)}$, where $\lambda$ is the passband and $a/b$ is the ratio of major to minor axes. The solid black lines in Fig.\ \ref{axrat} are the best fits, where the slope is $\gamma$ and where $A_{i}^{V}$ is assumed to be $(V-K')-(V-K')_{fit}$. We find smaller deviations from the mean color at lower galaxy luminosities, in agreement with the findings of Tully et al. In the lowest luminosity bin, we find a best fit with a negative slope, which is nonphysical, but is also formally consistent with zero. For that bin, we therefore adopt a slope of $0.00 \pm 0.40$. In Fig.\ \ref{gamma} we display the best-fit slope for each bin versus the absolute $V-$band magnitude associated with the median luminosity of each bin, corrected for inclination and Galactic extinction. We assume a linear function for the magnitude dependence of the inclination-dependent extinction correction, and find $\gamma_{V} = (-0.39 \pm 0.14)M_{V}^{b,i} - (6.91 \pm 2.79)$. We find a negligible change if we instead employ the median $M_{K'}$ of each bin and adopt the $(V-K')$ value of the fit in Fig.\ \ref{cm_diagram} to predict $M_{V}^{b,i}$. With this method, we find $\gamma_{V} (-0.36 \pm 0.13)M_{V}^{b,i} - (6.31 \pm 2.60)$, where the slopes are nearly identical and the intercept has shifted slightly, but is well within the uncertainties.

The last step in the analysis of \cite{tully1998} involves applying the inclination corrections to the absolute magnitudes in order to re-derive the luminosity-HI line width relation. The definition of the line width parameter used here and in \cite{tully1998} predates the newer definition of \cite{courtois2009}, so we denote this previous version as $W_{\textsc{R}}^{i}$ for consistency. We adopt the same line widths and uncertainties from \cite{tully1998}. We apply our derived inclination corrections to the $V-$band absolute magnitudes of the galaxies in the Pisces and Ursa Major clusters, and plot them against the corresponding $W_{\textsc{R}}^{i}$ values in Fig.\ \ref{final_v}. The solid line is the best fit, ${M_{V}^{b,i} = (-20.44 \pm 0.01) - (7.57 \pm 0.01)(\log W_{\textsc{R}}^{i} - 2.5)}$. Lastly, our equations for $\gamma_{V}$ and $M_{V}^{b,i}$ can now be used to derive $\gamma_{V}$ solely in terms of the distance-independent variable $W_{\textsc{R}}^{i}$, which we find to be ${\gamma_{V} = (1.01 \pm 4.06) + (2.94 \pm 1.09)(\log W_{\textsc{R}}^{i} - 2.5)}$. 

With the inclination-dependent extinction correction defined, we next turn to the $k-$correction. With no reference for their calibrating data set, we have assumed the data used to constrain the prescriptions of \cite{tp2000} may come from the well-cited work of \cite{frei1994}. Their study reports $k-$corrections in the $B-$band ($k_{B}$) for various colors for galaxies with Hubble types E, Sbc, Scd, and Sm that are located between $0<z<0.6$. When used with their prescription ${k_{B}(z) = B(z)-B(z=0)-2.5\log(1+z)}$, the $k-$correction in any filter can be derived by taking the difference between $B$ and another magnitude, using the corresponding color information they provide. We tested our assumption by comparing the data from \cite{frei1994} to the prescriptions of \cite{tp2000}. The results are shown in Fig.\ \ref{kb}. The left panel contains the $k-$corrections for the $B$ band, and the right for the $R$ and $I$ bands. The lines in each are best-fits to the data, where the $B$ and $R-I$ fits are from \cite{tp2000}, given as ${A_{k}^{B} = (3.6 - 0.36T)z}$ and ${A_{k}^{R} = [4.24(R - I) - 1.10]z}$, respectively. We found excellent agreement for the $B-$band data, however only if we shift each Hubble type $T$ to a lower integer. For example, Hubble type Sbc corresponds to $T=4$, yet we only find a satisfactory fit to the Sbc $k-$corrections if $T=3$ is used. The same is true for Scd, where we find a matching fit if $T=5$ is used, as opposed to $T=6$. For the Sm data, which normally corresponds to $T=9$, we find a good fit if $T=7$ is used. For the $R$ and $I$ $k-$corrections, \cite{tp2000} prescribe one fit to characterize both sets of data. We find good agreement with the data from \cite{frei1994} when $R-I$ values of 0.45, 0.35, and 0.25 are adopted for Sbc, Scd, and Sm types.

Given that our assumption of \cite{tully1998} basing their work on the data of \cite{frei1994} seems to hold, we therefore determined the $k-$correction in a similar way for the $V$ band, and display it in the center panel of Fig.\ \ref{kb}. The corrections in the $B$ band are highly dependent on the morphological classification, while the $R$ and $I$ bands have much less of a spread as a function of morphology. This is reflected in the formalisms of each correction, as $T$ is a factor in the $B-$band calibration yet is absent in the $R-I$ calibration. While the $V$ band has less of a dependence on morphology than the $B$ band, we nevertheless sought to arrive at a morphologically-dependent calibration given the obvious spread as a function of $T$. We assume, as we did with the $B$ band, that the Sbc, Scd, and Sm classifications correspond to $T =$ 4, 6, and 9 respectively, and that $k_{V}=0$ at $z=0$. We fit the data for the 3 Hubble types, and solved for a universal fit such that $T$ was a scale factor which yielded the 3 solutions, and arrived at $A_{k}^{V} = (2.23 - 0.22T)z$.

\begin{figure}
\includegraphics[trim={2.8cm 11.2cm 0cm 2.3cm},clip,scale=1]{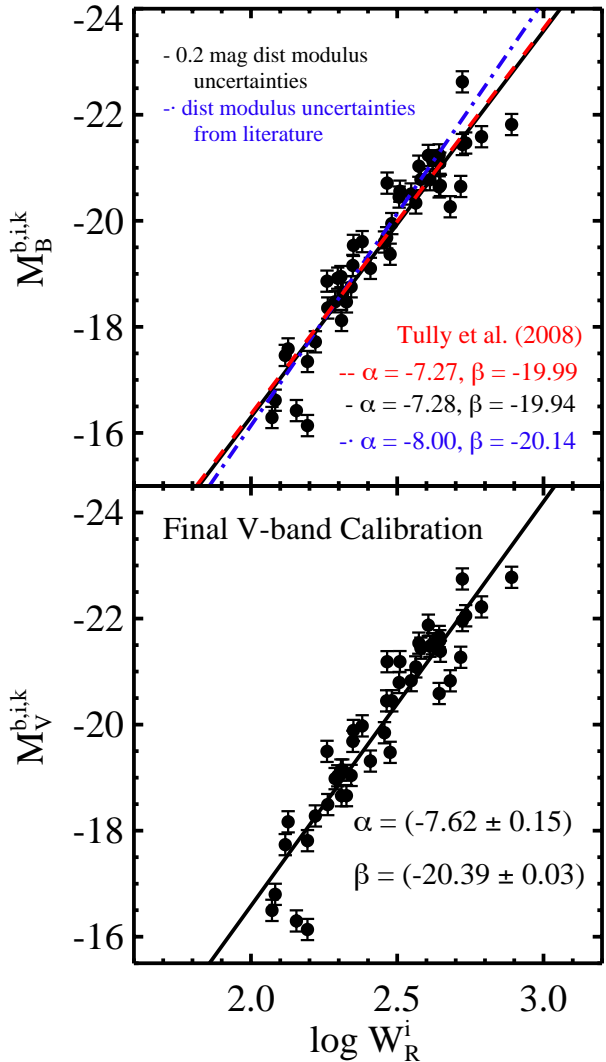}
\caption{TF $B-$band calibration (top) and our final TF $V-$band calibration (bottom). $B-$band magnitudes, widths, and distance moduli were retrieved from \cite{tully2008} for galaxies which had primary distance measurements from either Cepheids, TRGB, and/or SBF. $V-$band magnitudes were retrieved from RC3. The red dashed line in the top panel is the fit reported by \cite{tully2008}. The blue dot-dashed line is our fit using published distance modulus uncertainties, which resulted in a steeper slope than that reported (the slopes and intercepts of which are displayed; the top is the fit given by \cite{tully2008}, the bottom corresponds to the blue dot-dashed fit). Adopting 0.2\,mag uncertainty in the moduli achieves a near perfect match to the fit by \cite{tully2008}, shown as the solid black line and displayed as the middle slope and intercept. The solid black line in the bottom panel is our best-fit to the $V-$band TF calibration using distance moduli with 0.2\,mag uncertainties, Galactic extinction corrections from \cite{schlegel1998}, and our calibrations for the $V-$band inclination correction (see equation \ref{v_inc}, Fig.\ \ref{final_v}) and $k-$correction (see equation \ref{v_k}, Fig.
 \ref{kb}.)}\label{tully_b+vband}
\end{figure}

While the $k-$corrections are derived using galaxies out to $z=0.6$, the TF method is only applicable to galaxies at $z$ $\lesssim$ 0.1 \citep{reyes2011}, and \cite{tp2000} describe the corrections as always < 0.08 mag in the $B$ band. The maximum expected $k-$correction in the $V$ band is 0.2, which would correspond to $T = 1$ at $z=0.1$. This correction is larger than the upper limit reported by \cite{tp2000} for the $B$ band, however $z\lesssim 0.03$ for their sample.

Finally, with the inclination and $k-$corrections in hand, we were able to derive the $V-$band TF relation.  We first identified all galaxies in the calibrating sample of \cite{tully2008} that had distances derived from Cepheids, TRGB, and SBF. From their VizieR table, we retrieved the $B-$band magnitudes, distance moduli, and $W_{\textsc{R}}^{i}$ values. In addition, we retrieved all available $B$ and $V-$band magnitudes from RC3, as well as all original distance moduli and uncertainties from the literature for the calibrating sample. The distances reported by \cite{tully2008} match the original published values and do not appear to have been updated in any way.

We first aimed to recreate the fit to the $B-$band relationship by \cite{tully2008}. Since the difference was negligible, we began by using the originally published values of the distance moduli and uncertainties, with $B-$band magnitudes from RC3. Galactic extinction values were retrieved from \cite{schlegel1998}, and inclination and $k-$corrections were applied with equations \ref{b_inc} and \ref{b_k}. The difference in fits using $B-$band apparent magnitudes from RC3 versus magnitudes from \cite{tully2008} was negligible. When the published uncertainties on the distance measurements were used to convert apparent magnitudes to absolute magnitudes, the fit resulted in a slightly steeper slope than that reported, as shown in the top panel of Fig.\ \ref{tully_b+vband}, where the reported fit from \cite{tully2008} is the red dashed line, and our best fit is the blue dot-dashed line.  \cite{tully2008} 
assert that galaxies with distances from Cepheids, TRGB, or SBF are assumed to have a 0.2\,mag uncertainty in their moduli. If we adopt this, we reproduce the $B-$band fit almost exactly, shown as the solid black line in Fig.\ \ref{tully_b+vband}.

Thus, the $V-$band apparent magnitudes from RC3 for the calibrating sample of \cite{tully2008} were corrected for Galactic extinction using values from \cite{schlegel1998}, and for inclination-dependent extinction and $k-$corrections using equations \ref{v_inc} and \ref{v_k}. The distance moduli were then used to convert the corrected apparent magnitudes to absolute magnitudes, with 0.2\,mag typical uncertainty adopted for the distance moduli. Our best fit to the relationship between $M^{b,i,k}_{V}$ and $W^{i}_{\textsc{R}}$ for the calibrating sample is shown in the bottom panel of Fig.\ \ref{tully_b+vband}, and is given by equation \ref{vband_eq} as ${M_{V}^{b,i,k} = (-20.39 \pm 0.03) - (7.62 \pm 0.15)(\log W_{mx}^{i} - 2.5)}$. 

\bibliography{references}

\end{document}